\documentclass[11pt]{article}
\pdfoutput=1
\usepackage{jheppub}
\usepackage{tabu}
\usepackage[vcentermath]{youngtab}
\usepackage[usenames,dvipsnames,table]{xcolor}
\usepackage{graphicx,amsmath,amssymb,amsthm,multirow,array,bm,bbm,esint}
\usepackage[mathscr]{eucal}
\usepackage[bbgreekl]{mathbbol}
\usepackage{epsf,amsfonts}
\usepackage{slashed}
\usepackage[numbers,sort&compress]{natbib}
\usepackage{minibox}
\usepackage{rotating}
\usepackage{pdflscape}
\usepackage{array,tikz-cd}
\usepackage{subcaption}
\usepackage{framed}
\usepackage{tikz} 
\usepackage[compat=1.1.0]{tikz-feynman}
\usepackage{feynmf}
\usepackage{float}

\usepackage{slashed}

\def \sgn{\mbox{sgn\,}}         

\def \ber{\mbox{Ber}\! }          

\def \ftn{{\sigma}}       

\def \Tr{\textbf{\mbox{Tr\,}}}         
\def \STr{\textbf{\mbox{STr\,}}}         
\def \tr{\mbox{tr\,}}        
\def \bitr{{\mathbb{tr} \,}}         
\def \str{\mbox{str\,}}         

\newcommand{\maPsi}{ {\boldsymbol{\Psi}} }  

\newcommand{\maeta}{ {\boldsymbol{\zeta}} }  
\newcommand{\mazeta}{ {\boldsymbol{\zeta}} }  

\newcommand{\maA}{ {\boldsymbol{A}} }  
\newcommand{\maB}{ {\boldsymbol{B}} }  

\newcommand{\smD}{{ \boldsymbol{\mathfrak{D}} }}   

\newcommand{\mk}{ {\boldsymbol{k}} }    
\newcommand{\mg}{ {\boldsymbol{g}} }    
\newcommand{\mh}{ {\boldsymbol{h}} }    

\newcommand{\maJ}{ {\boldsymbol{\mathcal{J}}} }    

\newcommand{\maT}{ {\boldsymbol{T}} }   

\newcommand{\mrho}{ {\boldsymbol{\rho}} }   

\newcommand{\fluca}{ {\zeta} }   
\newcommand{\flucb}{ {\zeta} }   

\newcommand{\singlet}{{\mathtt{S}}}   
\newcommand{\sym}{ {\mathtt{ST}} }   
\newcommand{\anti}{ {\mathtt{A}} }  

\newcommand{\dyn}{ {\mathtt{x}} }   

\newcommand{\idm}{{\boldsymbol{I}}}		

\newcommand{\coeff}{{ \Lambda }}        

\newcommand{\level}{{ \mathtt{k} }}        

\newcommand{\const}{{\mathtt{\Xi}}}      

\newcommand{\channel}{{\mathfrak{C}}}      

\newcommand{\sdiffeo}{{ \text{SDiff} }}       
\newcommand{\tsdiffeo}{{ \text{\tiny SDiff} }}       

\newcommand{\soq}{{ SO(q)} }       
\newcommand{\hsoq}{{ \widehat{SO}(q)} }       
\newcommand{\tsoq}{{ \text{\tiny SO(q)} }}       

\newcommand{\msstar}{ \circledast }		

\newcommand{\sstar}{ \star }		

\newcommand{\bistar}{ \ast }		

\newcommand{\mstar}{ \circ }  

\newcommand{\sD}{{  \mathtt{D} }}       

\newcommand{\ie}{\emph{i.e.}, }
\newcommand{\eg}{\emph{e.g.}, }

\newcommand{\vs}{\emph{vice versa}}

\makeatletter
\newcommand{\doubletilde}[1]{{%
  \mathpalette\double@tilde{#1}%
}}
\newcommand{\double@tilde}[2]{%
  \sbox\z@{$\m@th#1\tilde{#2}$}%
  \ht\z@=.9\ht\z@
  \tilde{\box\z@}%
}
\makeatother

\title{Supersymmetric SYK Model with Global Symmetry}

\author{Prithvi Narayan}
\author{\!, Junggi Yoon}
\affiliation{International Centre for Theoretical Sciences (ICTS-TIFR), \\
Shivakote, Hesaraghatta Hobli, Bengaluru 560089, India.}
\emailAdd{prithvi.narayan@gmail.com}
\emailAdd{junggi.yoon@icts.res.in}

\abstract{In this paper, we introduce an $\mathcal{N}=1$ supersymmetric SYK model with $SO(q)$ global symmetry. We study the large $N$ expansion of the bi-local collective action of our model. At strong coupling limit, this model exhibits a super-reparametrization symmetry, and the $SO(q)$ global symmetry is enhanced to a $\widehat{SO}(q)$ local symmetry. The corresponding symmetry algebra is the semi-direct product of the super-Virasoro and the super-Kac-Moody algebras. These emergent symmetries are spontaneously and explicitly broken, which leads to a low energy effective action: super-Schwarzian action plus an action of a super-particle on the $SO(q)$ group manifold. We analyze the zero mode contributions to the chaotic behavior of four point functions in various $SO(q)$ channels. In singlet channel, we show that the out-of-time-ordered correlators related to bosonic bi-locals exhibit the saturation of the chaos bound as in the non-SUSY SYK model. On the other hand, we find that the ones with fermionic bi-locals in the singlet channel have ${\pi\over\beta}$ Lyapunov exponent. In the anti-symmetric channel, we demonstrate that the out-of-time-ordered correlator related to a $SO(q)$ generator grows linearly in time. We also compute the non-zero mode contributions which give consistent corrections to the leading Lyapunov exponents from the zero modes.

}

\begin{document}
\maketitle


\section{Introduction}
\label{sec: introduction}

The Sachdev-Ye-Kitaev~(SYK) model proposed in \cite{Sachdev:1992fk,KitaevTalks} consists of $N$ Majorana fermions with disorderd interactions. The model exhibits emergent reparametrization symmetry at strong coupling limit~\cite{KitaevTalks,Polchinski:2016xgd,Jevicki:2016bwu,Maldacena:2016hyu,Jevicki:2016ito} and  saturates~\cite{KitaevTalks,Maldacena:2016hyu} the chaos bound~\cite{Maldacena:2015waa}. The consequent connection to black hole physics has generated great interest recently~\cite{Garcia-Garcia:2016mno,Banerjee:2016ncu,Cotler:2016fpe,Nishinaka:2016nxg,Garcia-Garcia:2017pzl,Gurau:2017xhf,Gurau:2017qna,Gross:2017vhb,Bhattacharya:2017vaz,Kitaev:2017awl}. There have been proposals for gravity duals which capture various features of the SYK model: dilaton gravity\footnote{See also various related works on 2D dilaton gravity~\cite{Almheiri:2014cka,Almheiri:2016fws,Mezei:2017kmw,Grumiller:2017qao}.}~\cite{Jensen:2016pah,Maldacena:2016upp,Engelsoy:2016xyb}, the supersymmetric version of Jackiw-Teitelboim model~\cite{Forste:2017kwy}, the Liouville theory~\cite{Mandal:2017thl} and the 3D gravity~\cite{Das:2017pif,Das:2017hrt}. Also, see~\cite{Gross:2017hcz,Gross:2017aos} for the implication of the higher point functions for the bulk duals. The saturation of the chaos bound also has been observed in unitary quantum mechanical models without disorder which are called ``SYK-like'' tensor models~\cite{Witten:2016iux,Gurau:2016lzk,Klebanov:2016xxf}. The various aspects of the tensor models has been explored~\cite{Ferrari:2017ryl,Narayan:2017qtw,Klebanov:2017nlk,Gurau:2017qya,Diaz:2017kub,Giombi:2017dtl,deMelloKoch:2017bvv,Bulycheva:2017ilt,Choudhury:2017tax} including finite $N$ numerical analysis~\cite{Krishnan:2016bvg,Krishnan:2017ztz,Chaudhuri:2017vrv,Krishnan:2017lra}. Also, $\mathcal{N}=1$ supersymmetric tensor model was introduced in~\cite{Peng:2016mxj}.

The original SYK model has been generalized in various directions. For example, the higher dimensional generalizations have been studied in~\cite{Berkooz:2016cvq,Berkooz:2017efq,Murugan:2017eto}, and the complex SYK model with $U(1)$ symmetry~\cite{Sachdev:2015efa,Davison:2016ngz,Bulycheva:2017uqj} has been discussed. The flavor generalization proposed by~\cite{Gross:2016kjj} enriched the structure of the model. In particular, the SYK model with non-abelian global symmetries was worked out in~\cite{Yoon:2017nig}. On the other hand, the supersymmetric generalization of SYK model~\cite{Fu:2016vas,Murugan:2017eto,Peng:2017spg,Yoon:2017gut} and its random matrix behavior was investigated in~\cite{Li:2017hdt,Sannomiya:2016mnj}.

In this paper, we introduce $\mathcal{N}=1$ supersymmetric SYK model with $\soq$ global symmetry, which can be thought either as the supersymmetric generalization of the SYK model with global symmetry~\cite{Yoon:2017nig} or as flavour symmetry generalization of the $\mathcal{N}=1$ supersymmetric SYK model~\cite{Fu:2016vas,Murugan:2017eto,Peng:2017spg,Yoon:2017gut}. At strong coupling limit, the global $\soq$ symmetry is enhanced to the local $\hsoq$ symmetry together with the emergent super-reparametrization symmetry which was already found in $\mathcal{N}=1$ SUSY SYK model~\cite{Fu:2016vas}. The corresponding symmetry algebra is the semi-direct product of the super-Virasoro algebra and the super-Kac-Moody algebra. These emergent symmetries are spontaneously broken by the large $N$ classical solution and are explicitly broken by the kinetic term at finite coupling, which leads to Pseudo-Nambu-Goldstone boson. We found that the effective action is the super-Schwarzian action plus an action of a superparticle on the $SO(q)$ group manifold. (Section~\ref{sec: effective action}) \ie 
\begin{equation}
    S_{\text{eff}}\equiv-{N \alpha_{\tsdiffeo}\over J} \int d\tau d\theta \;2\; \mbox{\textsf{SSch}}[f,y;\tau,\theta] -{N\alpha_\tsoq \over J} \int d\tau d\theta \; {1\over 2\level} \tr \left[ \maJ\sD \maJ +{1\over \level} \maJ^3\right]
\end{equation}
where $\mbox{\textsf{SSch}}[f,y;\tau,\theta]\equiv\left[ {\sD^4 y\over  \sD y}- 2{ \sD^2 y \sD^3 y\over   [\sD y]^2} \right]$ is the super-Schwarzian derivative and $\mathcal{J}=-\level \sD\mg\mg^{-1}$ is the $\soq$ super-current. Although the previous works~\cite{Fu:2016vas,Murugan:2017eto,Peng:2016mxj} have studied the SUSY SYK model, the full analysis of the chaotic behavior\footnote{See~\cite{Peng:2016mxj} for the analysis of the chaotic behavior related to the bosonic zero mode.} has not been carried out. In particular, the contribution of the fermionic zero mode in the super-Schwarzian effective action to the Lyapunov exponent has not been worked out. In this paper, we will perform a more complete analysis on the chaotic behaviors of our model of which the singlet channel corresponds to $\mathcal{N}=1$ SUSY SYK model. We will now outline main results regarding Lyapunov exponent below.

Using the effective action, we evaluate~(see Section~\ref{sec: leading contribution zero modes}) the large time behavior of the out-of-time-ordered correlators, which give the leading Lyapunov exponents. In addition, we  also calculate~(see Section~\ref{sec: subleading contribution}) the contribution of the non-zero modes to the out-of-time-ordered correlators in order to find the correction to the Lyapunov exponent. We present the summary of the result in the Table~\ref{tab: result}. 
{
\renewcommand{\arraystretch}{2}
\begin{table}[t!]
\centering
\begin{tabular}{c |c | c |c}
\multicolumn{2}{c|}{Channel}                  &\hspace{6mm} Zero Mode \hspace{8mm}   & \hspace{5mm}Non-zero Mode\hspace{5mm}   \\ \hline
\multirow{2}{*}{\hspace{3mm}Singlet\hspace{3mm}} & \hspace{2mm}${\scriptstyle\langle \chi^i\chi^i\;\chi^j\chi^j\rangle,\langle \chi^i\chi^i\;b^jb^j\rangle,\langle b^ib^i\;b^jb^j\rangle } $\hspace{2mm}  &       $\displaystyle \beta J e^{{2\pi\over \beta}t}$       &      $\displaystyle {t\over \beta} e^{{2\pi\over \beta}t}$           \\ \cline{2-4} 
                         & ${\scriptstyle\langle b^i\chi^i \;b^j\chi^j\rangle}$ &      $\displaystyle \beta J e^{{\pi\over \beta}t}$        &       $\displaystyle {t\over \beta} e^{{\pi\over \beta}t}$          \\ \hline
\multirow{2}{*}{Anti}    & ${\scriptstyle\langle \chi^i\chi^i\;\chi^j\chi^j\rangle } $  &       $\displaystyle J t$       &     $\displaystyle {t\over \beta} $            \\ \cline{2-4} 
                         & ${\scriptstyle\langle \chi^i\chi^i\;b^jb^j\rangle,\langle b^ib^i\;b^jb^j\rangle,\langle b^i\chi^i \;b^j\chi^j\rangle}$ & \multicolumn{2}{c}{No Growth} \\ \hline
\multicolumn{2}{c|}{Symmetric-traceless}                      & \multicolumn{2}{c}{No Growth} \\ \hline
\end{tabular}
\caption{Summary of the large time behavior of the out-of-time-ordered correlators in each $SO(q)$ channel. Here, we omitted the $SO(q)$ indices in the four point functions.}
\label{tab: result}
\end{table}
}

The bosonic zero mode of the super-reparametrization~(which would correspond to the boundary graviton in the bulk dual.) is coupled to the bosonic bi-locals (\eg $\chi\chi$) in the singlet channel, which gives the maximal Lyapunov exponent $\lambda_L^{(2)}={2\pi \over \beta}$. On the other hand, the fermionic zero mode of the super-reparametrization~(boundary gravitino in the bulk dual) is coupled only to the fermionic bi-locals (\eg $b\chi$) in the singlet channel because of the fermi statistics. This leads to the Lyapunov exponent $\lambda_L^{({3\over 2})}={\pi\over \beta}$. The bosonic zero mode of the $\hsoq$ local symmetry is coupled to the bi-local field $\chi\chi$ in the anti-symmetric channel, and it would be a boundary gauge field in the bulk dual. This zero mode gives the linear growth in time ($\lambda_L^{(1)}=0$) of the out-of-time-ordered correlator. This is analogous to the linear growth found in~\cite{Yoon:2017nig,Choudhury:2017tax}. The fermionic zero mode~(gaugino in the bulk dual) of the local symmetry is coupled to the fermionic bi-locals in the anti-symmetric channel, but it does not lead to the exponential growth in large $t$. Assuming holographic duals, our result is analogous to the formula for the Lyapunov exponent of higher spin current of integer spin $s$ in CFT$_2$~\cite{Perlmutter:2016pkf}:
\begin{equation}
    \lambda_L^{(s)}={2\pi \over \beta}(s-1) \label{eq: lapunov exp for spin s}
\end{equation}
where $s=2,{3\over 2}, 1, {1\over 2}$ for our case.

These zero mode contributions are consistent with those of the non-zero modes. Namely, if and only if a four point function is coupled to the zero mode to have an exponential growth, it also gets the contribution from the non-zero modes with the same exponential growth rate. Moreover, this non-zero mode contribution gives the correction to the Lyapunov exponent from the zero mode.

The outline of this paper is as follows. \textbf{In Section~\ref{sec: review}}, we revise the generalized SYK model with flavor~\cite{Gross:2016kjj}. In particular, we focus on two type of the SYK models with $\soq$ global symmetry. We also review the $\mathcal{N}=1$ supersymmetric SYK model~\cite{Fu:2016vas} and its supermatrix formulation~\cite{Yoon:2017gut}.

In~\textbf{Section~\ref{sec: n=1 colored susy syk model}}, we extend the supermatrix formulation with the flavor space in which bi-local fields becomes a matrix in the extended bi-local superspace $(\tau_1,\theta_1,\alpha_1;\tau_2,\theta_2,\alpha_2)$. Then, we introduce the $\mathcal{N}=1$ supersymmetric SYK model with $\soq$ global symmetry. After disorder average, we derive the bi-local collective action of our model in large $N$. We discuss the emergent super-reparametrization and the enhanced $\hsoq$ local symmetry at strong coupling limit, and we show that its symmetry algebral is the semi-direct product of super-Virasoro algebra and super-Kac-Moody algebra. These emergent symmetries are spontaneously and explicitly broken to lead to the low energy effective action: super-Schwarzian action and an action of a super-particle on the $\soq$ group manifold.

In~\textbf{Section~\ref{sec: four point functions}}, we discuss the large $N$ expansion of the bi-local collective action. Expanding the bi-local superfield around the large $N$ classical solution, we obtain the quadratic action for fluctuations. We derive the Schwinger-Dyson equation for the two point functions of the bi-local fluctuations, which corresponds to the Schwinger-Dyson equation for the four point function of the SYK superfields. Following~\cite{Murugan:2017eto}, we discuss the conformal eigenfunctions for the four point functions of our model, and we expand the four point functions in terms of the conformal eigenfunctions.

In~\textbf{Section~\ref{sec: spectrum}}, we analyze the spectrum and OPE coefficients of our model.

In~\textbf{Section~\ref{sec: chaotic behavior}}, from the quadratic low energy effective action, we evaluate the zero mode contribution to the large time behavior of the out-of-time-ordered correlators. Also, we calculate the non-zero mode contribution to the out-of-time-ordered correlators, which gives the ${1\over \beta J}$ correction to the zero mode contribution.

In~\textbf{Section~\ref{sec:conclusion}}, we make concluding remarks and present the future directions.

In~\textbf{Appendix~\ref{app: notations}}, we provide a summary of the notation and the convention in this paper. In~\textbf{Appendix~\ref{app: shadow representation}}, we review the shadow representation for the conformal eigenfunctions for four point functions of SYK models. In~\textbf{Appendix~\ref{app: Effective Action}}, we derive the effective action for the zero modes from the broken super-reparametrization and broken $\hsoq$ local symmetry by using $\epsilon$-expansion. In~\textbf{Appendix~\ref{app: zero modes}}, we present the zero mode eigenfunction and their inner products.

\section{Review}
\label{sec: review}

\subsection{Generalized SYK Model with Flavor Revisited}
\label{sec: colored SYK model}

We begin with the generalized SYK model with flavor~\cite{Gross:2016kjj,Gurau:2017xhf,Yoon:2017nig}. We consider $q$ flavors of $N$ Majorana fermions:
\begin{equation}
    \chi^{i \alpha}(\tau)\hspace{10mm}(\; i=1,2,\cdots, N\hspace{3mm}\mbox{and}\hspace{3mm} \alpha=1,2,\cdots,q\; )\ .
\end{equation}
The Majorna fermion is transformed in the fundamental representation of $SO(q)$
\begin{equation}
    \chi^{i \alpha}(\tau)\hspace{5mm}\longrightarrow \hspace{5mm} \mg^{\alpha \alpha'}\chi^{i \alpha'}(\tau)\hspace{8mm}(\;\mg \in SO(q)\;) \ .
\end{equation}
One can construct $SO(q)$ invariant interaction for the generalized SYK model by using the two $SO(q)$ invariant tensors: $\epsilon_{\alpha_1 \alpha_2 \cdots \alpha_q }$ and $\delta_{\alpha_1\alpha_2}$. The simplest $SO(q)$ invariant actions\footnote{In general, one can also build $SO(q)$ invariant interactions by using $S_{q/2}$ character and $\delta$'s~\cite{Yoon:2017nig}. Furthermore, one may mix both $\delta$'s and $\epsilon$'s for the interaction.} are given by
\begin{align}
     S^{\epsilon}=& \int d\tau \;\left[ {1\over 2} \chi^{i \alpha}\partial_\tau \chi^{i \alpha} +  i^{q\over 2}   J^\epsilon_{i_1 \dots i_{q}} \chi^{i_1 \alpha_1} \chi^{i_2 \alpha_2}  \dots    \chi^{i_{q} \alpha_q} \epsilon_{\alpha_1\alpha_2\cdots \alpha_q}   \right]\ ,\\
     S^{\delta}=& \int d\tau \;\left[ {1\over 2} \chi^{i \alpha}\partial_\tau \chi^{i \alpha} +  i^{q\over 2}   J^\delta_{i_1 \dots i_{q}} \chi^{i_1 \alpha_1} \chi^{i_2 \alpha_2}  \dots   \chi^{i_{q} \alpha_{q}}\delta_{\alpha_1\alpha_2}\cdots \delta_{\alpha_{q-1}\alpha_q}    \right]
\end{align}
where $J^\epsilon_{i_1\cdots i_q}$ and $J^\delta_{i_1\cdots i_q}$ is a random coupling constant drawn from the Gaussian distribution
\begin{align}
    \mathcal{P}^{\epsilon}=&\exp\left[-{q! N^{q-1}\over J^2}J^\epsilon_{i_1\cdots i_q}J^\epsilon_{i_1\cdots i_q}\right]\ ,\\
    \mathcal{P}^{\delta}=&\exp\left[-{q^{{q\over 2}}N^{q-1} \over 2 J^2 }\sum_{i_1,\cdots, i_q=1}^N J^\delta_{i_1\cdots i_q} J^\delta_{i_1\cdots i_q}\right]\ ,
\end{align}
respectively. Note that we do not restrict the symmetry of the random coupling constants. The random coupling constant $J_{i_1\cdots i_q}$ can be decomposed by $S_q$ symmetry of the indices $i_1,\cdots, i_q$. Depending on the tensors $\delta$'s and $\epsilon$'s in the interaction, only particular $S_q$ representations of $J_{i_1\cdots i_q}$ give a contribution to the action. For example, the symmetric part of $J^\epsilon_{i_1\cdots i_q}$ in the indices give a contribution to the action. We define the bi-local field $\maPsi(\tau_1,\tau_2)$ by
\begin{equation}
    [\maPsi(\tau_1,\tau_2)]^{\alpha_1\alpha_2} \equiv {1\over N}\sum_{i=1}^N \chi^{i \alpha_1 }(\tau_1)\chi^{i \alpha_2 }(\tau_2)\ ,
\end{equation}
One may treat the bi-local field $[\maPsi(\tau_1,\tau_2)]^{\alpha_1\alpha_2}$ as a matrix in $(\tau,\alpha)$, and the corresponding matrix product is given by
\begin{equation}
    (A\mstar B)(\tau_1,\alpha_1;\tau_2,\alpha_2)\equiv\sum_{\alpha_3=1}^q\int d\tau_3 \; A(\tau_1,\alpha_1;\tau_3,\alpha_3)B(\tau_3,\alpha_3;\tau_2,\alpha_2)\ .\label{def: bi-local matrix product}
\end{equation}
Also, it is sometimes convenient to think of the bi-local field $\maPsi(\tau_1,\tau_2)$ as a $q\times q$ matrix of bi-local fields. After the disorder average\footnote{ One has to perform the quenched disorder average by using the replica trick. For this, one can apply the bi-local replica collective field theory to this model~\cite{Jevicki:2016bwu}, and take replica symmetry ansatz. Or, one can treat the random coupling constant as a (non-dynamical) additional field~\cite{Michel:2016kwn,Nishinaka:2016nxg}. In large $N$, the quenched average and the annealed average for the SYK model are shown to be equivalent~\cite{Gurau:2017xhf}.}, one can derive the collective actions~\cite{Jevicki:2016bwu,Yoon:2017nig}:
\begin{align}
   S_{col}^{\epsilon}=&{N\over 2}\Tr\left[-D\mstar \maPsi +\log \maPsi \right]  - {NJ^2\over 2} \int d\tau_1d\tau_2 \;\det \left[\maPsi(\tau_1,\tau_2)\right] \label{eq: nonsusy coll action1}\\
   S_{col}^{\delta}=&{N\over 2}\Tr\left[-D\mstar \maPsi +\log \maPsi \right]  - {NJ^2\over 2 q^{{q\over 2}}} \int d\tau_1d\tau_2 \; \left[\tr\left(- \maPsi(\tau_1,\tau_2)\maPsi(\tau_2,\tau_1) \right)\right]^{q/2 }\label{eq: nonsusy coll action2}
\end{align}
where $\Tr$ and $\log$ is the trace and the log of a matrix in the $(\tau,\theta)$ space, respectively. On the other hand, $\det$ and $\tr$ in the interaction terms is the determinant and the trace of a $q\times q$ matrix, respectively. In addition, the bi-local derivative is defined by
\begin{equation}
    D(\tau_1,\alpha_1;\tau_2,\alpha_2)\equiv\idm \partial_{\tau_1} \delta(\tau_1-\tau_2)
\end{equation}
where $\idm$ is the $q\times q$ identity matrix.
%
%
Note that the second term ${N\over 2}|\Tr\log \maPsi$ corresponds to the Jacobian in the Hubbard-Stratonovich type transformation from the fundamental fermion to the bi-local field~\cite{Jevicki:1980zg,deMelloKoch:1996mj,Jevicki:2016bwu,Yoon:2017nig}. In large $N$, one can derive the large $N$ saddle point equation which corresponds to the Schwinger-Dyson equation for the two point function of fermions~\cite{Yoon:2017nig}. At strong coupling limit~$|J\tau|\gg 1$, the saddle point equation has emergent reparametrization symmetry, and the $SO(q)$ invariant classical solution is found to be
\begin{equation}
    \maPsi_{cl}(\tau_1,\tau_2)=\Psi_{cl}(\tau_1,\tau_2) \idm= \Lambda{\sgn(\tau_{12})\over |\tau_{12}|^{2\over q}}\idm
\end{equation}
where the coefficient $\Lambda$ is given by
\begin{equation}
    J^2 \coeff^q \pi =\left({1\over 2}-{1\over q}\right) \tan {\pi \over q}
\end{equation}
Expanding the bi-local field $\maPsi$ around the classical solution, one can derive the quadratic action for the fluctuations~\cite{Yoon:2017nig}. Since the fermion is transformed in the fundamental representation of $SO(q)$, the bi-local fluctuation can be decomposed into singlet, anti-symmetric and symmetric-traceless representations:
\begin{equation}
     \yng(1)\; \otimes \; \yng(1) = \singlet \; \oplus \; \anti \; \oplus \; \sym\ .
\end{equation}
At quadratic level, the fluctuations of different representations are decoupled, and one can easily derive the Schwinger-Dyson equations for the two point function $\mathcal{F}_\channel$ ($\channel=\singlet, \anti, \sym$) of the bi-local fluctuations, which is four point function of fermions.
\begin{equation}
    \mathcal{F}_\channel - \mathcal{F}_{0,-\ftn\!(\channel)} = K_\channel \bistar \mathcal{F}_\channel \hspace{8mm} (\channel=\singlet\;,\; \anti\;,\; \sym)
\end{equation}
where $\ftn(\channel)$ is the sign of the representation $\channel$. (\ie $\ftn(\singlet)=\ftn(\sym)=+$, $\ftn(\anti)=-$) Then, the four point function $\mathcal{F}_\channel$ is a geometric series of which the common ratio $K_\channel$ and the first term~$\mathcal{F}_{0,-\ftn\!(\channel)}$ are given by
\begin{align}
    K_\channel\equiv&-\const_\channel J^2 \Psi_{cl}(\tau_1,\tau_3)\Psi_{cl}(\tau_2,\tau_4)[\Psi_{cl}(\tau_3,\tau_4)]^{q-2}\hspace{8mm} (\channel=\singlet\;,\; \anti\;,\; \sym)\\
    \mathcal{F}_{0,\mp}\equiv&\mp \Psi_{cl}(\tau_1,\tau_3)\Psi_{cl}(\tau_2,\tau_4)+ \Psi_{cl}(\tau_1,\tau_4)\Psi_{cl}(\tau_2,\tau_3)
\end{align}
where the coefficient $\const_\channel$ is found to be
\begin{alignat}{3}
    &\const^\epsilon_\singlet=(q-1)\hspace{4mm},\hspace{4mm} &&\const^\epsilon_\anti=1 \hspace{4mm},\hspace{4mm}&&\const^\epsilon_\sym=-1\\ 
    &\const^\delta_\singlet=(q-1)\hspace{4mm},\hspace{4mm} &&\const^\delta_\anti=1 \hspace{4mm},\hspace{4mm}&&\const^\delta_\sym=1
\end{alignat}
Note that the coefficient of the common ratio $K_\channel$ plays an important role in determining the spectrum and the Lyapunov exponent. In the singlet and the anti-symmetric representation, the spectrum and the Lyapunov exponent of the interaction with $\delta$'s is identical to that of the interaction with $\epsilon$. On the other hand, the spectrum of the symmetric traceless representation of $\delta$ interaction is different from that of $\epsilon$ interaction. Nevertheless, the out-of-time-ordered correlators of both interactions exponentially decay. We refer readers to~\cite{Yoon:2017nig} for details of the $\delta$ interaction.

\subsection{Supersymmetric SYK Model and Supermatrix Formulation}
\label{sec: review susy syk}

In this section, we briefly review $\mathcal{N}=1$ SUSY SYK model~\cite{Fu:2016vas} and the supermatrix formulation of SUSY vector models~\cite{Yoon:2017gut}. The action of the $\mathcal{N}=1$ SUSY SYK model can be written simply in the superspace as follow.
\begin{equation}
S=\int d\tau d\theta\; \left[-{1\over 2} \psi^{i } \sD \psi^{i} + i^{q-1\over 2}J_{i_1\cdots i_q}\psi^{i_1 }\cdots \psi^{i_q }\right]
\end{equation}
where $\sD\equiv\partial_\theta +\theta\partial_\tau$ is super-derivative, and we define superfield $\psi^i(\tau,\theta)$ with $N$ Majorana fermions $\chi^i$ and $N$ non-dynamical auxiliary bosons $b^i$ by
\begin{equation}
    \psi^i(\tau,\theta)\equiv\chi^i(\tau)+\theta b^i(\tau)\hspace{8mm}(i=1,2,\cdots, N) \ .
\end{equation}
Also, the random coupling constant $J_{i_1\cdots i_q}$ is drawn from the Gaussian distribution:
\begin{equation}
    \mathcal{P}=\exp\left[-{q N^{q-1}\over J}J_{i_1\cdots i_q}J_{i_1\cdots i_q}\right]\ .
\end{equation}
In large $N$, it is useful to define a bi-local superfield by
\begin{equation}
    \Psi(\tau_1,\theta_1;\tau_2,\theta_2)\equiv {1\over N} \sum_{i=1}^N \psi^i(\tau_1,\theta_1)\psi^i(\tau_2,\theta_2)
\end{equation}
This bi-local superfield can be considered as a matrix in the superspace $(\tau,\theta)$, and the matrix multiplication is defined by~\cite{Yoon:2017gut} by
\begin{equation}
    (A\sstar B)(\tau_1,\theta_1;\tau_2,\theta_2)\equiv \int A(\tau_1,\theta_1;\tau_3,\theta_3)\; d\tau_3 d\theta_3\; B(\tau_3,\theta_3;\tau_2,\theta_2)\label{def: superfield matrix product}
\end{equation}
Note that the position of the measure is important since the bi-local superfield $A$ and $B$ could be either Grassmannian even or odd. Our convention on the position of measure is simple and convenient the for matrix product. In derivation of the collective action and the low energy effective action, it is convenient to utilize the matrix structure of the bi-local superfield, which naturally leads to the supermatrix formulation~\cite{Yoon:2017gut}. Let us expand a bi-local superfield $A$ in components:
\begin{equation}
    A(\tau_1,\theta_1;\tau_2,\theta_2)=A_0(\tau_1,\tau_2)+\theta_1 A_1(\tau_1,\tau_2)- A_2(\tau_1,\tau_2)\theta_2 - \theta_1 A_3(\tau_1,\tau_2)\theta_2\ .
\end{equation}
Then, we define a supermatrix corresponding to the superfield $A$ as follow.
\begin{equation}
    A\equiv\begin{pmatrix}
    A_1 & A_3\\
    A_0 & A_2\\
    \end{pmatrix}\ .
\end{equation}
The bi-local superfield $A$ can be either Grassmannian even or odd so that the ordering of the Grassmannian variables $\theta_1, \theta_2$ and the component fields $A_i$ ($i=1,2,3$) is important in defining the supermatrix. It is crucial to note that the Grassmannian even (odd) superfield corresponds to the Grassmannian odd (even) supermatrix, repsectively. 

In the supermatrix notation, the matrix product defined in \eqref{def: superfield matrix product} is simplified as a matrix product of $2\times 2$ matrix:
\begin{equation}
    (A\sstar B)=\begin{pmatrix}
    A_1\bistar B_1+ A_3\bistar B_0 & A_1\bistar B_3+ A_3\bistar B_2\\
    A_0\bistar B_1+ A_2\bistar B_0 & A_0\bistar B_3+ A_2\bistar B_2\\
    \end{pmatrix}
\end{equation}
where the multiplication $\bistar$ of the components is nothing but the matrix product of non-SUSY bi-local fields:
\begin{equation}
    (A_i\bistar B_j)(\tau_1,\tau_2)\equiv \int d\tau_3 \; A_i(\tau_1,\tau_3)B_j(\tau_3,\tau_2)\hspace{8mm} (i,j=0,1,2,3)
\end{equation}
Note that the matrix product of two Grassmannian even superfield gives Grassmannian odd superfield due to the Grassmannian odd measure. This turns out to be more natural in supermatrix formulation because the matrix product of the two Grassmannian odd supermatrix is a Grassmannian even supermatrix and \vs. Hence, from now on, we refer to $|A|$ as the Grassmann signature of $A$ as a supermatrix. \ie $|A|=+1$ if $A$ is Grassmannian even supermatrix and \vs. Now, one can utilize all operations in the supermatrix. For example, the super-trace is defined by
\begin{equation}
    \str(A)=\bitr(A_1)+(-1)^{|A|}\bitr (A_2)=\int d\tau_1d\theta_1\; A(\tau_1,\theta_1;\tau_1,\theta_1)
\end{equation}
where $\bitr$ denotes the trace in the bi-local space $(\tau_1,\tau_2)$ \ie $\bitr(A_i)=\int d\tau_1 \; A_i(\tau_1,\tau_1)$ $(i=0,1,2,3)$. After the disorder average, one can derive the collective action for the $\mathcal{N}=1$ SUSY SYK model~\cite{Yoon:2017gut}:
\begin{equation}
S_{col}=-{N\over 2} \str \left[\mathfrak{D} \sstar  \Psi\right]+{N\over 2}\str \log \Psi - {J N\over 2q}\int  d\tau_1d\theta_1 d\tau_2 d\theta_2 [\Psi(\tau_1,\theta_1;\tau_2,\theta_2)]^q
\end{equation}
where the bi-local super-derivative is defined by
\begin{equation}
    \mathfrak{D}(\tau_1,\theta_1;\tau_2,\theta_2)=\begin{pmatrix}
    0 & \partial_1\delta(\tau_1-\tau_2) \\
\delta(\tau_1-\tau_2) & 0\\
\end{pmatrix}=\sD_{\theta_1}(\theta_1-\theta_2)\delta(\tau_1-\tau_2)
\end{equation}
Note that ${N\over 2}\str \log \Psi$ comes from the Jacobian of the transformation from the $N$ superfield $\psi^i$ to the bi-local superfield $\Psi$ (See~\cite{Yoon:2017gut} for details).

\section{$\mathcal{N}=1$ SUSY SYK Model with Global Symmetry}
\label{sec: n=1 colored susy syk model}

\subsection{Extended Supermatrix Formulation}
\label{sec: extended supermatrix formulation}

We will generalize the $\mathcal{N}=1$ supersymmetric SYK model by adding flavor to the $N$ superfield. For this, we first extend the bi-local superspace $(\tau_1,\theta_1;\tau_2,\theta_2)$ to bi-local superspace with extra flavor space:
\begin{equation}
    (\tau_1,\theta;\tau_2,\theta_2)\hspace{5mm}\longrightarrow \hspace{5mm} (\tau_1,\theta_1,\alpha_1;\tau_2,\theta_2,\alpha_2)
\end{equation}
where $\theta$'s are the Grassmannian coordinate and $\alpha$'s are the flavour index $(\alpha=1,2,\cdots, q)$. In this extended bi-local superspace, one can consider bi-local superfield $\maA^{\alpha_1\alpha_2}(\tau_1,\theta_1;\tau_2,\theta_2)$, and it is convenient to think of it as a $q\times q$ matrix $\maA(\tau_1,\theta_1;\tau_2,\theta_2)$. \ie
\begin{equation}
    [\maA(\tau_1,\theta_1;\tau_2,\theta_2) ]^{\alpha_1 \alpha_2}\equiv\maA^{\alpha_1\alpha_2}(\tau_1,\theta_1;\tau_2,\theta_2)\ .
\end{equation}
We use the convention that an object written in bold (\eg $\maA$) is a $q \times q$ matrix. We define a matrix product $\msstar$ for the extended bi-local superspace such that 
\begin{equation}
    (\maA\msstar \maB)^{\alpha_1\alpha_2}(\tau_1,\theta_1;\tau_2,\theta_2)\equiv\sum_{\alpha_3=1}^q\int (\maA)^{\alpha_1\alpha_3}(\tau_1,\theta_1;\tau_3,\theta_3)d\tau_3d\theta_3(\maB)^{\alpha_3\alpha_2}(\tau_3,\theta_3;\tau_2,\theta_2) \ .\label{def: bilocal super matrix product}
\end{equation}
Expanding a bi-local superfield into component fields, one can represent it as a supermatrix like the $\mathcal{N}=1$ SUSY SYK model in Section~\ref{sec: review susy syk}:
\begin{align}
    \maA(\tau_1,\theta_1;\tau_2,\theta_2)=&\maA_0(\tau_1,\tau_2)+\theta_1\maA_1(\tau_1,\tau_2)-\maA_2(\tau_1,\tau_2)\theta_2-\theta_1\maA_3(\tau_1,\tau_2)\theta_2\cr
    =&\begin{pmatrix}
    \maA_1(\tau_1,\tau_2) & \maA_3(\tau_1,\tau_2)\\
    \maA_0(\tau_1,\tau_2) & \maA_2(\tau_1,\tau_2)\\
    \end{pmatrix}
\end{align}
where the lowest component $\maA_0$ could be either Grassmannian even or odd. This choice of the signs and the ordering of Grassmann variables will lead to a natural definition of a supermatrix and its multiplication. Note that a Grassmannian odd superfield is mapped to a Grassmannian even supermatrix, and vice versa as before. In this supermatrix representation, the matrix product of two bi-local superfield in~\eqref{def: bilocal super matrix product} becomes the usual $(2q\times 2q)$ matrix product:
\begin{align}
    &(\maA\msstar \maB)=\begin{pmatrix}
    \maA_1& \maA_3\\
    \maA_0 & \maA_2\\
    \end{pmatrix}\msstar \begin{pmatrix}
    \maB_1 & \maB_3\\
    \maB_0 & \maB_2\\
    \end{pmatrix}\cr
    =&\begin{pmatrix}
    \maA_1\mstar \maB_1+ \maA_3\mstar \maB_0 & \maA_1\mstar \maB_3+ \maA_3\mstar \maB_2\\
    \maA_0\mstar \maB_1+ \maA_2\mstar \maB_0 & \maA_0\mstar \maB_3+ \maA_2\mstar \maB_2\\
    \end{pmatrix}
\end{align}
where the matrix product $\mstar$ for the component bi-local fields is defined in~\eqref{def: bi-local matrix product}. \ie
\begin{equation}
    (\maA_i\mstar \maB_j)^{\alpha_1\alpha_2}(\tau_1,\tau_2)\equiv\sum_{\alpha_3=1}^q\int d\tau_3[\maA_i(\tau_1,\tau_3)]^{\alpha_1\alpha_3}[\maB_j(\tau_3,\tau_2)]^{\alpha_3\alpha_2}
\end{equation}
One can easily see that the identity supermatrix gives the expected delta function in the extended bi-local superspace. \ie
\begin{equation}
\pmb{\mathbb{I}}^{\alpha_1\alpha_2}(\tau_1,\theta_1;\tau_2,\theta_2)\equiv \begin{pmatrix}
  \delta^{\alpha_1\alpha_2}\delta(\tau_1-\tau_2) & 0 \\
0 & \delta^{\alpha_1\alpha_2} \delta(\tau_1-\tau_2)\\
\end{pmatrix}=(\theta_1-\theta_2) \delta(\tau_1-\tau_2) \delta^{\alpha_1,\alpha_2}
\end{equation}
Furthermore, the natural definition of the trace in the extended bi-local superspace is consistent with the supertrace of a supermatrix. \ie
\begin{align}
&\sum_\alpha \int  d\tau_1 d\theta_1d\tau_2 \delta(\tau_{12}) \left[ \maA_0(\tau_1,\tau_2)+ \theta_1  \maA_1(\tau_1,\tau_2)-  \maA_2(\tau_1,\tau_2)\theta_1  \right]^{\alpha \alpha}\cr
=& \Tr \maA_1 - (-1)^{|\maA|}\Tr \maA_2=\STr \maA\label{eq:Define Super Trace}
\end{align}
where $(-1)^{|\maA|}$ is $1$ if $\maA$ is Grassmannian even supermatrix and $(-1)^{|\maA|}$ is $-1$ if $\maA$ is Grassmannian odd supermatrix. Note that the properties of the supermatrix will automatically holds. \eg
\begin{equation}
    \STr (\boldsymbol{A}\msstar \boldsymbol{B} )=(-1)^{|\boldsymbol{A}|\cdot |\boldsymbol{B}|}\STr (\boldsymbol{B}\msstar \boldsymbol{A} )
\end{equation}

%
%
%
%
%
%

\subsection{$\mathcal{N}=1$ SUSY SYK Model with Global Symmetry}
\label{sec: description of models}

The action of $\mathcal{N}=1$ SUSY SYK model with global symmetry is given by
\begin{equation}
S=\int d\tau d\theta\; \left[-{1\over 2} \psi^{i \alpha} \sD \psi^{i\alpha} + i^{q-1\over 2}J_{i_1\cdots i_q}\psi^{i_1 \alpha_1}\cdots \psi^{i_q \alpha_q}\epsilon_{\alpha_1 \cdots \alpha_q}\right]
\end{equation}
where $q$ is odd integer ($q\geqq 3$). The superfield $\psi^{i \alpha}(\tau,\theta)$ is defined by
\begin{equation}
    \psi^{i \alpha}(\tau,\theta) \equiv \chi^{i \alpha}(\tau)+ \theta b^{i \alpha}(\tau)
\end{equation}
where $\chi^{i\alpha}(\tau)$ and $b^{i\alpha}(\tau)$ are $qN$ Majorana fermions and $qN$ auxiliary bosons, respectively $(i=1,2,\cdots,N$ and $\alpha=1,2,\cdots, q)$. The superfield $\psi^{i\alpha}(\tau)$ transforms in the fundamental representation of $O(N)$ and $SO(q)$. \ie
\begin{align}
    \psi^{i\alpha}(\tau,\theta)\hspace{5mm} &\longrightarrow \hspace{5mm} \boldsymbol{\mathcal{O}}^{ij}\psi^{j\alpha}(\tau,\theta)\cr
    \psi^{i\alpha_1}(\tau,\theta)\hspace{5mm} &\longrightarrow \hspace{5mm}\mg^{\alpha_1\alpha_2}\psi^{i\alpha_2}(\tau,\theta)
\end{align}
where $\boldsymbol{\mathcal{O}}^{ij}$ is an $O(N)$ matrix, and $\mg$ is an $SO(q)$ matrix. Note that the action is invariant under the (global) $SO(q)$ transformation. $J_{i_1\cdots i_q}$ is the random coupling constant drawn from the Gaussian ensemble
\begin{equation}
    \exp\left[ -{ q!  N^{q-1} \over J}J_{i_1\cdots i_q}J_{i_1\cdots i_q}\right]\ .
\end{equation}
We define a bi-local superfield by
\begin{equation}
    \maPsi^{\alpha_1 \alpha_2}(\tau_1,\theta_1;\tau_2,\theta_2)\equiv{1\over N}\sum_{i=1}^N \psi^{i \alpha_1}(\tau_1,\theta_1)\psi^{i \alpha_2}(\tau_2,\theta_2)
\end{equation}
Note that the bi-local superfield $\maPsi(\tau_1,\theta_1;\tau_2,\theta_2)$ is anti-symmetric in the extended bi-local superspace. \ie
\begin{equation}
    \maPsi^{\alpha_1\alpha_2}(\tau_1,\theta_1;\tau_2,\theta_2)=-\maPsi^{\alpha_2\alpha_1}(\tau_2,\theta_2;\tau_1,\theta_1)\ .\label{eq: antisym of bilocal}
\end{equation}
In the supermatrix representation, the bi-local superfield $\maPsi$ reads
\begin{equation}
    \maPsi^{\alpha_1 \alpha_2}(\tau_1,\theta_1;\tau_2,\theta_2)={1 \over N }\sum_{i=1}^N \begin{pmatrix}
b^{i \alpha_1 }(\tau_1) \chi^{i \alpha_2}(\tau_2) &  - b^{i \alpha_1}(\tau_1) b^{i \alpha_2}(\tau_2)  \\
\chi^{i \alpha_1 }(\tau_1) \chi^{i \alpha_2 }(\tau_2)  & - \chi^{i \alpha_1 }(\tau_1) b^{i \alpha_2 }(\tau_2) \\
\end{pmatrix}
\end{equation}
The Jacobian which takes from the superfield $\psi^i$ to the bi-local field $\maPsi$ is given by~\cite{Yoon:2017gut}
\begin{equation}
    \log \mathcal{J} = - {N-1 \over 2}  \STr \log \maPsi 
\end{equation}
where $\STr$ is defined in \eqref{eq:Define Super Trace}. It is useful to define a super-derivative matrix:
\begin{align}
[\smD]^{\alpha_1 \alpha_2}(\tau_1,\theta_1;\tau_2,\theta_2) \equiv& \delta^{\alpha_1 \alpha_2} \sD_{\theta_1}(\theta_1-\theta_2)\delta(\tau_1-\tau_2)\cr
 =& \begin{pmatrix}
0 & \delta^{\alpha_1 \alpha_2} \partial_1\delta(\tau_1-\tau_2) \\
\delta^{\alpha_1 \alpha_2} \delta(\tau_1-\tau_2) & 0\\
\end{pmatrix}\label{def:bi-local superderivative}
\end{align}
where $\sD_{\theta}=\partial_{\theta} + \theta\partial_{\tau}$ is the super-derivative. Note that the super-derivative matrix $\smD$ is Grassmannian odd supermatrix. Using the super-derivative matrix, one can easily check that
\begin{align}
\smD\msstar \maA (\tau_1,\theta_1;\tau_2,\theta_2)=\begin{pmatrix}
\partial_{\tau_1}\maA_0(\tau_1,\tau_2) & \partial_{\tau_1}\maA_2(\tau_1,\tau_2)\\
\maA_1 & \maA_3\\
\end{pmatrix}\ ,
\end{align}
%
%
and therefore, its supertrace leads to the kinetic term:
\begin{equation}\begin{split}
N \STr ( \smD\msstar \Psi)
= & \sum_{\alpha} \int d\tau \left[ - \chi^{i\alpha} (\tau) \partial_{\tau}\chi^{i\alpha} (\tau) + b^{i\alpha}(\tau) b^{i\alpha}(\tau)\right]\cr
=& \sum_{\alpha} \int d\tau d\theta \psi^{i\alpha}(\tau,\theta) \sD_\theta \psi^{i\alpha}(\tau,\theta)
\end{split} \end{equation}
After disorder average, the collective action can be written as
\begin{equation}
S_{col}=-{N\over 2} \STr \left[\smD\msstar  \maPsi\right]+{N\over 2}\STr \log \maPsi + {JN\over 2  }\int  d\tau_1d\theta_1 d\tau_2 d\theta_2 \det[\maPsi(\tau_1,\theta_1;\tau_2,\theta_2)]\label{eq: collective action}
\end{equation}
where $\det[\maPsi]$ is a determinant of the $q\times q$ matrix $\maPsi$.

\subsection{Emergent Symmetry}
\label{sec: emergent symmetry}

In strong coupling limit $|J\tau|\gg 1$, the collective aciton in~\eqref{eq: collective action} has emergent symmetries including the super-reparametrization symmetry in the $\mathcal{N}=1$ supersymmetric SYK model. We define a critical collective action $S_{\text{\tiny critical}}$ as follows.
\begin{equation}
    S_{\text{\tiny critical}}\equiv - {N\over 2}\STr \log \maPsi + {JN\over 2 }\int  d\tau_1d\theta_1 d\tau_2 d\theta_2 \det[\maPsi(\tau_1,\theta_1;\tau_2,\theta_2)]\label{def: critical action}
\end{equation}
First, we consider the super-reparametrization $(\tau,\theta)\longrightarrow (\tau',\theta')$ given by~\cite{Fu:2016vas}
\begin{equation}
    \tau'=f(\tau+\theta\eta(\tau))\hspace{4mm},\hspace{4mm} \theta'=\sqrt{\partial_\tau f(\tau)} \left[\theta+\eta(\tau)+{1\over 2} \theta \eta(\tau)\partial_\tau \eta(\tau)\right]\label{def: super reparametrization parametrization}
\end{equation}
where $f(\tau)$ and $\eta(\tau)$ is an arbitrary bosonic and fermionic function of $\tau$, respectively. This transformation satisfies
\begin{equation}
    \sD_\theta=\sD_\theta \theta' \; \sD_{\theta'}\label{eq: super repara constraint}
\end{equation}
and, the Jacobian can be simplified by
\begin{equation}
    \ber\begin{pmatrix}
    \partial_\tau \tau' & \partial_\tau \theta'\\
    \partial_\theta \tau' & \partial_\theta \theta'\\
    \end{pmatrix}= \sD_\theta \theta'
\end{equation}
For now, it is convenient to parametrize the super-reparametrization as $\tau'=f(\tau,\theta)$ and $\theta'=y(\tau,\theta)$ with constraint~\eqref{eq: super repara constraint} so that the Jacobian is still $\sD_\theta y$. Then, the critical collective action in~\eqref{def: critical action} is invariant under the following transformation of the bi-local superfield $\maPsi$:
\begin{equation}
    \maPsi(\tau_1,\theta_1;\tau_2,\theta_2) \hspace{3mm}\longrightarrow\hspace{3mm} \maPsi_{(f,y)}(\tau_1,\theta_1;\tau_2,\theta_2) \equiv   [\sD_1 y_1 ]^{1\over q} \maPsi(f_1,y_1;f_2,y_2 ) [\sD_2 y_2 ]^{1\over q}
\end{equation}
where $f_i\equiv f(\tau_i,\theta_i),y_i\equiv y(\tau_i,\theta_i)$ ($i=1,2$). In addition to the super-reparametrization, the global $\soq$ symmetry of the collective action is enhanced to the local $\hsoq$ symmetry in strong coupling limit. This is analogous to the local symmetry in~\cite{Yoon:2017nig}, but in this case, the $\hsoq$ local transformation is parametrized not only by $\tau$ but also by the Grassmannian coordinate $\theta$. For this, let us introduce a $q\times q$ matrix $\mg(\tau,\theta)$ given by
\begin{equation}
    \mg(\tau,\theta)\equiv\mh(\tau) +\theta \mk(\tau)\mh(\tau)=e^{\theta \mk(\tau)}\mh(\tau)
\end{equation}
where $\mh(\tau,\theta)$ is a (bosonic) $SO(q)$ matrix. \ie
\begin{equation}
    \mh(\tau) \mh^t (\tau)=\idm\hspace{5mm},\hspace{5mm} \det \mh=1
\end{equation}
where $\idm$ is the $q\times q$ identity matrix. On the other hand, $\mk(\tau)$ is a $q\times q$ (fermionic) anti-symmetric matrix. \ie
\begin{equation}
    \mk^t(\tau)=-\mk(\tau)
\end{equation}
Hence, it is easy to see that the matrix $\mg(\tau,\theta)$ is also $SO(q)$ matrix:
\begin{equation}
    \mg^t(\tau,\theta)=\mh^t(\idm +\theta \mk^t)=\mh^{-1}(\idm -\theta \mk)=\mg^{-1}
\end{equation}
and
\begin{equation}
    \det\mg
    =\exp\left[\tr \log(\idm +\theta \mk)\right]=1
\end{equation}
Then, one can consider a local transformation of the superfield $\psi^{i\alpha}$ by the matrix $\mg(\tau,\theta)$:
\begin{equation}
    \psi^{i\alpha_1}(\tau,\theta)\hspace{5mm} \longrightarrow \hspace{5mm} \sum_{\alpha_2=1}^q\mg^{\alpha_1\alpha_2}(\tau,\theta)\psi^{i\alpha_2}(\tau,\theta)
\end{equation}
Accordingly, the bi-local field is transformed as follows.
\begin{equation}
    \maPsi(\tau_1,\theta_1;\tau_2,\theta_2)\quad\longrightarrow \quad  \mg(\tau_1,\theta_1)\maPsi(\tau_1,\theta_1;\tau_2,\theta_2 )\mg^{-1}(\tau_2,\theta_2) 
\end{equation}
Under this $SO(q)$ local transformation, the critical collective action~\eqref{def: critical action} is also invariant because of the $q\times q$ determinant and the supertrace $\STr$ for the extended bi-local superspace. Therefore, together with the super-reparametrization, the critical collective action~\eqref{def: critical action} is invariant under the transformation:
\begin{align}
    &\maPsi(\tau_1,\theta_1;\tau_2,\theta_2)\cr
    \longrightarrow\hspace{5mm}& \maPsi_{[(f,y),\mg]}(\tau_1,\theta_1;\tau_2,\theta_2) \equiv   [\sD_1 y_1 ]^{1\over q} \mg(\tau_1,\theta_1)\maPsi(f_1,y_1;f_2,y_2 )\mg^{-1}(\tau_2,\theta_2) [\sD_2 y_2 ]^{1\over q}\label{eq:Local Symmetry Transformation}
\end{align}
where $f_i\equiv f(\tau_i,\theta_i),y_i\equiv y(\tau_i,\theta_i)$ ($i=1,2$).

\subsection{Super-Reparametrization and Super Kac-Moody Algebra}
\label{sec: sr and skm}

We will now determine the algebra obeyed by the group of transformations given in \eqref{eq:Local Symmetry Transformation}, which we will call $\sdiffeo\ltimes \hsoq$. To do this, first note that the infinitesimal action of the transformations are parametrized by 
\begin{equation}\label{eq:Infinitesimal Transformations}
    f(\tau) = \tau + \epsilon(\tau) + \theta \eta(\tau) \hspace{5mm},\hspace{5mm} y = \theta +  \eta(\tau) + {\theta  \over 2} \partial_\tau \epsilon(\tau) \hspace{5mm},\hspace{5mm} \mg(\tau,\theta) = \idm +  i \pmb{\rho}(\tau) +    \theta \pmb{k}(\tau)  
\end{equation}
Note that $\pmb \rho$ is obviously an element in the algebra of $so(N)$, whereas $\pmb k^t = - \pmb k$ and hence can be expanded in the basis of $so(N)$ adjoint generator matrices. If the corresponding modes are labelled by 
\begin{alignat}{3}
&\epsilon(\tau) = \sum_{n \in \mathbb Z} \epsilon_n \tau^{n+1} \hspace{5mm} ,\hspace{5mm}  
&&\eta(\tau) = \sum_{r \in {1 \over 2}+ \mathbb Z} \eta_r \tau^{r+{1 \over 2}} \hspace{5mm} ,\hspace{5mm} 
\cr
&\pmb \rho(\tau) = \sum_{n \in \mathbb Z}\rho^a_n \maT^a \tau^n \ ,\      &&\pmb k(\tau) = \sum_{r \in {1 \over 2}+ \mathbb Z} k^a_r \maT^a \tau^{r - {1 \over 2} }\label{eq:Mode Expansion}
\end{alignat}
and define the infinitesimal generators $L_n,G_r,J_n^a,F^a_r$ via 
\begin{equation}\label{eq:Define Generators}
\delta_{\epsilon,\eta} =-\sum_{n \in \mathbb Z} \epsilon_n L_n + \sum_{r \in {1 \over 2}+ \mathbb Z} \eta_r G_r + i \sum_{n \in \mathbb Z} \rho^a_n J^a_n +  \sum_{n \in \mathbb {1 \over 2} +Z} k^a_r F^a_r 
\end{equation}
then the action given in \eqref{eq:Infinitesimal Transformations}) can be realized as
\begin{alignat}{4}
    &L_n  = - \left( \tau^{n+1} \partial_\tau + {(n+1)\over 2} \tau^n \theta \partial_\theta \right) \hspace{5mm },\hspace{5mm} &&G_r  = \tau^{r + {1 \over 2}} \left( \partial_\theta - \theta \partial_\tau \right) \\
    &J^a_n = \maT^a \tau^n  \hspace{5mm },\hspace{5mm} &&F^a_r = - \theta \maT^a \tau^{ r -	 {1 \over 2} }
\end{alignat}
where $T^a$ are the algebra generators which obey $[\maT^a,\maT^b] = i f^{abc} \maT^c$. One can check that these generators satisfy the algebra
\begin{alignat}{6}
    &[L_m , L_n ] = (m-n) L_{m+n} \hspace{10mm} &&\{ G_r , G_s \} = 2 L_{r+s} \hspace{10mm} &&[L_m, G_s] = \left( {m\over 2} - s\right) G_{m+s} \cr
    &[J^a_n , J^b_m ]  = i f^{abc} J^c_{n+m} \hspace{10mm} &&[L_n , J^a_m ] = - m J^a_{n+m} \hspace{10mm} &&[L_n , F^a_r ] = - (r + {n \over 2} ) F^a_{n+r}  \label{eq:The Algebra}\\
    &[J^a_n , F^b_r] = i f^{abc} F^c_{n+r} \hspace{10mm} && \{ G_s , F^a_r\} = -J^a_{r+s} \hspace{10mm} && [ G_s , J^a_n ] = n  F^a_{n+s}\cr
    &\{ F^a_r , F^b_s\} = 0 \hspace{10mm}\nonumber
\end{alignat}
with all the other commutator/anti-commutators are vanishing. Note that the modes $L_{0}$, $L_{\pm}$, $G_{\pm{1 \over 2}}$ together with  $J_0^a$ generates $osp(1|2) \times so(q)$ subalgebra which forms the global symmetry. For completeness, we record below the group composition action on the $\maPsi$.  
\begin{equation}\begin{split}
   \maPsi(\tau_1,\theta_1;\tau_2,\theta_2) &  \xrightarrow{[(f',y'),\mg']\cdot [(f,y),\mg]} \maPsi_{[(f',y'),\mg']\cdot [(f,y),\mg]}(\tau_1,\theta_1;\tau_2,\theta_2) \\
     \maPsi_{[(f',y'),\mg']\cdot [(f,y),\mg]}(\tau_1,\theta_1;\tau_2,\theta_2) & \equiv 
   [\sD_{\theta_1} y'_1  \sD_{y'_1} y_1 ]^{1\over q} [\mg'(\tau_1,\theta_1) \mg(f'_1,y'_1)] \\ 
   & \times \maPsi(f(f'_1,y'_1), y(f'_1,y'_1);f(f'_2,y'_2),y(f'_2,y'_2))    \\
   &\hspace{10mm} \times [\mg'(\tau_2,\theta_2) \mg(f_2,y_2)]^{-1}  [\sD_{\theta_2} y'_1 \sD_{y'_2} y_2 ]^{1\over q} 
\end{split}\end{equation}
where all objects are thought of as functions of $\tau,\theta$. We then have the following transformation law
 \begin{equation}\label{eq:Composition Law Finite}
     [(f',y'),\mg']\cdot [(f,y),\mg] = [(f,y) \circ (f', y'), \mg'(\tau,\theta)\mg(f',y')]
 \end{equation}
where $(\tau,\theta) \xrightarrow{ (f,y) \circ (f', y')} ( \  f(f'(\tau,\theta),y'(\tau,\theta))\ ,\ y(f'(\tau,\theta),y'(\tau,\theta))  \ ) $. One can also use this to obtain the algebra 
\footnote{The commutator of two transformations is
\begin{equation}\begin{split}
\label{eq:Commutation Operation}
    \delta_{[\epsilon_2,\eta_2, \pmb \rho_2, \pmb k_2],[\epsilon_1,\eta_1, \pmb \rho_1, \pmb k_1]} & = 
    \delta_{[ \epsilon_2 \epsilon_1'-\epsilon_1 \epsilon_2' + 2 \eta_2 \eta_1\ , \   {1 \over 2} \eta_2 \epsilon_1'+ \epsilon_2 \eta_1'  -   {1 \over 2} \epsilon_2' \eta_1 - \epsilon_1 \eta_2' } \\
   &  \hspace{-5mm}
\ _{,-i  \eta_2 \pmb k_1 +i \eta_1 \pmb k_2 + \epsilon_2 \pmb \rho_1' - \epsilon_1 \pmb \rho_2' +i [ \pmb \rho_2 , \pmb \rho_1]  \ , \ 
    \epsilon_2\pmb  k_1' - \epsilon_1 \pmb k_2' + { \epsilon_2' \pmb k_1 - \epsilon_1' \pmb k_2  \over 2} + i \eta_2 \pmb \rho_1' - i\eta_1 \pmb \rho_2' + i [\pmb k_2 ,\pmb \rho_1]  -i [\pmb k_1 , \pmb \rho_2]
    }
\end{split}\end{equation}
which upon using the mode expansion \eqref{eq:Mode Expansion} and the generators \eqref{eq:Define Generators} results in the same algebra as in \eqref{eq:The Algebra}.
}

\subsection{Large $N$ Classical Solution}
\label{sec: large n expansion}

By varying the collective action~\eqref{eq: collective action} with respect to the bi-local superfield~$\maPsi$, we obtain the large $N$ saddle point equation. Then, multiplying another bi-local field $\maPsi$ to the saddle point equation, we have
%
%
%
%
%
\begin{align}
&(\smD \msstar \maPsi)^{\alpha \gamma}(1,2)+J{ \epsilon^{\alpha\alpha_2\cdots \alpha_q}\epsilon^{\mu  \gamma_2\cdots \gamma_q} \over (q-1)!}\int d\mu_3\; \Psi^{\alpha_2\gamma_2  }(1,3)\cdots  \Psi^{\alpha_q \gamma_q }(1,3)\Psi^{\mu \gamma}(3,2)\cr
&=\delta^{\alpha \gamma}(\theta_1-\theta_2)\delta(\tau_{12}) \label{eq: saddle point equation}\ .
\end{align}
where the measure $d\mu_3$ is defined by $d\tau_3d\theta_3$. 
%
%
Note that the large $N$ saddle point equation~\eqref{eq: saddle point equation} corresponds to the Schwinger-Dyson equation for the two point function of the fermion superfield. In strong coupling limit $|J\tau|\gg 1$, one can drop the first term of the saddle point equation~\eqref{eq: saddle point equation} which comes from the kinetic term. To solve the saddle point equation in the strong coupling limit, we take an $SO(q)$ invariant ansatz:
\begin{equation}
	\maPsi_{cl}(\tau_1,\theta_1;\tau_2,\theta_2)=\Psi_{cl}(\tau_1,\theta_1;\tau_2,\theta_2) \;\idm\ .
\end{equation}
where $\idm$ is the $q\times q$ identity matrix. Then, the saddle point equation~\eqref{eq: saddle point equation} is reduced to
\begin{equation}
	(\theta_1-\theta_2)\delta(\tau_{12}) - J\int d\tau_3d\theta_3 [\Psi_{cl}(\tau_1,\theta_1;\tau_3,\theta_3)]^{q-1}\Psi_{cl}(\tau_3,\theta_3;\tau_2,\theta_2)=0 
\end{equation}
This equation is identical to the Schwinger-Dyson equation for the $\mathcal{N}=1$ SUSY SYK model, and the solution is given by~\cite{Fu:2016vas}
\begin{equation}
    \Psi_{cl}(\tau_1,\theta_1;\tau_2,\theta_2) = \Lambda { \sgn(\tau_{12})  \over |\tau_{12} - \theta_1 \theta_2|^{2\Delta}} = \Lambda \left\lbrace {\sgn(\tau_{12}) \over |\tau_{12}|^{2\Delta} } + {2\Delta \theta_1 \theta_2 \over  |\tau_{12}|^{{2\Delta}+1} }\right\rbrace \label{eq: classical solution in susy syk} 
\end{equation}
where $\Delta\equiv{1\over 2q} $ and the coefficient $\Lambda$ is
\begin{equation}
    \Lambda^q = {\tan\pi \Delta \over 2 \pi J}
\end{equation}
In supermatrix notation, the classical solution $\maPsi$ is Grassmannian odd supermatrix represented by
\begin{equation}
   \maPsi_{cl} = \Lambda \left( \begin{array}{cc} 
    0     &  -{2\Delta \over  |\tau_{12}|^{{2\Delta}+1 }} \idm \\
     {\sgn(\tau_{12})  \over |\tau_{12}|^{2\Delta} }  \idm & 0
    \end{array} \right)\ . \label{eq: classical solution supermatrix}
\end{equation}

\subsection{Low Energy Effective Action}
\label{sec: effective action}

In the strong coupling limit $|J\tau|\gg 1$, the collective action has emergent super-reparametrization and the $\hsoq$ local symmetry. This emergent symmetries are broken spontaneously by the classical solution. Furthermore, moving away from the infinite coupling, the kinetic term, which we have ignored in the strict strong coupling limit, also explicitly breaks the emergent symmetries. Hence, this leads to zero modes (Pseudo-Nambu-Goldstone bosons) for the broken symmetries. In order to derive the effective action for those low energy modes, we will transform the classical solution by the super-reparametrization and the $\hsoq$ local transformation parametrized by $(f(\tau,\theta),y(\tau,\theta))$ and $\mg(\tau,\theta)$, respectively:
\begin{equation}
    \maPsi_{cl}(\tau_1,\theta_1;\tau_2,\theta_2)\quad\longrightarrow \quad [\sD_1 y_1 ]^{1\over q} \mg(\tau_1,\theta_1)\maPsi_{cl}(f_1,y_1;f_2,y_2 )\mg^{-1}(\tau_2,\theta_2) [\sD_2 y_2 ]^{1\over q}
\end{equation}
where $f_i\equiv f(\tau_i,\theta_i),y_i\equiv y(\tau_i,\theta_i)$ ($i=1,2$). Then, we substitute the transformed classical solution into the kinetic term which explicitly breaks the super-reparametrization and the $\hsoq$ local transformation, which leads to the effective action for the zero modes. For this, it is useful to use the $\epsilon$-expansion of $q$ given by~\cite{Jevicki:2016bwu,Jevicki:2016ito} 
\begin{equation}
    q\equiv {1\over 1-\epsilon}\hspace{10mm} (0<\epsilon<1)\ .
\end{equation}
In Appendix~\ref{app: Effective Action}, we showed that the $\epsilon$-expansion of the kinetic term with the transformed classical solution gives
\begin{align}
    &-{N\over 2}\STr\left[\smD\msstar \maPsi_{cl}\right]_{\epsilon\text{-expansion}}
    \cr
    =&-{N\over 4(q-1)^{1\over q}\alpha_0^{1\over q}}\int d\tau d\theta \left({\sD^4 y\over  \sD y}- 2{ \sD^2 y \sD^3 y\over   [\sD y]^2}\right)-{N\over 2(q-1)^{1\over q}\alpha_0^{1\over q}}\int d\tau d\theta  \tr (\sD^3\mg\cdot \mg^{-1})\ .\label{eq: kinetic term with classical sol}
\end{align}
where $\alpha_0$ is defined by
\begin{equation}
    \alpha_0\equiv{2\pi \over (q-1) \tan {\pi \over 2q} }\ .\label{def: alpha0}
\end{equation}
Note that the leading and the subleading terms in the $\epsilon$-expansion contribute to \eqref{eq: kinetic term with classical sol} while the higher order terms vanishes. Hence, the low energy effective action reads 
\begin{equation}
    S_{\text{eff}}\equiv-{N \alpha_{\tsdiffeo}\over J} \int d\tau d\theta \;2\left[ {\sD^4 y\over  \sD y}- 2{ \sD^2 y \sD^3 y\over   [\sD y]^2} \right]-{N\alpha_\tsoq \over J} \int d\tau d\theta \; {1\over 2\level} \tr \left[ \maJ\sD \maJ +{1\over \level} \maJ^3\right]\label{def: low energy effective action}
\end{equation}
where the super-current $\maJ(\tau,\theta)$ is defined\footnote{We include the level $\level$ in the definition of the super-current as a bookkeeping parameter, and it does not play any role in this paper. } by
\begin{equation}
    \maJ \equiv -\level\; \sD \mg \cdot \mg^{-1}=-\level\mk+\theta \; (-\level\partial_\tau \mh \cdot \mh^{-1} -\level\mk \mk )
\end{equation}
where $\mg=e^{\theta \mk}\mh$. In addition, the coefficients $\alpha_\tsdiffeo$ and $\alpha_\tsoq$ are defined\footnote{Note that the numerical coefficients $\alpha_\tsdiffeo$ and $\alpha_\tsoq$ should be evaluated more precisely by the large $q$ expansion~\cite{Maldacena:2016hyu} or $s$-regularization~\cite{Jevicki:2016ito}.} by
\begin{equation}
    \alpha_\tsdiffeo\equiv {1\over 8(q-1)^{1\over q} \alpha_0^{1\over q}} \hspace{8mm},\hspace{8mm}
    \alpha_\tsoq\equiv{1\over (q-1)^{1\over q} \alpha_0^{1\over q}}
\end{equation}
The first term of \eqref{def: low energy effective action} is the super-Schwarzian action which appears in the $\mathcal{N}=1$ SUSY SYK model~\cite{Fu:2016vas}. Using~\eqref{def: super reparametrization parametrization}, the super-Schwarzian effective action reads
\begin{equation}
    S_{\text{\tiny eff}, \tsdiffeo}=-{N\alpha_\tsdiffeo\over J} \int d\tau \left[\{f(\tau),\tau\}+2\partial_\tau \eta(\tau) \partial_\tau^2\eta(\tau)- \{f(\tau),\tau\} \eta(\tau)\partial_\tau \eta(\tau)\right]\label{eq: sdiffeo effective action}
\end{equation}
On the other hand, the second term in \eqref{def: low energy effective action} is an action of a super-particle on the $SO(q)$ group manifold which is analogous to the effective action in the (non-SUSY) SYK model with global symmetry~\cite{Yoon:2017nig}. In particular, in terms of the component fields of the $\hsoq$ matrix
\begin{equation}
    \mg(\tau,\theta)=e^{\theta \mk(\tau)}\mh(\tau)=\mh(\tau)+\theta \mk(\tau)\mh(\tau)\ ,
\end{equation}
one can express the $\soq$ effective action as
\begin{equation}
    S_{\text{\tiny eff},\tsoq}=-{N\alpha_\tsoq \over J}\int d\tau \; \tr \left[ {1\over 2}\partial^2_\tau \mh \cdot \boldsymbol{h}^{-1} + {1\over 2}\partial_\tau \boldsymbol{k} \cdot \boldsymbol{k} + {1\over 2}\boldsymbol{k} \cdot \partial_\tau \boldsymbol{h} \cdot \boldsymbol{h}^{-1}\cdot \boldsymbol{k} \right]\ .\label{eq: soq effective action}
\end{equation}
Furthermore, defining a matrix $M(\tau)$ by
\begin{equation}
    \boldsymbol{M}(\tau)\equiv\mathcal{P}\exp\left[-\int^\tau \mh(\tau')d\tau'\right]\ , 
\end{equation}
one has
\begin{align}
S_{\text{\tiny eff},\tsoq}
=-{N\alpha_\tsoq \over J}\int d\tau \; \tr \left[ {1\over 2}\partial_\tau \boldsymbol{M}\partial_\tau \boldsymbol{M} -{1\over 2}  \boldsymbol{k} \partial_\tau \boldsymbol{k} - {1\over 4}  \boldsymbol{k} \;[\partial_\tau \boldsymbol{M},\boldsymbol{k}]  \right]
\end{align}
This is a supersymmetric $q\times q$ matrix model, and it would be interesting to study it as the supersymmetric generalization of $c=1$ matrix model.

\section{Four Point Functions}
\label{sec: four point functions}

\subsection{Quadratic Action}
\label{sec: quadratic action}

Recall that the collective action was
\begin{equation}
S_{col}=-{N\over 2} \STr \left[\smD\msstar  \maPsi\right]+{N\over 2}\STr \log \maPsi + {JN\over 2  }\int  d\tau_1d\theta_1 d\tau_2 d\theta_2 \det[\maPsi(\tau_1,\theta_1;\tau_2,\theta_2)]
\end{equation}
In this section, we will expand this collective action to quadratic order in fluctuations around the large $N$ classical solution presented in Section~\eqref{sec: large n expansion}. Since $\maPsi$ lies in the product of two fundamental representation, it is natural to decompose the fluctuations into singlet, anti-symmetric and symmetric-traceless representations of $SO(N)$:
\begin{equation}
     \yng(1)\; \otimes \; \yng(1) = \singlet \; \oplus \; \anti \; \oplus \; \sym\ .
\end{equation}
Accordingly, we expand the bi-local field $\maPsi$ around the large $N$ classical solution as follows:
\begin{equation}
    \maPsi=\maPsi_{cl}+{\sqrt{2\over N}}\maeta_\singlet+{\sqrt{2\over N}}\mazeta_\anti+{\sqrt{2\over N}}\mazeta_\sym\label{eq: fluctuation around classical sol}
\end{equation}
where the bi-local fluctuation $\flucb_\channel$ ($\channel=\singlet,\anti,\sym$) are given by 
\begin{align}
    \maeta_\singlet(\tau_1,\theta_1;\tau_2,\theta_2)=&{1\over \sqrt{q}} \eta(\tau_1,\theta_1;\tau_2,\theta_2) \idm\\
    \mazeta_\anti(\tau_1,\theta_1;\tau_2,\theta_2)=&{1\over \sqrt{2\dyn_\anti}} \zeta_\anti^a(\tau_1,\theta_1;\tau_2,\theta_2) \maT^a_\anti\\
    \mazeta_\sym(\tau_1,\theta_1;\tau_2,\theta_2)=&{1\over \sqrt{2\dyn_\sym}} \zeta_\sym^a(\tau_1,\theta_1;\tau_2,\theta_2) \maT^a_\sym
\end{align}
where $\maT^a_\anti$ and $\maT^a_\sym$ are the generators in the anti-symmetric and symmetric-traceless representations, respectively. Also, $\dyn_{\text{\tiny R}}$ denotes the Dynkin index of the representation $\text{R}$ of $SO(q)$. \ie
\begin{equation}
	\dyn_{\text{\tiny R}}\equiv{\dim (\text{R}) C_2(\text{R})\over 2\dim(\anti)}
\end{equation}
where $C_2(\text{R})$ is the Casimir of the representation $\text{R}$. Note that the generator $\maT^a_\anti$ and $\maT^a_\sym$ are normalized in a way that 
\begin{equation}
    \tr (\maT^a_\channel\maT^b_\channel)=2\dyn_\channel\delta^{ab}\hspace{10mm}(\channel=\anti,\sym)
\end{equation}
Like the bi-local superfield $\maPsi$ (See~\eqref{eq: antisym of bilocal}), the fluctuations $\mazeta_\channel$ ($\channel=\singlet,\anti,\sym$) are also anti-symmetric in the extended bi-local superspace. But, because of the properties of the generators (\ie $(\maT^a_\anti)^t = -\maT^a_\anti,(\maT^a_\sym)^t = \maT^a_\sym$), the symmetry of the component fluctuations $\flucb_\channel^a$ becomes
\begin{equation}
    \fluca_\singlet(1,2)=-\fluca_\singlet(2,1)\hspace{4mm},\hspace{4mm}\flucb_\anti^a(1,2)=\flucb_\anti^a(2,1)\hspace{4mm},\hspace{4mm}\flucb_\sym^a(1,2)=-\flucb_\sym^a(2,1)
\end{equation}
where we use a short hand notation $\flucb_\channel(1,2)\equiv\flucb_\channel(\tau_1,\theta_1;\tau_2,\theta_2)$ ($\channel=\singlet,\anti,\sym$). The $SO(q)$ symmetry ensures that the bi-local fluctuations $\fluca_\singlet, \flucb_\anti^a$ and $\flucb_\sym^a$ do not couple to each other in the quadratic action. Indeed, expanding the collective action
\begin{equation}
    S_{col}\left[\maPsi_{cl}+  \sqrt{{2/N}}( \mazeta_\singlet + \mazeta_\anti + \mazeta_\sym) \right]= N S^{(0)}+ S^{(2)}[\mazeta_\singlet,\mazeta_\anti,\mazeta_\sym]+ {1\over \sqrt{N}} S^{(3)}[\mazeta_\singlet,\mazeta_\anti,\mazeta_\sym]+\cdots\ ,
\end{equation}
we find the quadratic collective action to be
%
%
\begin{align}
    S^{(2)} =& -{1\over 2}\sum_{\channel=\singlet,\anti,\sym}\str (\Psi_{cl}^{-1}\star\flucb_\channel \star\Psi_{cl}^{-1} \star \flucb_\channel ) \cr
    &+ \sum_{\channel=\singlet,\anti,\sym} {\const_\channel J\over 2} \int d\mu_1 d\mu_2 \; [\Psi_{cl}(1,2)]^{q-2}\flucb_\channel(1,2)\flucb_\channel(1,2)
\end{align}
where the measure is defined by $d\mu_i\equiv d\tau_i d\theta_i$ $(i=1,2)$ and $\Psi_{cl}$ has been given in~\eqref{eq: classical solution in susy syk} (\ie $\maPsi_{cl}=\Psi_{cl}\idm$). In addition, the numerical constants $\const_\channel$ ($\channel=\singlet,\anti,\sym$) are 
\begin{equation}
    \const_\singlet=(q-1)\hspace{4mm},\hspace{4mm}\const_\channel=-1\hspace{4mm},\hspace{4mm}\const_\channel=-1
\end{equation}
These numerical constants are important in analyzing the spectrum and the chaotic behavior.

Now, we will derive the Schwinger-Dyson equation for two point functions of the bi-local fluctuation defined\footnote{Here, we included the factor $2$ due to the definition of the fluctuation in~\eqref{eq: fluctuation around classical sol}.} by
\begin{align}
    \mathcal{F}_\singlet(\tau_1,\theta_1,\tau_2,\theta_2;\tau_3,\theta_3,\tau_4,\theta_4)\equiv&2\left\langle \fluca_\singlet(\tau_1,\theta_1;\tau_2,\theta_2)\fluca_\singlet(\tau_3,\theta_3;\tau_4,\theta_4) \right\rangle\ , \label{def: def four point function 1}\\
  \mathcal{F}_{\anti}^{ab} (\tau_1,\theta_1,\tau_2,\theta_2;\tau_3,\theta_3,\tau_4,\theta_4)\equiv&2\left\langle \flucb^a_{\anti}(\tau_1,\theta_1;\tau_2,\theta_2)\flucb^b_{\anti} (\tau_3,\theta_3;\tau_4,\theta_4) \right\rangle \ ,\label{def: def four point function 2}\\
    \mathcal{F}_{\sym}^{ab} (\tau_1,\theta_1,\tau_2,\theta_2;\tau_3,\theta_3,\tau_4,\theta_4)\equiv&2\left\langle \flucb^a_{\sym}(\tau_1,\theta_1;\tau_2,\theta_2)\flucb^b_{\sym} (\tau_3,\theta_3;\tau_4,\theta_4) \right\rangle\ ,\label{def: def four point function 3}
\end{align}
which corresponds to the Schwinger-Dyson equation for four point functions of the fermi superfield $\psi^{i\alpha}$'s. Note that the four point function $\mathcal{F}^{ab}_{\anti/\sym}$ is proportional to $\delta^{ab}$ because of $\soq$ symmetry. \ie
\begin{equation}
    \mathcal{F}_{\anti}^{ab}=\delta^{ab}\mathcal{F}_{\anti}\hspace{5mm},\hspace{5mm}\mathcal{F}_{\sym}^{ab}=\delta^{ab}\mathcal{F}_{\sym}
\end{equation}
Hence, we will omit the index $a$ in the derivation of the Schwinger-Dyson equation for simplicity. Let us consider the following functional identity for $\channel=\singlet,\anti,\sym$:
\begin{align}
    \int \mathcal{D}\flucb_\singlet\mathcal{D}\flucb_\anti\mathcal{D}\flucb_\sym \;{\delta\over \flucb_\channel(6,5)}\left( \flucb_\channel(3,4)e^{-S_{col}^{(2)}[\fluca_\singlet,\flucb_\anti,\flucb_\sym]} \right) =0\ .\label{eq: functional id}
\end{align}
To derive the Schwinger-Dyson equation from this identity, it is useful to evaluate the following functional derivatives\footnote{See Appendix~\ref{app: notations} for the convention of the bi-local functional derivatives.} with respect to the bi-locals:
\begin{equation}
     { \delta S^{(2)} \over  \delta \flucb_\channel(6,5) } =   -(\Psi_{cl}^{-1}\star\flucb_\channel \star\Psi_{cl}^{-1})(5,6)  - \const_\channel J [\Psi_{cl}(5,6)]^{q-2}\flucb_\channel(5,6)  \hspace{8mm}(\channel=\singlet,\anti,\sym)
\end{equation}
and
\begin{alignat}{6}
    {\delta \zeta_{\singlet/\sym}(3,4)\over \delta \zeta_{\singlet/\sym}(6,5)}&= -&&\theta_{63}\theta_{54}\delta(\tau_{63})\delta(\tau_{54})+\theta_{64}\theta_{53}\delta(\tau_{64})\delta(\tau_{53})\\
    {\delta \zeta_{\anti}(3,4)\over \delta \zeta_{\anti}(6,5)}&= &&\theta_{63}\theta_{54}\delta(\tau_{63})\delta(\tau_{54}) +\theta_{64}\theta_{53}\delta(\tau_{64})\delta(\tau_{53})
\end{alignat}
where $\theta_{ij}\equiv\theta_i-\theta_j$. Together with them, we multiply $\Psi_{cl}(1,5)d\mu_5\;\Psi_{cl}(2,6)d\mu_6$ to \eqref{eq: functional id} and integrate it over $\mu_5$ and $\mu_6$, then we have the Schwinger-Dyson equation for the four point function $\mathcal{F}_\channel$:
\begin{equation}
\mathcal{F}_\channel(1,2;3,4) = \mathcal{F}_{-\ftn(\channel), 0}( 1,2;3,4) + \int d\mu_5 d\mu_6 K_\channel ( 1, 2; 5,6) \mathcal{F}_\channel (5,6;3,4) \label{eq: eq for geometric series}
\end{equation}
where $\ftn(R)$ is the sign of the representation $R$ (\ie $\sigma(\singlet)=\sigma(\sym)\equiv +$ and $\sigma(\anti)\equiv-$). This equation expresses the four point function as a geometric series of which the common ratio is given by
\begin{alignat}{4}
        &\fluca_\singlet \hspace{5mm} &&: \hspace{5mm}K_\singlet(1, 2; 3,4) &&= (q-1)&&J  \Psi_{cl}( 1,3) \Psi_{cl}(2,4)  \Psi_{cl}^{q-2}(3,4)\\
        &\flucb_\anti \hspace{5mm} &&: \hspace{5mm}  K_\anti(1,2;3,4) &&= \hspace{2mm} &&J\Psi_{cl}( 1,3) \Psi_{cl}(2,4)  \Psi_{cl}^{q-2}(3,4)\\
        &\flucb_\sym \hspace{5mm} &&: \hspace{5mm} K_\sym(1,2;3,4) &&= \hspace{8mm} - &&J  \Psi_{cl}( 1,3) \Psi_{cl}(2,4)  \Psi_{cl}^{q-2}(3,4)
\end{alignat}
and its first term is
\begin{alignat}{6}
    &\fluca_\singlet,\flucb_\anti \hspace{5mm} &&:\hspace{5mm}\mathcal{F}_{-,0}\equiv\;&&-\;&&\Psi_{cl}(1;3)\Psi_{cl}(2;4)+\Psi_{cl}(1;4)\Psi_{cl}(2;3)\\
    &\flucb_\sym \hspace{5mm} &&:\hspace{5mm}\mathcal{F}_{+,0}\equiv\;&& &&\Psi_{cl}(1;3)\Psi_{cl}(2;4)+\Psi_{cl}(1;4)\Psi_{cl}(2;3)
\end{alignat}
\subsection{Conformal Eigenfunctions}
\label{sec: conformal eigenfunctions}

In this section, we will study two types of super-conformal eigenfunctions by using supersymmetric shadow representation following~\cite{Murugan:2017eto}. As shown in~\cite{Murugan:2017eto}, one of the super-conformal eigenfunction is turned out to be proportional to conformal eigenfunction found in~\cite{Maldacena:2016hyu}. The other super-conformal eigenfunction is also proportional to the other non-SUSY one in~\cite{Peng:2017spg,Bulycheva:2017uqj,Yoon:2017nig}. Hence, the properties of both super conformal eigenfunctions follows from the non-SUSY ones~\cite{Peng:2017spg,Bulycheva:2017uqj}.

The basic idea of the supersymmetric shadow representation~\cite{Murugan:2017eto} is that a super-conformal four point function can be constructed by deforming two decoupled CFTs by
\begin{equation}
    \varepsilon \int dyd\theta_y\; \mathcal{V}_h(y,\theta_y) \mathcal{V}'_{{1\over 2}-h}(y,\theta_y) 
\end{equation}
where the operators $\mathcal{V}_h$ and $\mathcal{V}'_{{1\over 2}-h}$ of conformal dimension $h$ and ${1\over 2}-h$ belong to different CFTs. Note that $\mathcal{V}_h$ has opposite Grassman signature to $\mathcal{V}'_h$ because the measure is Grassmannian odd. Then, one can write the (connected) four point function as
\begin{equation}
    {\left\langle \psi(1)\psi(2)\psi(3)\psi(4) \right\rangle\over  \langle \psi(1)\psi(2)\rangle \langle\psi(3)\psi(4) \rangle } =\epsilon\int dyd\theta_y \; {\left\langle \psi(1)\psi(2)\mathcal{V}_h(y) \right\rangle\over  \langle \psi(1)\psi(2)\rangle }  {\langle \mathcal{V}'_{{1/ 2}-h}(y)\psi(3)\psi(4) \rangle\over   \langle \psi(3)\psi(4)\rangle}\label{eq: shadow representation decomposition}
\end{equation}
where we omit the $O(N)$ indices and the flavor indices of the superfield. In this super-shadow representation, one can obtain four types of superconformal eigenfunctions depending on the Grassmann signature of $\mathcal{V}_h$ and the symmetry under the exchange $1\,\leftrightarrow\, 2$ (See Appendix~\ref{app: shadow representation} for the details):
\begin{equation}
    \Upsilon^B_{\mp,h}\hspace{4mm},\hspace{4mm} \Upsilon^F_{\mp,h}\ .
\end{equation}
The eigenvalue of the super-conformal Casimir corresponding to $\Upsilon^{B/F}_{\mp,h}$ is $h(h-{1\over 2})$~\cite{Murugan:2017eto}. The superscript $B$ (or, $F$) correpsonds to the case $\mathcal{V}_h$ is Grassmannian even (or, odd) and $\mathcal{V}_h'$ is Grassmannian odd (or, even, respectively) in \eqref{eq: shadow representation decomposition}. The subscript $\mp$ denotes the symmetry under the exchange $1\leftrightarrow 2$ in the four point function (and, the corresponding eigenfunction) without the denominator in \eqref{eq: shadow representation decomposition}. For example, $\Upsilon^{B/F}_-$ is symmetric under exchange $1\leftrightarrow 2$ because of the anti-symmetry of the two point function in the denominator\footnote{The notation for the eigenfunction in this paper is opposite to that of \cite{Peng:2017spg,Bulycheva:2017uqj} where the label of the eigenfunction denotes of its symmetry including the denominator. \eg $\Upsilon_-$ in this paper corresponds to $\Phi^s$ in \cite{Peng:2017spg,Bulycheva:2017uqj}.}. Furthermore, by construction of the shadow representation~\eqref{eq: shadow representation decomposition}, the exchange~$(\tau_1,\theta_1;\tau_2,\theta_2)\leftrightarrow (\tau_3,\theta_3;\tau_4,\theta_4)$ maps $\Upsilon^B$ to $\Upsilon^F$ and \vs. \ie
\begin{equation}
    \Upsilon^B_{\mp,h}(1,2,3,4)=\Upsilon^F_{\mp,{1\over 2}- h }(3,4,1,2)\label{eq: upsilon b and f} \ .
\end{equation}
Hence, we will mainly work with $\Upsilon^B_{\mp,h}$. In Appendix~\ref{app: shadow representation}, $\Upsilon^B_{+,h}$ is obtained in the same way as $\Upsilon^{B}_{-,h}$ in~\cite{Murugan:2017eto}:
\begin{align}
    \Upsilon^B_{-,h}(1,2,3,4)=&{1\over 2}\int dy d\theta_y\;{ |\langle 1,2 \rangle|^h |\langle 3,4 \rangle|^{1/2-h}\sgn(\tau_{34}) P(3,4,y) \over |\langle 1,y \rangle|^h|\langle 2,y \rangle|^h|\langle 3,y \rangle|^{1/2-h} |\langle 4,y \rangle|^{1/2-h}}\ ,\label{def: eigenfunction Bm}\\
    \Upsilon^B_{+,h}(1,2,3,4)=&-{1\over 2}\int dy d\theta_y\;{ |\langle 1,2 \rangle|^h |\langle 3,4 \rangle|^{1/2-h}\sgn(\tau_{12}) P(3,4,y) \over |\langle 1,y \rangle|^h|\langle 2,y \rangle|^h|\langle 3,y \rangle|^{1/2-h} |\langle 4,y \rangle|^{1/2-h}}\cr
    &\hspace{12mm}\times\sgn(\tau_1-y)\sgn(\tau_2-y)\sgn(\tau_3-y)\sgn(\tau_4-y)\label{def: eigenfunction Bp}
\end{align}
where $\langle i,j\rangle\equiv \tau_i-\tau_j -\theta_i\theta_j$ ($i,j=1,2,3,4$) and $P(1,2,3)$ is defined by
\begin{equation}
    P(1,2,3)\equiv{\theta_1\tau_{23}+\theta_2\tau_{31}+\theta_3\tau_{12}-2\theta_1\theta_2\theta_3\over |\langle1,2\rangle \langle2,3\rangle \langle 3,1\rangle |^{1\over 2} }\ .
\end{equation}
Using $OSp(1|2)$ analogous to the global conformal group $SL(2,R)$ in the non-SUSY case, the super-conformal eigenfunctions can be expressed in terms of the super-conformal cross ratio $\chi$ and $\zeta$ given by
\begin{align}
    \chi\equiv&{\langle 1,2\rangle\langle 3,4\rangle\over \langle 1,3\rangle\langle 2,4\rangle }-\zeta\ ,\label{def: cross ratio1}\\
    \zeta\equiv&{\langle 1,2\rangle\langle 3,4\rangle+\langle 2,3\rangle\langle 1,4\rangle+\langle 3,1\rangle\langle 2,4\rangle\over \langle 1,3\rangle\langle 2,4\rangle }\ .\label{def: cross ratio2}
\end{align}
In particular, it is convenient to fix
\begin{equation}
    \tau_1=0\hspace{4mm},\hspace{4mm}\tau_2=\chi\hspace{4mm},\hspace{4mm}\tau_3=1\hspace{4mm},\hspace{4mm}\tau_4=\infty \hspace{4mm},\hspace{4mm} \theta_3=\theta_4=0
\end{equation}
and, the super-conformal cross ratio becomes
\begin{equation}
    \chi=\tau_2\hspace{4mm},\hspace{4mm} \zeta=\theta_1\theta_2\label{eq: gauge1}\ .
\end{equation}
Then, the super-conformal eigenfunctions can be expressed as
%
%
%
\begin{equation}
    \Upsilon^B_{\mp,h}(\chi,\zeta)=\left(1+{h\zeta\over \chi}\right)\Phi_{\mp,h}(\chi) \ .\label{eq: upsilon phi}
\end{equation}
%
%
where the conformal eigenfunctions $\Phi_{\mp,h}$ are given by
\begin{align}
    \Phi_{-,h}(\chi)\equiv&{1\over 2} \int_{-\infty}^\infty dy { |\chi|^h \over |y|^h |y-1|^{1-h}|y-\chi|^h }\\
    \Phi_{+,h}(\chi)\equiv&{\sgn(\chi)\over 2} \int_{-\infty}^\infty dy { |\chi|^h \sgn(y)\sgn(y-1)\sgn(y-\chi) \over |y|^h |y-1|^{1-h}|y-\chi|^h }\ .
\end{align}
Note that the exchange $1\leftrightarrow 2$ corresponds to the following transformation of the super-conformal cross ratio
\begin{equation}
    (\chi,\zeta)\hspace{4mm}\longrightarrow \hspace{4mm} \left({\chi\over \chi-1},{\zeta \over \chi-1}\right)\ .
\end{equation}
Under this transformation, one can easily confirm the symmetry of $\Upsilon^B_{\mp,h}(\chi,\zeta)$ from the analogous symmetry of $\Phi_{\mp,h}$. \ie
\begin{equation}
    \Phi_{\mp,h}\left({\chi\over \chi-1}\right)=\pm \Phi_{\mp,h}(\chi)\hspace{5mm}\Longrightarrow \hspace{5mm} \Upsilon^B_{\mp,h}\left({\chi\over \chi-1},{\zeta\over \chi-1}\right)=\pm \Upsilon^B_{\mp,h}(\chi,\zeta)\ .\label{eq: 12 exchange symmetry}
\end{equation}
and, similar for $\Upsilon^F_{\mp}$. In addition, note that the super-conformal cross ratios in~\eqref{def: cross ratio1} and \eqref{def: cross ratio2} are invariant under the exchange $(1,2)\leftrightarrow(3,4)$. Hence, in terms of the super-conformal cross ratio, the relation between $\Upsilon^B$ and $\Upsilon^F$ in~\eqref{eq: upsilon b and f} becomes 
\begin{equation}
    \Upsilon^F_{\mp,h}(\chi,\zeta)=\Upsilon^B_{\mp,{1\over 2}-h}(\chi,\zeta)\label{eq: 12 34 exchange symmetry}
\end{equation}

Now, we will present the inner products of the super-conformal eigenfunctions. The inner product in the non-SUSY SYK model was defined by \cite{Maldacena:2016hyu}
\begin{equation}
    (f_1,f_2)\equiv\int_0^2 {d\chi \over \chi^2} f_1(\chi)f_2(\chi)
\end{equation}
And, \cite{Murugan:2017eto} generalized it into the supersymmetric inner product:
\begin{equation}
    \langle F,G \rangle\equiv-\int_0^2 {d\chi d\zeta\over \chi+\zeta}F(\chi,\zeta)G(\chi,\zeta)
\end{equation}
where $d\zeta\equiv d\theta_2 d\theta_1$ for the choice in \eqref{eq: gauge1} so that $\int d\zeta \; \zeta=1$.

As shown in~\cite{Murugan:2017eto} for the case of $\Upsilon^B_{-,h}$, \eqref{eq: upsilon phi} enable us to reduce the inner product of the super-conformal eigenfunctions into that of the corresponding conformal eigenfunctions:
\begin{equation}
    \langle \Upsilon^B_{\mp,h},\Upsilon^B_{\mp, h'} \rangle
    =(1-h-h')(\Phi_{\mp,h},\Phi_{\mp, h'})
\end{equation}

The complete set of the conformal eigenfunctions consists of continuous and discrete states. Note that the conformal eigenfunction $\Phi_{\mp,h}$ has symmetry analogous to \eqref{eq: 12 34 exchange symmetry}:
\begin{equation}
    \Phi_{\mp, h}(\chi)=\Phi_{\mp, 1-h}(\chi)\ ,
\end{equation}
This symmetry of the conformal eigenfunction can restrict $h$ for the complete set. Therefore, the continuous states are given by
\begin{equation}
    \Phi_{\mp,h}(\chi) \hspace{5mm}:\hspace{5mm} h= {1\over 2} +is \hspace{8mm}  (s\geqq 0)\ ,
\end{equation}
and their inner product is found to be~\cite{Peng:2017spg,Bulycheva:2017uqj}
\begin{align}
    (\Phi_{\mp,h},\Phi_{\mp, h'})={\pi \tan {\pi h}\over 4h-2} 2\pi [\delta(s-s')+\delta(s+s')]
\end{align}
where $h={1\over 2}+is$ and $h'={1\over 2}+is'$. On the other hand, the discrete states are
\begin{alignat}{3}
    &\Phi_{-,h}(\chi)\hspace{5mm}:&&\hspace{5mm}  h= 2n\hspace{8mm} &&(n=1,2,\cdots)\ ,\\
    &\Phi_{+,h}(\chi)\hspace{5mm}:&&\hspace{5mm}  h= 2n-1\hspace{8mm} &&(n=1,2,\cdots)\ ,
\end{alignat}
and, the inner product is given by
\begin{align}
    (\Phi_{\mp, h},\Phi_{\mp,h})={\delta_{n,n'}\pi^2\over 4h-2}
\end{align}
This leads to the complete set of the super-conformal eigenfunctions via~\eqref{eq: upsilon phi}. However, since \eqref{eq: 12 34 exchange symmetry} is not the symmetry of $\Upsilon^B_{\mp,h}$ but gives the relation to $\Upsilon^F_{\mp,{1\over 2}-h}$, there is no restriction in $h$ for the super-conformal eigenfunctions. Hence, we have
\begin{align}
    \langle \Upsilon^B_{\mp,h},\Upsilon^B_{\mp, h'} \rangle=-2\pi \delta(s-s'){\pi \tan \pi h\over 2}\label{eq: inner prod 1}
\end{align}
where $h={1\over 2}+is$ and $h'={1\over 2}+is'$ $(s,s'\in\mathbb{R})$ for continuous states, and
\begin{alignat}{3}
    &\langle \Upsilon^B_{-,2n},\Upsilon^B_{-, 2n'} \rangle=- {\delta_{n,n'}\pi^2\over 2}\hspace{5mm}&&,\hspace{5mm}\langle\Upsilon^B_{+, 2n-1},\Upsilon^B_{+, 2n'-1} \rangle= -{\delta_{n,n'}\pi^2\over 2}\label{eq: inner prod 2}\\
    &\langle \Upsilon^B_{-,1-2n},\Upsilon^B_{-, 1-2n'} \rangle= {\delta_{n,n'}\pi^2\over 2}\hspace{5mm}&&,\hspace{5mm}\langle \Upsilon^B_{+,2-2n},\Upsilon^B_{+, 2-2n'} \rangle= {\delta_{n,n'}\pi^2\over 2}\label{eq: inner prod 3}
\end{alignat}
where $n, n'\in \mathbb{Z}$ for discrete states. Note that the inner product between $\Upsilon^B_{-,h}$ and $\Upsilon^B_{+,h'}$ vanishes because it is reduced to the inner product of $\Phi_{-,h}$ and $\Phi_{+,h}$, which is zero due to the symmetry in~\eqref{eq: 12 exchange symmetry}. All other inner products between also vanishes because they have different eigenvalues of super-conformal Casimir.

The eigenfunctions $\Upsilon^B_{-,{1\over 2}+is}$ $(s\in\mathbb{R})$, $\Upsilon^B_{-,2n}$ and $\Upsilon^B_{-,1-2n}$ $(n=1,2,\cdots)$ form a complete set of states which is anti-symmetric under the exchange $1\leftrightarrow 2$~\cite{Murugan:2017eto}. Likewise, another complete set of states which are symmetric under the exchange $1\leftrightarrow 2$ is composed of $\Upsilon^B_{+,{1\over 2}+is}$ $(s\in\mathbb{R})$, $\Upsilon^B_{+,2n-1}$ and $\Upsilon^B_{+,2-2n}$ $(n=1,2,\cdots)$. From the inner products of the super-conformal eigenfunctions in~\eqref{eq: inner prod 1}$\sim$\eqref{eq: inner prod 3}, the corresponding completeness relations reads
\begin{align}
    &-\sum_{n=1}^\infty \left[\Upsilon^B_{-,2n}(\chi,\zeta)\Upsilon^B_{-,2n}(\chi',\zeta')-\Upsilon^B_{-,1-2n}(\chi,\zeta)\Upsilon^B_{-,1-2n}(\chi',\zeta')\right]\cr
    &-\left.\int_{-\infty}^\infty {ds\over \pi^2 \tan \pi h}\Upsilon^B_{-,h}(\chi,\zeta)\Upsilon^B_{-,h}(\chi',\zeta') \right|_{h={1\over 2}+is}=-(\chi+\zeta)(\chi'+\zeta')\delta(\chi-\chi')\ ,\\
    &-\sum_{n=1}^\infty \left[\Upsilon^B_{+,2n-1}(\chi,\zeta)\Upsilon^B_{+,2n-1}(\chi',\zeta')-\Upsilon^B_{+,2-2n}(\chi,\zeta)\Upsilon^B_{+,2-2n}(\chi',\zeta')\right]\cr
    &-\left.\int_{-\infty}^\infty {ds\over \pi^2 \tan \pi h}\Upsilon^B_{+,h}(\chi,\zeta)\Upsilon^B_{+,h}(\chi',\zeta') \right|_{h={1\over 2}+is}=-(\chi+\zeta)(\zeta+\zeta')\delta(\chi-\chi')   \ . 
\end{align}
%

\subsection{Diagonalization}
\label{sec: diagonalization}

Now, one can expand the four point functions $\mathcal{F}_{\channel}$ $(\channel=\singlet,\anti,\sym)$ in terms of the complete set of states which has the same symmetry under the exchange $1\leftrightarrow 2$:
\begin{align}
    \mathcal{F}_{\singlet/\sym}=&-\left.\int_{-\infty}^\infty {ds\over \pi^2\tan \pi h}\Upsilon^B_{-,h}(\chi,\zeta) \langle \Upsilon^B_{-,h}, \mathcal{F}_{\singlet/\sym}\rangle \right|_{h={1\over 2}+is}\cr
    &-\sum_{n=1}^\infty {2\over \pi^2}\left[ \Upsilon^B_{-,2n}(\chi,\zeta) \langle \Upsilon^B_{-,2n}, \mathcal{F}_{\singlet/\sym}\rangle-\Upsilon^B_{-,1-2n}(\chi,\zeta) \langle \Upsilon^B_{-,1-2n}, \mathcal{F}_{\singlet/\sym}\rangle \right]\ , \label{eq: four point function expansion 1}\\
    \mathcal{F}_{\anti}=&-\left.\int_{-\infty}^\infty {ds\over \pi^2\tan \pi h}\Upsilon^B_{+,h}(\chi,\zeta) \langle \Upsilon^B_{+,h}, \mathcal{F}_{\anti}\rangle\right|_{h={1\over 2}+is} \cr
    &-\sum_{n=1}^\infty {2\over \pi^2}\left[ \Upsilon^B_{+,2n-1}(\chi,\zeta) \langle \Upsilon^B_{+,2n-1}, \mathcal{F}_{\anti}\rangle-\Upsilon^B_{+,2-2n}(\chi,\zeta) \langle \Upsilon^B_{+,2-2n}, \mathcal{F}_{\anti}\rangle \right]\ .   \label{eq: four point function expansion 2} 
\end{align}
Since the bi-local super-conformal Casimir~\cite{Murugan:2017eto,Yoon:2017gut} commutes with the kernel, the super-conformal eigenfunction also diagonalize the kernel. Noting that \eqref{eq: eq for geometric series} can be written as
\begin{equation}
    \mathcal{F}_{\channel}={1\over 1-K_\channel}\mathcal{F}_{-\ftn(\channel),0}\ ,
\end{equation}
we need to evaluate the eigenvalue of the kernel $K_\channel$ corresponding to the super-confomal eigenfunctions as well as the overlap between the eigenfunction and the first term $\mathcal{F}_{\channel,0}$ of the geometric series in order to compute the inner product of $\mathcal{F}_{\channel}$ and the super-conformal eigenfunction. \ie
\begin{equation}
	\langle \mathcal{F}_{\channel},\Upsilon^{B/F}_{\mp} \rangle= {1\over1-k^{B/F}_{\channel }}\langle \mathcal{F}_{-\ftn(\channel),0}, \Upsilon^{B/F}_{\mp} \rangle
\end{equation}
By construction of the super-conformal eigenfunction with the super-shadow representation, the kernel will be diagonalized by the three point function $ \langle \psi(3)\psi(4) \mathcal{V}_h^{B/F}(y,\theta_y) \rangle$ in the super-shadow representation~\eqref{eq: shadow representation decomposition} for each channel.
%
%
For simplicity, one can take $y\rightarrow \infty$ limit where the three point function behaves as
\begin{equation}
    \langle \psi(3)\psi(4) \mathcal{V}_h^{B/F}(y,\theta_y) \rangle\sim y^{-2h}S^{B/F}(3,4)
\end{equation}
Then, the simpler function $S^{B/F}(3,4)$ will also diagonalize the kernel with the same eigenvalue. The flavor structure of $\psi(3)\psi(4)$ in the three point function will determine the symmetry\footnote{Here, in contrast to the super-conformal eigenfunctions $\Upsilon^B_{\mp,h}$, we did not include the two point function in the denominator. Therefore, $S^{B/F}_{-,h}$ is anti-symmetric under the exhange $3\leftrightarrow 4$ and \vs.} of $S^{B/F}(3,4)$ under the exchange $3\leftrightarrow 4$, and therefore, we have four types of eigenfunction for the kernel:
\begin{align}
    S^B_{-,h}(t_3,\theta_3,t_4,\theta_4)\equiv&{\sgn(\tau_{34})\over |\langle 3,4\rangle|^{2\Delta-h}}\ , \label{def: sBm function}\\
    S^F_{-,h}(t_3,\theta_3,t_4,\theta_4)\equiv&{(\theta_{3}-\theta_4)\over | \tau_{34}|^{2\Delta-h+{1\over 2} }}\ , \label{def: sFm function}\\
    S^B_{+,h}(t_3,\theta_3,t_4,\theta_4)\equiv&{1\over | \langle 3,4\rangle |^{2\Delta-h}}\ , \label{def: sBp function}\\
    S^F_{+,h}(t_3,\theta_3,t_4,\theta_4)\equiv&{\sgn(\tau_{34})(\theta_{3}-\theta_4) \over |\tau_{34} |^{2\Delta-h+{1\over 2} }}\ .\label{def: sFp function}
\end{align}
Using the function $S^{B/F}_{\mp,h}$, one can easily evaluate the eigenvalues of the kernel for each channel. For example, the eigenvalues of the kernel for the singlet channel can be evaluated by $S^{B/F}_{-,h}$ because of the symmetry under the $1 \leftrightarrow 2$ exchange:
\begin{align}
    \int d\tau_3 d\theta_3 d\tau_4 d\theta_4\; K_\singlet(1,2,3,4) S_{-,h}^B (3,4)= & k^B_\singlet (h) S^B_{-,h}(1,2)\label{eq: eigenvaleu eq1}\ ,\\
    \int d\tau_3 d\theta_3 d\tau_4d\theta_4\; K_\singlet(1,2,3,4) S_{-,h}^F (3,4)=&k^F_\singlet (h) S^F_{-,h}(1,2)    \label{eq: eigenvaleu eq2}\ .
\end{align}
where the kernel in the singlet channgel is given by
\begin{equation}
    K_\singlet(1,2,3,4)=(q-1)J\Psi_{cl}(1,3)\Psi_{cl}(2,4)[\Psi_{cl}(3,4)]^{q-2}
\end{equation}
The eigenvalue equation can be further simplified by choosing the particular points
\begin{equation}
    \tau_1=1\hspace{4mm},\hspace{4mm} \tau_2=0\hspace{4mm},\hspace{4mm} \theta_1=\theta_2=0\ .
\end{equation}
Then, the RHS of \eqref{eq: eigenvaleu eq1} is simply reduced to the eigenvalue $k^B_\singlet(h)$, and the LHS become a simple integral:
\begin{align}
    k^B_-(h)
    =&(q-1) {\tan {\pi \over 2q}\over 2\pi }\int d\tau_3 d\theta_3 d\tau_4 d\theta_4 \; {\sgn(1-\tau_3)\sgn(-\tau_4) \over |1-\tau_3|^{2\Delta}|\tau_4|^{2\Delta}|\langle 3,4 \rangle|^{1-2\Delta-h}}\cr
    =&-(q-1){\sin 2\pi \Delta -\sin \pi h \over \sin 2\pi \Delta } {\Gamma(-h+2\Delta )\Gamma(h+2\Delta )\over[\Gamma(2\Delta)]^2}\ .
\end{align}
The other eigenvalue $k^F_\singlet(h)$ immediately follows\footnote{One can repeat the same calculation for \eqref{eq: eigenvaleu eq2} and confirm this result.} from the relation between the eigenfunctions $\Upsilon^B_{-,h}$ and $\Upsilon^F_{-,{1\over 2}-h}$ in~\eqref{eq: 12 34 exchange symmetry}:
\begin{equation}
k^F_\singlet(h)=k^B_\singlet\left({1\over 2} -h\right)\label{eq: ks BF relation}
\end{equation}
%
%
%
In the same way, one can evaluate the eigenvalue $k^{B/F}_{\anti}(h)$ of the kernel in the anti-symmetric channel given by
\begin{equation}
    K_\anti(1,2,3,4)=J\Psi_{cl}(1,3)\Psi_{cl}(2,4)[\Psi_{cl}(3,4)]^{q-2}
\end{equation}
and, we have
\begin{align}
    k^B_\anti(h)
    =&{\sin 2\pi \Delta +\sin \pi h \over \sin 2\pi \Delta } {\Gamma(-h+2\Delta )\Gamma(h+2\Delta )\over[\Gamma(2\Delta)]^2}\\
    k^F_\anti(h)=&k^B_\anti\left({1\over 2}-h\right)
\end{align}
%
%
%
Note that it is enough to evaluate for the singlet and anti-symmetric channel because the eigenfunctions of the singlet and the symmetric-traceless channel are identical, and the two kernels are proportional each other:
\begin{equation}
    K_{\sym}(1,2,3,4)=-{1\over q-1} K_{\singlet}(1,2,3,4)
\end{equation}

Now, we compute the overlap $\langle \Upsilon^B_{\channel ,h} , \mathcal{F}_{-\ftn(\channel),0} \rangle$. For this purpose, we will employ the trick of \cite{Murugan:2017eto}. We start with a $OSp(1|2)$ invariant integral of the form
\begin{align}
    {1\over 2}\int_{-\infty}^\infty {d\tau_1 d\theta_1 d\tau_2 d\theta_2\over \tau_{12}-\theta_1\theta_2} {d\tau_3 d\theta_3 d\tau_4 d\theta_4\over \tau_{34}-\theta_3\theta_4}dy d\theta_y {|\langle 1,2\rangle|^h \sgn(\tau_{12}) P(1,2,y) |\langle 3,4\rangle |^{{1\over 2} -h}\over |\langle 1,y\rangle |^h |\langle 2,y\rangle|^{h} |\langle 3,y\rangle|^{{1\over 2}-h}|\langle 4,y\rangle|^{{1\over 2}-h} }\mathcal{F}'_{-,0}\label{eq: integral for 1st term}
\end{align}
where $\mathcal{F}'_{-,0}$ is defined by
\begin{equation}
    \mathcal{F}'_{-,0}\equiv -{\Psi_{cl}(1,3)\Psi_{cl}(2,4)\over \Psi_{cl}(1,2)\Psi_{cl}(3,4)}
\end{equation}
We can evaluate this integral by using $OSp(1|2)$ to set any three of $\tau_1,\cdots, \tau_4,y$ to $a,b,c$ and the fermionic partners of $a,b$ to $0$ provided we insert in the integral $-\sgn(a-b)|(a-c)(b-c)|$. We will now perform this integral by the following two choices:
\begin{itemize}
    \item First, we choose 
\begin{align}
    \tau_1=0(=c)\quad,\quad \tau_3=1\quad,\quad\theta_3=0\quad,\quad \tau_4=\infty\quad ,\quad \theta_4=0
\end{align}
and we insert $|(\tau_4-0)|$ in the integral. Recalling the form of the $\Upsilon^F_{-,h}\left(=\Upsilon^B_{-,{1/2}-h}\right)$ in \eqref{def: eigenfunction Bm} and the symmetry 
\begin{align}
    \Upsilon^F_{-,h}(\chi,\zeta)=\Upsilon^F_{-,h}\left({\chi\over \chi-1},{\zeta\over \chi-1}\right)
\end{align}
which is analogous to \eqref{eq: 12 exchange symmetry}, one has the integral in~\eqref{eq: integral for 1st term} to be
%
%
%
\begin{equation}
    -\int_{-\infty}^\infty { d\chi d\zeta \over \chi+\zeta} \; \Upsilon^F_{-,h}(1,2,3,4)\mathcal{F}'_{-,0}(1,2,3,4)= \langle \Upsilon^F_{-,h},\mathcal{F}_{-,0} \rangle
\end{equation}
%
%
%
%
%

\item On the other hand, one can also choose 
\begin{align}
    \tau_1=1\quad,\quad \tau_2=0\quad,\quad y=\infty(=c)\quad,\quad \theta_1=\theta_2=0\ .
\end{align}
In this choice, the integral~\eqref{eq: integral for 1st term} can be evaluated to be
\begin{align}
    &{1\over 2}\int_{-\infty}^\infty {d\tau_3 d\theta_3 d\tau_4 d\theta_4\over \tau_{34}-\theta_3\theta_4}   |\langle 3,4\rangle |^{{1\over 2} -h} {\Psi_{cl}(1,3)\Psi_{cl}(2,4)\over \Psi_{cl}(1,2)\Psi_{cl}(3,4)}\cr
    =&{1\over 2\Lambda^q J (q-1)}\int_{-\infty}^\infty d\tau_3 d\theta_3 d\tau_4 d\theta_4 \; K_\singlet((1,0),(0,0);3,4)  S^B_{-,{1\over 2}-h}(3,4)\cr
    =&{1\over 2(q-1)J\Lambda^q}k^F_\singlet(h)
\end{align}
where we used \eqref{def: sBm function} and \eqref{eq: ks BF relation}.
%
%
%
\end{itemize}
Comparing the two ways of evaluating the same integral~\eqref{eq: integral for 1st term}, we have~\cite{Murugan:2017eto}
\begin{align}
    \langle \Upsilon^F_{-,h},\mathcal{F}_{-,0} \rangle =& {\alpha_0\over 2}k^F_\singlet(h)
\end{align}
where the constant $\alpha_0$ is given by
\begin{equation}
    \alpha_0={2\pi \over (q-1) \tan {\pi \over 2q} }\ .
\end{equation}
Recalling the relation between $\Upsilon^B$ and $\Upsilon^F$ in~\eqref{eq: 12 34 exchange symmetry} as well as the relation $k^B_\singlet(h)$ and $k^F_\singlet(h)$ in \eqref{eq: ks BF relation}, we obtain
\begin{align}
    \langle \Upsilon^B_{-,h},\mathcal{F}_{-,0} \rangle =\langle \Upsilon^F_{-,{1\over 2}-h},\mathcal{F}_{-,0} \rangle={\alpha_0\over 2}k^B_\singlet(h)\ .
\end{align}
For the anti-symmetric channel, one can repeat the same trick with the following $OSp(1|2)$-invariant integral
\begin{align}
    &-{1\over 2}\int_{-\infty}^\infty {d\tau_1 d\theta_1 d\tau_2 d\theta_2\over \tau_{12}-\theta_1\theta_2} {d\tau_3 d\theta_3 d\tau_4 d\theta_4\over \tau_{34}-\theta_3\theta_4}dy d\theta_y {|\langle 1,2\rangle|^h \sgn(\tau_{34}) P(1,2,y) |\langle 3,4\rangle |^{{1\over 2} -h}\over |\langle 1,y\rangle |^h |\langle 2,y\rangle|^{h} |\langle 3,y\rangle|^{{1\over 2}-h}|\langle 4,y\rangle|^{{1\over 2}-h} }\mathcal{F}'_{+,0}\cr
    &\hspace{8mm}\times \sgn(\tau_1-y)\sgn(\tau_2-y)\sgn(\tau_3-y)\sgn(\tau_4-y)\ .
\end{align}
where we define
\begin{equation}
    \mathcal{F}'_{+,0}={\Psi_{cl}(1,3)\Psi_{cl}(2,4)\over \Psi_{cl}(1,2)\Psi_{cl}(3,4) }\ .
\end{equation}
From the same two choices of ``gauge'', we have
\begin{equation}
    \langle \Upsilon^B_{+,h},\mathcal{F}_{+,0}\rangle=(q-1){\alpha_0\over 2}k^B_{\anti}(h)
\end{equation}
Note that we do not need to repeat the calculation for the case of the symmetric channel because it is the same as that of the singlet channel. Now, using the eigenvalues of the kernels and the overlaps, one can express the four point functions in~\eqref{eq: four point function expansion 1} and \eqref{eq: four point function expansion 2} in terms of the super-conformal eigenfunctions.

%
%
%
%

\vspace{3mm}

\noindent
\textbf{Summary:} Here, we summarize the result for the four point functions and the eigenvalues of the kernels.
\begin{itemize}
    \item \textbf{Four Point Function:}
\begin{align}
    {\mathcal{F}_{\singlet}\over \alpha_0}=&-\sum_{n=1}^\infty {1\over \pi^2}\left[ \Upsilon^B_{-,2n}(\chi,\zeta) {k^B_\singlet(2n)\over 1-k^B_\singlet(2n)}-\Upsilon^B_{-,1-2n}(\chi,\zeta) {k^B_\singlet(1-2n)\over 1-k^B_\singlet(1-2n)} \right]\cr
    &-\left.\int_{-\infty}^\infty {ds\over 2\pi^2\tan \pi h}\Upsilon^B_{-,h}(\chi,\zeta) {k^B_\singlet(h)\over 1-k^B_\singlet(h)} \right|_{h={1\over 2}+is}\label{eq: four point function S expansion}\\
    {\mathcal{F}_{\anti}\over (q-1)\alpha_0}=&-\sum_{n=1}^\infty {1\over \pi^2}\left[ \Upsilon^B_{+,2n-1}(\chi,\zeta) {k^B_\anti(2n-1)\over 1-k^B_\anti(2n-1)}-\Upsilon^B_{+,2-2n}(\chi,\zeta) {k^B_\anti(2-2n)\over 1-k^B_\anti(2-2n)} \right]\cr
    &-\left.\int_{-\infty}^\infty {ds\over 2\pi^2\tan \pi h}\Upsilon^B_{+,h}(\chi,\zeta) {k^B_\anti(h)\over 1-k^B_\anti(h)} \right|_{h={1\over 2}+is}\label{eq: four point function A expansion}\\
    {\mathcal{F}_{\sym}\over (q-1)\alpha_0}=&-\sum_{n=1}^\infty {1\over \pi^2}\left[ \Upsilon^B_{-,2n}(\chi,\zeta) {k^B_\sym(2n)\over 1- k^B_\sym(2n)}-\Upsilon^B_{-,1-2n}(\chi,\zeta) {k^B_\sym(1-2n)\over 1 - k^B_\sym(1-2n)} \right]\cr
    &-\left.\int_{-\infty}^\infty {ds\over 2\pi^2\tan \pi h}\Upsilon^B_{-,h}(\chi,\zeta) {k^B_\sym(h)\over 1 - k^B_\sym(h)} \right|_{h={1\over 2}+is}\label{eq: four point function ST expansion}
\end{align}

    \item \textbf{Eigenvalue} for $\channel=\singlet,\anti,\sym$
\begin{align}
    k^B_\singlet(h)=&-(q-1){\sin {2\pi \Delta} -\sin (\pi h)\over \sin{2\pi \Delta}}{\Gamma\left(-h+2\Delta\right)\Gamma\left(h+2\Delta\right)\over \left[\Gamma\left(2\Delta\right)\right]^2}\\
    k^B_\anti(h)=&{\sin 2\pi \Delta +\sin \pi h \over \sin 2\pi \Delta } {\Gamma(-h+2\Delta )\Gamma(h+2\Delta )\over[\Gamma(2\Delta)]^2}=-{1\over q-1} k^B_\singlet(-h)\\
    k^B_\sym(h)=&-{1\over q-1}k^B_\singlet(h)
\end{align}
%
%
%
%

\end{itemize}

\section{Spectrum}
\label{sec: spectrum}

In this section, we will take OPE limit to read off the spectrum. Using the expression of the conformal eigenfunction $\Phi_{\mp,h}(\chi)$ for $0<\chi<1$~\cite{Peng:2017spg,Bulycheva:2017uqj}, the super-conformal eigenfunctions for~$0<\chi<1$ can be written as
\begin{align}
    \Upsilon^B_{-,h}(\chi,\zeta)=&\left(1+{h\zeta\over \chi}\right)\Phi_{-,h}(\chi)\equiv\left(1+{h\zeta\over \chi}\right)\left[A(h) F_{h}(\chi) +B(h) F_{1-h}(\chi)\right]\\
    \Upsilon^B_{+,h}(\chi,\zeta)=&\left(1+{h\zeta\over \chi}\right)\Phi_{+,h}(\chi)\equiv\left(1+{h\zeta\over \chi}\right)\left[B(h) F_{h}(\chi) +A(h) F_{1-h}(\chi)\right]    
\end{align}
where the function $F_h(\chi)$ is defined\footnote{Note that the function $F_h(\chi)$ in this paper differs from that of \cite{Murugan:2017eto} by the factor ${[\Gamma(h)]^2\over \Gamma(2h)}$.} by 
\begin{align}
    F_{h}(\chi)=&{ [\Gamma(h)]^2\over \Gamma(2h)} \chi^h \; {}_2 F_1(h,h;2h;\chi)
\end{align}
and the coefficient $A(h)$ and $B(h)$ are 
\begin{align}
    A(h)\equiv {1\over 2} \tan (\pi h) \cot {\pi h\over 2}\hspace{8mm},\hspace{8mm}   B(h)\equiv -{1\over 2} \tan (\pi h) \tan {\pi h\over 2}
\end{align}
Following~\cite{Murugan:2017eto}, we will manipulate \eqref{eq: four point function S expansion}$\sim$\eqref{eq: four point function ST expansion} in order to find the spectrum and OPE coefficient in each channel. In this calculation, it is useful to note that
\begin{equation}
    A(h)=B(1-h)\label{eq: AB relation1}
\end{equation}
and
\begin{equation}
    A(2n)=B(1-2n)=1\hspace{5mm},\hspace{5mm} A(2n-1)=B(2-2n)=0\hspace{8mm}(n\in\mathbb{Z})\label{eq: AB relation2}
\end{equation}

\hspace{3mm}

\noindent
\textbf{Singelet Channel:} Using \eqref{eq: AB relation1} and \eqref{eq: AB relation2}, the singlet channel four point function~\eqref{eq: four point function S expansion} becomes
\begin{align}
    &{\mathcal{F}_{\singlet}\over \alpha_0}=-\sum_{n=1}^\infty {A(2n)\over \pi^2}F_{2n}(\chi)\left[ \left(1+{2n\zeta\over \chi}\right) {k^B_\singlet(2n)\over 1-k^B_\singlet(2n)}-\left(1+{(1-2n)\zeta\over \chi}\right) {k^B_\singlet(1-2n)\over 1-k^B_\singlet(1-2n)} \right]\cr
    &\hspace{3mm}-\left.\int_{-\infty}^\infty {ds\over 2\pi^2\tan \pi h}\left(1+{h\zeta\over \chi}\right)\left[ A(h) F_h(\chi)+A(1-h) F_{1-h}(\chi) \right] {k^B_\singlet(h)\over 1-k^B_\singlet(h)} \right|_{h={1\over 2}+is}
\end{align}
By change $h$ into $1-h$ in the second term of the contour integral, one has
\begin{align}
    &{\mathcal{F}_{\singlet}\over \alpha_0}=-\sum_{n=1}^\infty {A(2n)\over \pi^2}F_{2n}(\chi)\left[ \left(1+{2n\zeta\over \chi}\right) {k^B_-(2n)\over 1-k^B_\singlet(2n)}-\left(1+{(1-2n)\zeta\over \chi}\right) {k^B_\singlet(1-2n)\over 1-k^B_\singlet(1-2n)} \right]\cr
    &-\left.\int_{-\infty}^\infty ds{A(h)F_h(\chi)\over 2\pi^2\tan \pi h}\left[ \left(1+{h\zeta\over \chi}\right){k^B_\singlet(h)\over 1-k^B_\singlet(h)}-\left(1+{(1-h)\zeta\over \chi}\right){k^B_\singlet(1-h)\over 1-k^B_\singlet(1-h)} \right]\right|_{h={1\over 2}+is}
\end{align}
Note that the second term in the discrete summation diverges at $h=2$ (\ie $k^B_\singlet(-1)=1$). This divergent contribution comes from the zero mode corresponding to the broken super-conformal symmetry. In this section, we drop the zero mode contribution in evaluating the spectrum and OPE coefficients. Expressing the discrete sum as a residue sum, we have
\begin{align}
    &\left.{\mathcal{F}_{\singlet}\over \alpha_0}\right|_{\text{\tiny non-zero}}=-\sum_{n=1}^\infty \underset{h=2n}{\text{Res}} {F_{h}(\chi)\over 2\pi \tan{\pi h\over 2} } \left(1+{h\zeta\over \chi}\right) {k^B_\singlet(h)\over 1-k^B_\singlet(h)}\cr
    &+\sum_{n=2}^\infty \underset{h=2n}{\text{Res}} {F_h(\chi)\over 2\pi \tan{\pi h\over 2} }\left(1+{(1-h)\zeta\over \chi}\right) {k^B_\singlet(1-h)\over 1-k^B_\singlet(1-h)}\cr
    &-{1\over 2\pi i}\int_{\mathcal{C}} dh{F_h(\chi)\over 2\pi\tan {\pi h\over 2}}\left[ \left(1+{h\zeta\over \chi}\right){k^B_\singlet(h)\over 1-k^B_\singlet(h)}-\left(1+{(1-h)\zeta\over \chi}\right){k^B_\singlet(1-h)\over 1-k^B_\singlet(1-h)} \right]\ .
\end{align}
By pulling the contour to infinity, the residue sum is cancelled. At the same time, this picks up poles at 
\begin{equation}
k^B_\singlet(h)=1\quad,\quad k^B_\singlet(1-h)=k^F_\singlet(h-{1\over 2})=1\hspace{8mm} (h>{1\over 2})
\end{equation}
and, we have
\begin{align}
    \left.{\mathcal{F}_{\singlet}\over \alpha_0}\right|_{\text{\tiny non-zero}}
    =& \underset{k_\singlet^B(h)=1\;,\; h>1/2}{\text{Res}}\;\; {F_{h}(\chi)\over 2\pi \tan{\pi h\over 2} } \left(1+{h\zeta\over \chi}\right) {k^B_\singlet(h)\over 1-k^B_\singlet(h)}\cr
    &-\underset{k_\singlet^F(h-{1\over 2})=1\;,\; h>1/2}{\text{Res}} \;\; {F_{h}(\chi)\over 2\pi \tan{\pi h\over 2} }\left(1+{(1-h)\zeta\over \chi}\right) {k^F_\singlet(h-1/2)\over 1-k^F_\singlet(h-1/2)}\label{eq: singlet four point function ope1}     
\end{align}
As pointed out in~\cite{Murugan:2017eto}, $h=1$ does not contribute to the residue sum because ${1\over \tan{\pi h\over 2}}$ cancels the simple pole of ${1\over 1-k^B_\singlet(h)}$ at $h=1$. Also, the residue sum includes the double pole at $h=2$, which has to be carefully analyzed together with the zero mode contributions. To read off OPE coefficients, it is convenient to express \eqref{eq: singlet four point function ope1} as
\begin{align}
    \left.{\mathcal{F}_{\singlet}\over \alpha_0}\right|_{\text{\tiny non-zero}}
    =& \sum_{k_\singlet^B(h)=1\;,\; h>1/2} (c^B_{\singlet,h})^2\left(1+{h\zeta\over \chi}\right) \chi^{h}\;{}_2F_1(h,h,2h;\chi)\cr
    &+\sum_{k_\singlet^F(h-{1\over 2})=1\;,\; h>1/2} (c^F_{\singlet,h})^2 \left(1+{(1-h)\zeta\over \chi}\right)\chi^{h}\;{}_2F_1(h,h,2h;\chi)\label{eq: singlet four point function ope2}
\end{align}
where the OPE coefficients are given by
\begin{alignat}{3}
    &(c^B_{\singlet,h})^2\equiv &&{1\over - 2 \pi \tan{\pi h\over 2} } {[\Gamma(h)]^2\over \Gamma(2h)} {1\over [k_\singlet^{B}(h)]'}>0 \hspace{8mm}&&\mbox{for}\hspace{4mm}k^B_\singlet(h)=1\\
    &(c^F_{\singlet,h})^2\equiv &&{1\over - 2 \pi \tan{\pi h\over 2} } {[\Gamma(h)]^2\over \Gamma(2h)} {1\over -[k_\singlet^{F}(h-1/2)]'}>0\hspace{8mm}&&\mbox{for}\hspace{4mm}k^F_\singlet(h-{1\over 2})=1
\end{alignat}
Note that $(1+{h\zeta\over \chi})\chi^h \; {}_2F_1(h,h,2h,\chi)$ in the first term of \eqref{eq: singlet four point function ope2} is the super-conformal block of (bosonic) primary of dimension $h$. On the other hand, $(1+{(1-h)\zeta\over \chi})\chi^{h}\;{}_2F_1(h,h,2h;\chi)$ in the second term of \eqref{eq: singlet four point function ope2} is an eigenfunction of the super-Casimir with eigenvalue $(h-{1\over 2})(h-1)$, which implies that it is related to the descendants of the fermionic primary of dimension $h-{1\over 2}$~\cite{Murugan:2017eto}.

Note that $k^B_\singlet(h)$ (and, $k^F_\singlet(h-1)$) is increasing function (and, decreasing function) around the zeros of $k^B_\singlet(h)=1$ (and, $k^F_\singlet(h-1)=1$, respectively). In addition, since $\tan {\pi h\over 2}$ is negative at these points, the square of the OPE coefficients are positive.

\hspace{3mm}

\noindent
\textbf{Anti-symmetric Channel:} In the same way, the four point function~\eqref{eq: four point function A expansion} in the anti-symmetric channel can be written as
%
%
\begin{align}
    &\left.{\mathcal{F}_{\anti}\over (q-1)\alpha_0}\right|_{\text{\tiny non-zero}}
    =+\sum_{n=1}^\infty \underset{h=2n-1}{\text{Res}} {F_{h}(\chi)\over 2\pi \cot{\pi h\over 2}} \left(1+{h\zeta\over \chi}\right) {k^B_\anti(h)\over 1-k^B_\anti(h)} \cr
    &-\sum_{n=2}^\infty \underset{h=2n-1}{\text{Res}} {F_{h}(\chi)\over 2\pi \cot{\pi h\over 2}}\left(1+{(1-h)\zeta\over \chi}\right) {k^B_\anti(1-h)\over 1-k^B_\anti(1-h)}\cr
    &+{1\over 2\pi i}\left.\int_{\mathcal{C}} dh{F_h(\chi)\over 2\pi\cot {\pi h\over 2}}\left[ \left(1+{h\zeta\over \chi}\right){k^B_\anti(h)\over 1-k^B_\anti(h)}-\left(1+{(1-h)\zeta\over \chi}\right){k^B_\anti(1-h)\over 1-k^B_\anti(1-h)} \right]\right|_{h={1\over 2}+is}
\end{align}
Note that we dropped $h=1$ term in the second discrete sum because it corresponds to the zero mode contribution and has to be carefully analyzed separately ($k^B_\anti(0)=1$). As in the singlet channel, we pull the contour to positive infinity, which cancels the discrete residue sum and leads to other residue sum including the double pole at $h=1$.
\begin{align}
    {\mathcal{F}_{\anti}\over (q-1)\alpha_0}
    =&-\underset{k_\anti^B(h)=1\;,\; h>1/2}{\text{Res}} {F_{h}(\chi)\over 2\pi \cot{\pi h\over 2} } \left(1+{h\zeta\over \chi}\right) {k^B_\anti(h)\over 1-k^B_\anti(h)}\cr
    &+\underset{k_\anti^F(h-{1\over 2})=1\;,\; h>1/2}{\text{Res}} \;\; {F_{h}(\chi)\over 2\pi \cot{\pi h\over 2} }\left(1+{(1-h)\zeta\over \chi}\right) {k^F_\anti(h-1/2)\over 1-k^F_\anti(h-1/2)}\label{eq: anti residue result}     
\end{align}
Expressing this in terms of the super-conformal blocks, one can read off the OPE coefficient:
\begin{align}
    {\mathcal{F}_{\anti}\over (q-1)\alpha_0}
    =&\sum_{k_\anti^B(h)=1\;,\; h>1/2} \left( c^B_{\anti,h} \right)^2 {1\over 2\pi \cot{\pi h\over 2} } \left(1+{h\zeta\over \chi}\right) \chi^{h}\; {}_2F_1(h,h,2h;\chi)\cr
    &+\sum_{k_\anti^F(h-{1\over 2})=1\;,\; h>1/2} \left(c^F_{\anti,h}\right)^2 \left(1+{(1-h)\zeta\over \chi}\right)\chi^{h}{}_2F_1(h,h,2h;\chi)
\end{align}
where the square of OPE coefficients are given by
\begin{alignat}{3}
    &(c^B_{\anti,h})^2\equiv&&{1\over  2 \pi \cot{\pi h\over 2} } {[\Gamma(h)]^2\over \Gamma(2h)} {1\over [k_\anti^{B}(h)]'}>0 \hspace{8mm}&&\mbox{for}\hspace{4mm}k^B_\anti(h)=1\\
    &(c^F_{\anti,m})^2\equiv&&{1\over  2 \pi \cot{\pi h\over 2} } {[\Gamma(h)]^2\over \Gamma(2h)} {1\over -[k_\anti^{F}(h-1/2)]'}>0\hspace{8mm}&&\mbox{for}\hspace{4mm}k^F_\anti(h)=1
\end{alignat}
Note that the square of the OPE coefficients $(c^{B/F}_{\anti,h})^2$ is positive because $k^B_\anti(h)$ (and, $k^F_\anti(h-{1\over 2})$) is an increasing function (and, a decreasing function) around the zeros of $k^B_\anti(h)=1$ (and, $k^F_\anti(h-1)=1$, respectively) and because $\cot {\pi h\over 2}$ is positive at the zeros.

\vspace{3mm}

\noindent
\textbf{Symmetric-traceless Channel:} Since the symmetric-traceless channel has the same conformal eigenfunctions as the singlet channel, the evaluation of the spectrum and OPE coefficients in the symmetric-traceless channel is parallel to that of the singlet channel. Recall that the eigenvalue of the kernel in the symmetric-traceless channel is
\begin{equation}
    k^{B/F}_{\sym}(h)=-{1\over q-1}k^{B/F}_{\singlet}(h)\ .
\end{equation}
Thus, the singlet-traceless channel does not have divergence in the four point function. This is consistent with the fact that the zero modes in the low energy effective action are not coupled to the symmetric-traceless channel (See Section~\ref{sec: leading contribution zero modes}). In the same ways as in the symmetric-traceless channel, one can immediately obtain
\begin{align}
    {\mathcal{F}_{\sym}\over (q-1)\alpha_0}
    =& \sum_{k_\sym^B(h)=1\;,\; h>1/2} \left(c^B_{\sym,h}\right)^2\left(1+{h\zeta\over \chi}\right) \chi^{h}{}_2F_1(h,h,2h;\chi)\cr
    &+\sum_{k_\sym^F(h-{1\over 2})=1\;,\; h>1/2} \left(c^F_{\sym,h}\right)^2 \left(1+{(1-h)\zeta\over \chi}\right)\chi^{h}{}_2F_1(h,h,2h;\chi)
\end{align}
where (the square of) the OPE coefficients are given by
\begin{alignat}{3}
    &(c^B_{\sym,h})^2=&&{1\over - 2 \pi \tan{\pi h\over 2} } {[\Gamma(h)]^2\over \Gamma(2h)} {1\over -[k_\sym^{B}(h)]'}>0 \hspace{8mm}&&\mbox{for}\hspace{4mm}k^B_\sym(h)=1\\
    &(c^F_{\sym,h})^2=&&{1\over - 2 \pi \tan{\pi h\over 2} } {[\Gamma(h)]^2\over \Gamma(2h)} {1\over [k_\sym^{F}(h-1/2)]'}>0\hspace{8mm}&&\mbox{for}\hspace{4mm}k^F_\sym(h)=1
\end{alignat}
One can confirm that they are positive.

%
%
%
%
%
%
%
%

\section{Out-of-time-ordered Correlators and Chaos}
\label{sec: chaotic behavior}

\subsection{Quadratic Low Energy Effective Action}
\label{sec: quadratic effective action}

For the contribution of the zero mode to four point function, we will first obtain the quadratic low energy effective action and derive the two point functions of the zero modes. Let us consider the infinitesimal super-reparametrization~\eqref{def: super reparametrization parametrization} with $f(\varphi)$ and $\eta(\varphi)$ around the finite temperature solution given by
\begin{align}
    f(\varphi)=&\tan\left[{(\varphi+\epsilon(\varphi))\over 2}\right]\hspace{5mm},\hspace{5mm} \eta(\varphi)\ .
\end{align}
where we use the dimensionless variable $\varphi\equiv{2\pi \over \beta}\tau$ for convenience. Together with the infinitesimal $\hsoq$ transformations given by
\begin{equation}
    \mg=\idm +i\boldsymbol{\rho}(\varphi) +\theta \mk(\varphi)\ ,
\end{equation}
we expanding the zero modes as follows:
\begin{align}
    \epsilon(\varphi)=&{1\over 2\pi}\sum_{n\in \mathbb{Z}}\epsilon_n e^{-in \varphi}\ ,\\
    \eta(\varphi)=&{1\over 2\pi}\sum_{n\in \mathbb{Z}+{1\over 2}}\eta_n e^{-in \varphi}\ ,\\
    \mrho(\varphi)=&{1\over 2\pi}\sum_{n\in \mathbb{Z}}\mrho_n e^{-in \varphi}={1\over 2\pi\sqrt{2\dyn_\anti}}\sum_{n\in \mathbb{Z}}\rho_n^a\maT^a_\anti e^{-in \varphi}\ ,\\
    \mk(\varphi)=&{1\over 2\pi}\sum_{n\in \mathbb{Z}+{1\over 2}}\mk_n e^{-in \varphi}={1\over 2\pi\sqrt{2\dyn_\anti}}\sum_{n\in \mathbb{Z}+{1\over 2}}k_n^a\maT^a_\anti e^{-in \varphi}\ ,
\end{align}
where $\maT^a_\anti$ is the $so(q)$ generator in the anti-symmetric representation. With these zero modes, the quadratic effective action can be derived from~\eqref{eq: sdiffeo effective action} and~\eqref{eq: soq effective action}:
%
%
%
%
%
\begin{align}
    S_{\text{\tiny eff}, \tsdiffeo}^{(2)}
    =&{N\alpha_\tsdiffeo\over \beta J}\sum_{n=2}^\infty  n^2(n^2-1)\epsilon_n\epsilon_{-n} - {4iN\alpha_\tsdiffeo\over \beta J}\sum_{n={3\over 2}}^\infty n\left(n^2-{1\over 4}\right)\eta_n\eta_{-n}\\
    S_{\text{\tiny eff}, \tsoq}^{(2)}
    =&{N\alpha_\tsdiffeo\over\beta J}\sum_{n=1}^\infty n^2 \rho_n^a\rho_{-n}^a+{iN\alpha_\tsdiffeo\over\beta J}\sum_{n={1\over 2}}^\infty nk_n^a  k_{-n}^a
\end{align}
From the quadratic effective actions, one can read off the two point functions of the zero modes:
\begin{align}
    \langle\epsilon_n \epsilon_{-n}\rangle=&{\beta J\over N\alpha_\tsdiffeo}{1\over n^2(n^2-1)}\label{eq: 2pt ftn zero mode 1}\\
    \langle\eta_n \eta_{-n}\rangle=&-{i\beta J\over 4 N\alpha_\tsdiffeo}{1\over n(n^2-{1\over 4})}\label{eq: 2pt ftn zero mode 2}\\
    \langle \rho^a_n \rho^b_{-n}\rangle=&\delta^{ab}{\beta J\over N\alpha_\tsoq}{1\over n^2}\label{eq: 2pt ftn zero mode 3}\\
    \langle k^a_n k^b_{-n}\rangle=&\delta^{ab}{i\beta J\over  N\alpha_\tsoq}{1\over n}\label{eq: 2pt ftn zero mode 4}
\end{align}
%
%
%
%
%
%

\subsection{Leading Contribution: Zero Modes}
\label{sec: leading contribution zero modes}

In this section, we will evaluate the zero mode contribution to the out-of-time-ordered correlators. For this, we consider the super-reparametrization and the $SO(q)$ local transformation of the classical solution at finite temperature. Using the super-conformal transformation
\begin{align}
    f(\varphi,\theta)=&f(\varphi+\theta \eta(\varphi))\ ,\\
    y(\varphi,\theta)=&\sqrt{\partial_\varphi f(\varphi)} \; \eta(\varphi) +\theta \sqrt{\partial_\varphi f(\varphi)}\left(1+{1\over 2}\eta(\varphi) \partial_\varphi\eta(\varphi)\right)\ ,
\end{align}
one can obtain the classical solution at finite temperature from the zero temperature one. For this, it is convenient to use the dimensionless variable $\varphi$ (equivalently, we choose $\beta=2\pi$):
\begin{align}
    \tau=&\tan {\varphi\over 2}\\
    \theta'=&{1\over \sqrt{2}}\sec {\varphi\over 2}\theta\ ,
\end{align}
where $(\tau,\theta')$ is the coordinate on the super-line at zero temperature. The corresponding super-Jacobian factor is given by
\begin{equation}
    \sD_\theta \theta'=\sqrt{\partial_\tau f(\tau)}={1\over \sqrt{2}}\sec{\varphi\over 2}\ .
\end{equation}
Hence, the classical solution at finite temperature is found to be
\begin{equation}
    \maPsi_{cl}(\varphi_1,\theta_1;\varphi_2,\theta_2)= {\Lambda\over 2} {\sgn(\varphi_{12})\over  \left|\sin{\varphi_{12}\over 2}-{1\over 2}\theta_1\theta_2\right|^{2\Delta} }\idm\ .\label{eq: classical solution at finite temperature}
\end{equation}
\vspace{3mm}

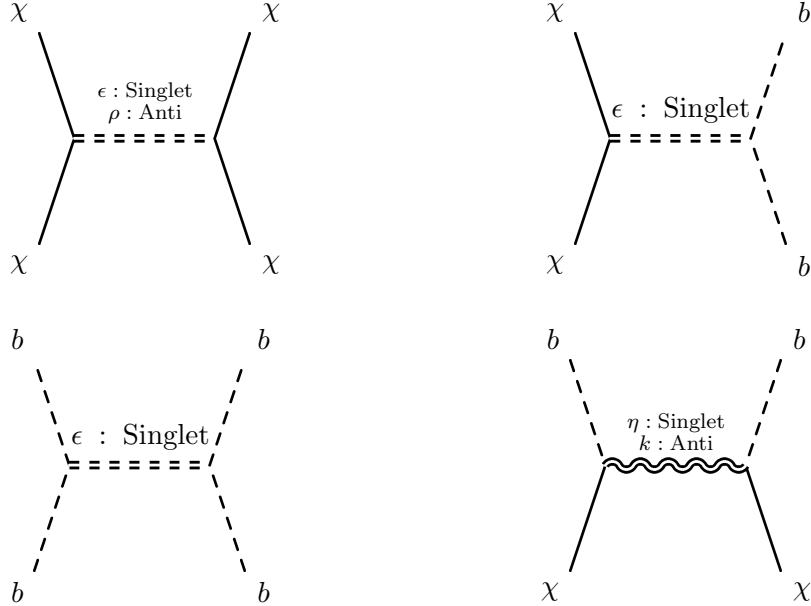
\begin{figure}[H]
\centering
\begin{subfigure}[t]{0.45\linewidth}
\centering
\begin{fmffile}{diagram1}
    \begin{fmfgraph*}(100,80)
        \fmfleft{a,b}
        \fmfright{c,d}
        \fmf{plain}{a,v1,b}
        \fmf{plain}{c,v2,d}
        \fmf{dbl_dashes,label=$\substack{\epsilon\;:\;\text{Singlet}\\\rho \;:\;\text{Anti}}$,tension=0.5}{v1,v2}
        \fmflabel{$\chi$}{a}
        \fmflabel{$\chi$}{b}
        \fmflabel{$\chi$}{c}
        \fmflabel{$\chi$}{d}
    \end{fmfgraph*}    
\end{fmffile}
\end{subfigure}
\begin{subfigure}[t]{0.45\linewidth}
\centering
\begin{fmffile}{diagram2}
    \begin{fmfgraph*}(100,80)
        \fmfleft{a,b}
        \fmfright{c,d}
        \fmf{plain}{a,v1,b}
        \fmf{dashes}{c,v2,d}
        \fmf{dbl_dashes,label=$\epsilon\;:\;\text{Singlet}$,tension=0.5}{v1,v2}
        \fmflabel{$\chi$}{a}
        \fmflabel{$\chi$}{b}
        \fmflabel{$b$}{c}
        \fmflabel{$b$}{d}
    \end{fmfgraph*}    
\end{fmffile}
\end{subfigure}

\vspace{15mm}

\begin{subfigure}[t]{0.45\linewidth}
\centering
\begin{fmffile}{diagram3}
    \begin{fmfgraph*}(100,80)
        \fmfleft{a,b}
        \fmfright{c,d}
        \fmf{dashes}{a,v1,b}
        \fmf{dashes}{c,v2,d}
        \fmf{dbl_dashes,label=$\epsilon\;:\;\text{Singlet}$,tension=0.5}{v1,v2}
        \fmflabel{$b$}{a}
        \fmflabel{$b$}{b}
        \fmflabel{$b$}{c}
        \fmflabel{$b$}{d}
    \end{fmfgraph*}    
\end{fmffile}
\end{subfigure}
\begin{subfigure}[t]{0.45\linewidth}
\centering
\begin{fmffile}{diagram4}
    \begin{fmfgraph*}(100,80)
        \fmfleft{a,b}
        \fmfright{c,d}
        \fmf{plain}{a,v1}
        \fmf{plain}{c,v2}
        \fmf{dashes}{b,v1}
        \fmf{dashes}{d,v2}
        \fmf{dbl_wiggly,label=$\substack{\eta\;:\;\text{Singlet}\\k \;:\;\text{Anti}}$,tension=0.5}{v1,v2}
        \fmflabel{$\chi$}{a}
        \fmflabel{$b$}{b}
        \fmflabel{$\chi$}{c}
        \fmflabel{$b$}{d}
    \end{fmfgraph*}    
\end{fmffile}
\end{subfigure}
\vspace{3mm}
\caption{Schematic diagrams for the contributions of the zero modes. The single solid line denotes the fermion $\chi$ while the single dashed line represents the auxiliary boson $b$. In addition, the double wavy line represents the $\eta$ zero mode for the singlet channel or the $k$ zero mode for the anti-symmetric channel. The double dashed line denotes the $\epsilon$ zero mode for the singlet channel or the $\rho$ zero mode for the anti-symmetric channel. However, the bi-locals $b^ib^i$ is not coupled to the zero mode $\rho$.}
\label{fig: scattering}
\end{figure}

\noindent
\textbf{Bosonic Reparametrization Zero Mode:} First, we evaluate the contribution of the bosonic zero mode of the broken super-reparametrization to the four point function 
\begin{equation}
    \langle \maPsi(1,2)\maPsi(3,4)\rangle = \maPsi_{cl}(1,2)\maPsi_{cl}(3,4)+\mathcal{O}({1\over N})\ .
\end{equation}
For this, we consider infinitesimal super-reparametrization on the circle by $\epsilon(\varphi)$:
\begin{alignat}{3}
    &\varphi\hspace{8mm}&&\longrightarrow \hspace{8mm}&&\varphi +\epsilon(\varphi)\ ,\\
    &\theta\hspace{8mm}&&\longrightarrow \hspace{8mm}&&\sqrt{1+\epsilon'}\theta=\left(1+{1\over 2}\epsilon'\right)\theta \ ,
\end{alignat}
and, the super-Jacobian factor is
\begin{align}
    \sD_\theta \theta'= \sqrt{f'}=1+{1\over 2}\epsilon'\ .
\end{align}
%
%
%
The variation of the classical solution~\eqref{eq: classical solution at finite temperature} with respect to~$\epsilon_n$ is found to be
\begin{equation}
    \delta_{\epsilon_n}\maPsi_{cl}
    ={2\Delta\over 2\pi} i \maPsi_{cl}  e^{-in{\varphi_1+\varphi_2\over 2}} \left( 1+ { {1\over 2} \theta_1\theta_2 \over  \sin {\varphi_{12}\over 2} } \right)\left[{\sin {n\varphi_{12}\over 2} \over \tan {\varphi_{12}\over 2}}-n \cos {n\varphi_{12}\over 2} \right]\label{eq: variation of classical solution zero mode1}
\end{equation}
where we define
\begin{equation}
    \epsilon(\varphi)\equiv{1\over 2\pi} \sum_{n}\epsilon_n e^{-in \varphi}\ .
\end{equation}
It is important to note that the variation $\delta_{\epsilon_n} \maPsi_{cl}$ is proportional to the $q\times q$ identity matrix (\ie $\maPsi_{cl}\sim \idm$). Therefore, the $\epsilon$ zero modes give a contribution only to the singlet channel $\mathcal{F}_\singlet$ up to leading order, which is given by
\begin{equation}
    \mathcal{F}^{\epsilon}_\singlet(1,2,3,4)\equiv{N\over q} \sum_{n}\left\langle  \epsilon_n\epsilon_{-n}\right\rangle \tr\left(\delta_{\epsilon_n} \maPsi_{cl}(1,2) \right) \tr \left(\delta_{\epsilon_{-n}} \maPsi_{cl}(3,4)\right)\label{eq: zero mode four point function sum1}
\end{equation}
where we chose the normalization ${N\over q}$ so that $\mathcal{F}^{\epsilon}_\singlet$ corresponds to the zero mode contribution to $\mathcal{F}_\singlet$ in~\eqref{def: def four point function 1}. Using the two point function of $\epsilon_n$ in~\eqref{eq: 2pt ftn zero mode 1}, we first evaluate $\mathcal{F}^\epsilon_\singlet$ in the particular ordering of $\varphi$'s on the circle as follow.
\begin{equation}
    \varphi_2<\varphi_3=0<\varphi_1<\varphi_4=\pi \label{def: configuration on circle}\ .
\end{equation}
Taking analytic continuation of $\varphi_1$ and $\varphi_2$
\begin{equation}
    \varphi_1={\pi \over 2}-{2\pi i\over \beta}t\hspace{5mm},\hspace{5mm}\varphi_2=-{\pi \over 2}-{2\pi i\over \beta}t\ ,\label{eq: analytic continuation of otoc}
\end{equation}
one can evaluate the out-of-time-ordered correlator from \eqref{eq: zero mode four point function sum1}~\cite{Maldacena:2016hyu,Yoon:2017nig}. At large $t$, we have
\begin{align}
    {\mathcal{F}^{\epsilon}_\singlet(1,2,3,4)\over \Psi_{cl}(1,2)\Psi_{cl}(3,4)}\simeq
     {\beta J\over 16\pi q \alpha_\tsdiffeo }\left(1+{1\over 2}\theta_1\theta_2\right)\left(1+{1\over 2}\theta_3\theta_4\right)e^{{2\pi \over \beta}t}\ .
\end{align}
%
%
%
where $\Psi_{cl}$ is given by
\begin{equation}
    \Psi_{cl}(\varphi_1,\theta_1;\varphi_2,\theta_2)\equiv{1\over q} \tr \left[\maPsi(\varphi_1,\theta_1;\varphi_2,\theta_2)\right]={\Lambda\over 2} {\sgn(\varphi_{12})\over  \left|\sin{\varphi_{12}\over 2}-{1\over 2}\theta_1\theta_2\right|^{2\Delta} }\ .
\end{equation}
Note that the $\epsilon$ zero mode contribution is composed of the terms such as $1, \theta_1\theta_2, \theta_3\theta_4$ or $\theta_1\theta_2\theta_3\theta_4$, which corresponds to the contribution to $\langle \chi^i\chi^i \chi^j \chi^j\rangle$, $\langle b^ib^i \chi^j \chi^j\rangle$, $\langle \chi^i\chi^i b^j b^j\rangle$ or $\langle b^ib^i b^j b^j\rangle$ in the singlet channel, respectively. They saturate the chaos bound, which is consistent with~\cite{Peng:2017spg}. Moreover, they are all related to the bosonic bi-local fields, and it is natural to get the $\epsilon$ zero mode contribution because of fermi statistics. (See Figure~\ref{fig: scattering}.)

\vspace{3mm}

\noindent
\textbf{Fermionic Reparametrization Zero Mode:} We analyze the contribution of the fermionic zero mode for the broken super-reparametrization. For this we consider the following infinitesimal super-reparametrization:
\begin{align}
    f(\varphi,\theta)=&\varphi+\theta \eta(\varphi) \ ,\\
    y(\varphi,\theta)=& \eta(\varphi) +\theta \left(1+{1\over 2}\eta(\varphi)\partial_\varphi \eta(\varphi)\right)\simeq\eta+\theta
\end{align}
where the corresponding super-Jacobian factor is 
\begin{align}
    \sD y=  \left( 1+{1\over 2} \eta \eta' \right)+\theta \eta'\simeq 1+\theta \eta'\ .
\end{align}
Under this infinitesimal transformation, the variation of the classical solution at finite temperature with respect to $\eta_n$ is found to be
\begin{align}
    \delta \maPsi_{cl}=&\maPsi_{cl} i {2\Delta\over 2\pi } e^{-in{\varphi_1+\varphi_2\over 2}}\left[ {\theta_1+\theta_2\over 2}\left(\cot{\varphi_{12}\over 4}\sin {n\varphi_{12}\over 2}-2n\cos {n\varphi_{12}\over 2} \right)\right. \cr
    &\hspace{35mm}\left.-  {\theta_1-\theta_2\over 2} i\left(\tan{\varphi_{12}\over 4}\cos {n\varphi_{12}\over 2}-2n\sin {n\varphi_{12}\over 2} \right)  \right]\delta \eta_n\label{eq: variation of classical solution zero mode2}
\end{align}
where we expand $\eta(\varphi)$ as follows.
\begin{equation}
    \eta(\varphi)={1\over 2\pi }\sum_{n\in \mathbb{Z}+{1\over 2} } \eta_n e^{-in\varphi}
\end{equation}
As in the $\epsilon$ zero mode, the variation of the classical solution is proportional to the $q\times q$ identity matrix. Hence, the $\eta$ zero mode also contributes only to the singlet channel up to leading order, which reads
\begin{equation}
    \mathcal{F}^\eta_\singlet(1,2,3,4)\equiv -{N\over q} \sum_{n}\left\langle  \eta_n\eta_{-n}\right\rangle \tr\left(\delta_{\eta_n} \maPsi_{cl}(1,2) \right) \tr \left(\delta_{\eta_{-n}} \maPsi_{cl}(3,4)\right)
\end{equation}
where the normalization ${N\over q}$ is chosen for its contribution to $\mathcal{F}_\singlet$ in~\eqref{def: def four point function 2}. In the same way, we evaluate $\mathcal{F}^\eta_\singlet$ with the particular ordering of $\varphi$'s on the Euclidean circle in~\eqref{def: configuration on circle}, and then we take analytic continuation $\varphi_1$ and $\varphi_2$ by~\eqref{eq: analytic continuation of otoc}. The large $t$ behavior of $\mathcal{F}^\eta_\singlet$ is found to be
\begin{equation}
    {\mathcal{F}^\eta_\singlet(1,2,3,4)\over \Psi_{cl}(1,2)\Psi_{cl}(3,4)}\simeq
    {\beta J\over 16\pi q^2 \alpha_\sdiffeo } {1+i\over\sqrt{2}}    (\theta_1-i\theta_2)(\theta_3-i\theta_4)e^{{\pi \over \beta}t}
\end{equation}
%
%
%
%
Note that this contribution consist of $\theta_1\theta_3$, $\theta_2\theta_3$, $\theta_1\theta_4$ or $\theta_2\theta_4$. They contribute to the four point functions such as $\langle b^i \chi^i b^j\chi^j \rangle$ in the singlet channel which are made of the fermionic bi-local fields. Also, the fermi statistics would allow such a contribution while the $\epsilon$ zero mode contribution is not allowed at the leading order (See~Figure~\ref{fig: scattering}). Hence, the Lyapunov exponent of those out-of-time-ordered correlators is ${\pi\over \beta}$ .

\vspace{3mm}

\noindent
\textbf{Bosonic Local $SO(q)$ Transformation Zero Mode:} Now, we consider the zero mode contribution from the broken local $SO(q)$ transformation. In particular, the bosoinc zero mode comes from the infinitesimal transformation $\mrho$:
\begin{align}
    \mg(\varphi,\theta)=\mh(\varphi)=\idm +i\mrho(\varphi)
\end{align}
%
%
%
The variation of the finite temperature classical solution with respect to $\rho$ is
\begin{equation}
    \delta_{\rho_n^a}\maPsi_{cl}
    ={1\over \pi \sqrt{2\dyn}} \maPsi_{cl} e^{-in{\varphi_1+\varphi_2\over 2}}\sin\left(n{\varphi_1-\varphi_2\over 2}\right) \maT^a_\anti\label{eq: variation of classical solution zero mode3}   
\end{equation}
where the expansion of the zero mode $\rho(\varphi)$ is given by
\begin{equation}
    \mrho(\varphi)={1\over 2\pi }\sum_{n} \mrho_n e^{-in\varphi}={1\over 2\pi \sqrt{2\dyn_\anti}} \sum_{n} \rho^a_n \maT^a_\anti e^{-in \varphi}\ .
\end{equation}
It is crucial to note that the variation is proportional to the generators $\maT_\anti^a$ of the anti-symmetric representation. Therefore, as in the non-SUSY SYK model with global symmetry~\cite{Yoon:2017nig}, the zero mode $\rho$ give a contribution only to the anti-symmetric channel. Thus, the zero mode contribution can be evaluated\footnote{Recall that $\mathcal{F}_{\anti/\sym}^{ab}=\delta^{ab}\mathcal{F}_{\anti/\sym}$} by
%
%
%
%
\begin{align}
    \delta^{ab}\mathcal{F}^\rho_\anti(1,2,3,4)\equiv{N\over 2\dyn_\anti}\sum_n \langle \rho_n^c \rho_{-n}^d \rangle \left\langle\tr\left[ \delta_{\rho_n^c}\maPsi(1,2)\maT^a_\anti\right]\tr \left[\delta_{\rho_{-n}^d}\maPsi(3,4)\maT^b_\anti\right] \right\rangle
\end{align}
where we also the normalization ${N\over 2\dyn_\anti}$ for the consistency with $\mathcal{F}_\anti$ in~\eqref{def: def four point function 3}. As before, the analytic continuation~\eqref{eq: analytic continuation of otoc} of the ordering in~\eqref{def: configuration on circle} gives the large time behavior of $\mathcal{F}^\rho_\anti$:
\begin{equation}
    {\delta^{ab}\mathcal{F}^\rho_\anti(1,2,3,4)\over \Psi_{cl}(1,2)\Psi_{cl}(3,4)}
    \simeq-i{\beta J\over 2\pi \alpha_\tsoq }\delta^{ab}{2\pi \over \beta}t\ .\label{eq: zero mode contribution h}
\end{equation}
%
%
The leading contribution of the $\rho$ zero mode is independent of Grassmannian variables. This implies that only the four point function $\langle \chi^i\chi^i \chi^j \chi^j \rangle$ in the anti-symmetric channel gets the $\rho$ zero mode contribution even though the fermi statistics  does not prevent the contribution to other four point functions such as $\langle \chi^i\chi^i b^j b^j \rangle$ or $\langle b^ib^i b^j b^j \rangle$. The corresponding Lyapunov exponent is $0$. But, the out-of-time-ordered correlator grows linearly in time, which was also observed in the non-SUSY SYK model with global symmetry~\cite{Yoon:2017nig} as well as the SYK-like tensor model~\cite{Choudhury:2017tax}.

\vspace{3mm}

\noindent
\textbf{Fermionic $SO(q)$ Local Transformation Zero Mode:} The fermionic $SO(q)$ transformation can be parametrized by
\begin{align}
    \mg(\varphi,\theta)=\idm+\theta \mk(\tau)\ ,
\end{align}
%
%
%
%
%
and, we expand the fermionic zero mode $\mk(\varphi)$ as follows.
\begin{equation}
    \mk(\varphi)={1\over 2\pi }\sum_{n\in \mathbb{Z}+{1\over 2}} \mk_n e^{-in\varphi}={1\over 2\pi \sqrt{2\dyn}} \sum_{n\in \mathbb{Z}+{1\over 2}} k^a_n \maT^a_\anti e^{-in \varphi}\ .
\end{equation}
One can eaily evaluate the variation of the classical solution at finite temperature with repsect to $k_n^a$:
\begin{align}
    \delta\maPsi_{cl}
    =&{1\over \pi \sqrt{2\dyn}}\maPsi_{cl} e^{-in{\varphi_1+\varphi_2\over 2}} \left[\cos\left({n\varphi_{12}\over 2}\right){\theta_1-\theta_2\over 2}-i\sin\left({n\varphi_{12}\over 2}\right){\theta_1+\theta_2\over 2}\right] \delta k^a_n \maT^a_\anti\ .\label{eq: variation of classical solution zero mode4}
\end{align}
Again, the variation is proportional to $\maT_\anti^a$, the anti-symmetric channel gets the contribution from the zero mode $\mk$ up to leading order:
%
%
%
%
\begin{align}
    \delta^{ab}\mathcal{F}^k_\anti(1,2,3,4)\equiv-{N\over 2\dyn_\anti}\sum_n \langle k_n^c k_{-n}^d \rangle \left\langle\tr\left[ \delta_{k_n^c}\maPsi(1,2)\maT^a_\anti\right]\tr \left[\delta_{k_{-n}^d}\maPsi(3,4)\maT^b_\anti\right] \right\rangle\ .
\end{align}
In the same way, the large time behavior of $\mathcal{F}^k_\anti$ is found to be
\begin{equation}
    {\delta^{ab}\mathcal{F}^k_\anti(1,2,3,4)\over \Psi_{cl}(1,2)\Psi_{cl}(3,4)}
    \simeq {\beta J\over 4\pi \alpha_\tsoq } \delta^{ab}(\theta_1\theta_3+\theta_2\theta_3+\theta_1\theta_4-\theta_2\theta_4 )+\mathcal{O}\left(e^{-{\pi\over \beta}t}\right)\ .\label{eq: zero mode contribution k}
\end{equation}
There is no growing term in large $t$, which would be expected from the bulk point of view if exist. Namely, this contribution could be related to a boundary gaugino which would give a negative (or, zero) Lyapunov exponent from \eqref{eq: lapunov exp for spin s}.

%
%

\vspace{3mm}

\noindent
\textbf{Summary:} Here, we present the summary of the zero mode contribution to the out-of-time-ordered correlators.

\vspace{3mm}

{
\renewcommand{\arraystretch}{2.3}
\begin{table}[H]
\centering
\begin{tabular}{c |c | c }
\multicolumn{2}{c|}{Channel}                  &\hspace{8mm} Zero Mode Contribution \hspace{8mm}   \\ \hline
\multirow{2}{*}{\hspace{3mm}Singlet\hspace{3mm}} & \hspace{2mm}${\scriptstyle\langle \chi^i\chi^i\chi^j\chi^j\rangle,\langle \chi^i\chi^ib^jb^j\rangle,\langle b^ib^ib^jb^j\rangle } $\hspace{2mm}  &       \hspace{6mm}$\beta J\left(1+{1\over 2}\theta_1\theta_2\right)\left(1+{1\over 2}\theta_3\theta_4\right)e^{{2\pi \over \beta}t}$\hspace{6mm}              \\ \cline{2-3} 
                         & ${\scriptstyle\langle b^i\chi^i b^j\chi^j\rangle}$ &    \hspace{6mm}  $  \beta J(\theta_1-i\theta_2)(\theta_3-i\theta_4)e^{{\pi \over \beta}t}$   \hspace{6mm}      \\ \hline
\multirow{2}{*}{Anti}    & ${\scriptstyle\langle \chi^i\chi^i\chi^j\chi^j\rangle } $  &       $J t $               \\ \cline{2-3} 
                         & ${\scriptstyle\langle \chi^i\chi^ib^jb^j\rangle,\langle b^ib^ib^jb^j\rangle,\langle b^i\chi^i b^j\chi^j\rangle}$ & No Growth \\ \hline
\multicolumn{2}{c|}{Symmetric-traceless}                      & No Growth \\ \hline
\end{tabular}
\caption{Summary of the zero mode contribution to the out-of-time-ordered correlators at large $t$. We omitted the $SO(q)$ indices in the four point functions.}
\label{tab: summary zero mode}
\end{table}
}

\subsection{Subleading Contribution: Non-zero Modes}
\label{sec: subleading contribution}

In this section, we will study the contributions of the non-zero mode to the out-of-time-ordered correlators following~\cite{Maldacena:2016hyu}. The key idea is to replace the eigenvalue $k^B_{\channel}$ in the residue sum of \eqref{eq: four point function S expansion}, \eqref{eq: four point function A expansion} and \eqref{eq: four point function ST expansion} with other function $k_{R,\channel}$ which we will define soon. 
%
%
In non-SUSY SYK model~\cite{Maldacena:2016hyu}, such a function~$k_{R}$ appeared in the diagonalization of the retarded kernel. In our model, we found that the eigenvalues of the retarded kernel found in~\cite{Peng:2017spg} play the same role in evaluation of the non-zero mode contributions. The eigenvalues of the retarded kernel in the singlet and anti-symmetric channels are given by~\cite{Peng:2017spg}
\begin{align}
    k_{R,\singlet}(h)\equiv&-{\Gamma\left(2-{1\over q}\right)\Gamma\left({1\over 2} -h+{1\over q}+{1\over 2}\right)\over \Gamma\left(1+{1\over q}\right)\Gamma\left({3\over 2} -h-{1\over q}+{1\over 2}\right) }\ ,\\
    k_{R,\anti}(h)\equiv&{\Gamma\left(2-{1\over q}\right)\Gamma\left({1\over 2} -h+{1\over q}-{1\over 2}\right)\over \Gamma\left(1+{1\over q}\right)\Gamma\left({3\over 2} -h-{1\over q}-{1\over 2}\right) }\ .
\end{align}
Note that we will not define $k_{R,\sym}(h)$ separately because the eigenvalue $k^B_\sym$ is proportional to $k^B_\anti$ (\ie $k^B_{\sym}(h)=-{1\over q-1}k^B_{\singlet}(h)$) and we can simply use $k_{R,\singlet}$ for the symmetric-traceless channel. The eigenvalue of the retarded kernel is related to the eigenvalues $k^B_{\singlet/\anti}(h)$ as follows.
\begin{align}
    {k_{R,\singlet}(1-h)\over k^B_\singlet(h)}=&{\cos\pi\left({1\over 2q}-{h\over 2}\right)\over \cos\pi\left({1\over 2q}+{h\over 2}\right) }\ ,\\
    {k_{R,\singlet}(h)\over k^B_\singlet(h)}=&{h-{1\over q}\over h-1+{1\over q}}{\sin\pi\left({1\over 2q}+{h\over 2}\right)\over \sin\pi\left({1\over 2q}-{h\over 2}\right) }\ ,\\
    {k_{R,\anti}(1-h)\over k^B_\anti(h)}=&(q-1){h-{1\over q}\over h-1+{1\over q}}{\sin\pi\left({1\over 2q}-{h\over 2}\right)\over \sin\pi\left({1\over 2q}+{h\over 2}\right) }\ ,\\
    {k_{R,\anti}(h)\over k^B_\anti(h)}=&(q-1){\cos\pi\left({1\over 2q}+{h\over 2}\right)\over \cos\pi\left({1\over 2q}-{h\over 2}\right) }\ .
\end{align}
At particular value of $h$, they becomes
\begin{alignat}{3}
    &k_\singlet^B(h)=&&k_{R,\singlet}(1-h)\hspace{10mm} &&(h=2,4,6,\cdots)\label{eq: kr identity1}\\
    &k_\singlet^B(h)=&&-{h-1+{1\over q} \over h-{1\over q}}k_{R,\singlet}(h)\hspace{10mm} &&(h=-1,-3,-5,\cdots)\label{eq: kr identity2}\\
    &k_\anti^B(h)=&&-{h-1+{1\over q} \over \left(q-1\right)\left(h-{1\over q}\right)}k_{R,\anti}(1-h)\hspace{10mm} &&(h=1,3,5,\cdots)\label{eq: kr identity3}\\
    &k_\anti^B(h)=&&{1\over q-1}k_{R,\anti}(h)\hspace{10mm} &&(h=-2,-4,-6,\cdots)\label{eq: kr identity4}
\end{alignat}
Unlike non-SUSY SYK model, since the four point function of the SUSY SYK model is expanded in terms of $\Upsilon^B_{\mp,h}$ and $\Upsilon^B_{\mp,1-h}$, we need not only the relation between $k_{R,\singlet/\anti}(1-h)$ and $k^B_{,\singlet/\anti}(h)$ but also the relation between $k_{R,,\singlet/\anti}(h)$ and $k^B_{,\singlet/\anti}(h)$.

We are interested in evaluating the following out-of-time-ordered correlator:
\begin{align}
     \sum_{i,j}\text{TR} \left[ y \psi^{i}(t,\theta_1)\;y\; \psi^{j }(0,\theta_2)\;y\; \psi^{i }(t,\theta_3)\;y\; \psi^{j }(0,\theta_4)\right]\label{def: adjoint otoc 2}\ ,
\end{align}
where $y^4$ is the thermal density matrix (\ie $y\equiv [\rho(\beta)]^{1\over 4}$). Note that we omit the $SO(q)$ indices, but one can easily consider this out-of-time-ordered correlator in each channel. In~\cite{Maldacena:2016hyu}, this out-of-time-ordered correlator can be evaluated at a particular cross ratio $\chi=2(1-i \sinh {2\pi \over \beta}t)^{-1}$ by analytic continuation from $\chi>1$ regime. In our case, we will follow the same calculation of the four point function at particular super-conformal cross ratios depending on whether we evaluate two point function of bosonic or fermionic bi-local fields.

Let us consider the same configuration of the four point function as in~\eqref{def: configuration on circle}.
\begin{equation}
    \varphi_1={\pi \over 2}-{2\pi i\over \beta}t\quad,\quad \varphi_2=-{\pi \over 2}-{2\pi i\over \beta}t\quad,\quad \varphi_3=0\quad,\quad \varphi_4=\pi
\end{equation}
%
%
%
At $t=0$, these four points are on the thermal circle. We map $(\varphi_i,\theta_i)$ to $(\tau_i,\theta_i')$ on the line $(i=1,2,3,4)$ by
\begin{align}
    \tau=&\tan {\varphi \over 2}\\
    \theta'=&\sqrt{{1\over 2}\sec^2{\varphi\over 2}}\theta={1\over \sqrt{2}}\sec {\varphi\over 2} \; \theta    
\end{align}
%
%
%
In particular, because $\tau_4=\infty$, the super-conformal cross ratio is simplified by
\begin{align}
    \chi=&
    {\tau_{12}\over \tau_{13}}+{\tau_{23}\over \tau_{13}^2}\theta'_3\theta'_1+{\theta'_2\theta'_3\over \tau_{13}}\\
    \zeta=&
    -{\theta'_1\theta'_2+\theta'_2\theta'_3+\theta'_3\theta'_1\over \tau_{13}}
\end{align}
%
%
%
%
Setting $\theta_4$ to be $0$, one can choose either $\theta_1,\theta_2$ or $\theta_3$ to be zero. We found that two choices $\theta_3=\theta_4=0$ and $\theta_2=\theta_4$ are useful for our purpose\footnote{$\theta_1=\theta_4=0$ is also straightforward, and it reproduce a part of results of $\theta_2=\theta_4=0$ case in large $t$. But, it does not seem to capture the ${\pi \over \beta}$ Lyapunov exponent.}.
\begin{itemize}
    \item $\boldsymbol{\theta_3=\theta_4=0}:$ For this choice, the super-conformal cross ratios are found to be
\begin{align}
    \chi=&{2\over 1-i\sinh{2\pi \over \beta}t}\label{def: cross ratio chi 34}\\
    \zeta=&i{1+ie^{{2\pi \over \beta}t}\over 1-ie^{{2\pi \over \beta}t}}{\theta_1\theta_2\over  \cosh {2\pi \over \beta }t }
\end{align}
At large $t$, they are asymptotic to
\begin{align}
    \chi\simeq& {4i\over e^{{2\pi \over \beta}t}}+{8\over e^{{4\pi \over \beta}t}}+\cdots\\
    \zeta\simeq& \left[-{2i\over e^{{2\pi \over \beta}t} }+\cdots\right]\theta_1\theta_2
\end{align}
From this cross ratio, one can read off the two point functions of bosonic bi-local fields such as $\chi^i\chi^i\chi^j\chi^j$ or $\chi^i\chi^i b^jb^j$. We present the large time behavior of two quantities which appear in the calculation of the four point function:
\begin{align}
    {\zeta\over \chi}\simeq&-{1\over 2}\theta_1\theta_2+\cdots \label{eq: cross ratio asymp 1}\\
    \log {1\over -i\chi}\simeq& {2\pi \over \beta}t+\cdots \label{eq: cross ratio asymp 2}\\
    {1\over -i\chi}\left(1-{\zeta\over \chi}\right)\simeq& {1\over 4}\left(1+{1\over 2}\theta_1\theta_2\right)e^{{2\pi \over \beta}t}+\cdots\label{eq: cross ratio asymp 3}
\end{align}

%
%
%
%
%
%
%
%
%
%

\item$\boldsymbol{\theta_2=\theta_4=0}:$ The super-conformal cross ratio are given by
\begin{align}
    \chi=&
    {2\over 1-i\sinh{2\pi \over \beta}t}-i \left({1+ie^{{2\pi \over \beta}t}\over 1-ie^{{2\pi \over \beta}t} }\right)^3{\theta_3\theta_1\over \sqrt{2}\left(\cosh {\pi \over \beta}t +i\sinh {\pi \over \beta}t\right)}\label{eq: cross ratio1 theta24} \\
    \zeta=&
    i{1+ie^{{2\pi \over \beta}t}\over 1-ie^{{2\pi \over \beta}t}}{\theta_3\theta_1\over \sqrt{2}\left(\cosh {\pi \over \beta}t +i\sinh {\pi \over \beta}t\right)}\label{eq: cross ratio2 theta24}
\end{align}
The large time behavior of the cross ratios are given by
\begin{align}
    \chi+\zeta\simeq&{4i\over e^{{2\pi\over \beta}t}}+{2\sqrt{2}(1-i)\over e^{{3\pi \over \beta}t}}\theta_3\theta_1\\
    \chi\simeq& {1\over \sqrt{2}}(1-i)e^{-{\pi\over \beta}t}\theta_1\theta_3+{4i\over e^{{2\pi \over \beta }t}}\\
    \zeta\simeq& -{1\over \sqrt{2}}(1-i)e^{-{\pi\over \beta}t}\theta_1\theta_3
\end{align}
%
%
%
Also, for the estimation of the contour integral, it is useful to note that
\begin{equation}
    \chi^{1\over 2} \left(1+{a\zeta\over \chi}\right)\simeq \sqrt{2}(1+i)e^{-{\pi \over \beta}t} + (1+i)\left(a-{1\over 2}\right)\theta_1\theta_3 +\cdots  \label{eq: cross ratio asymp 5}
\end{equation}
where $a$ is a constant. This cross ratio gives two point functions between fermionic bi-local fields such as $\langle b^i\chi^i b^j\chi^j\rangle$ or $\langle \chi^i\chi^i\chi^j\chi^j\rangle$. 

\end{itemize}

\noindent
Now, using these super-conformal cross ratios, we will evaluate the out-of-time-ordered correlators in each channel.

\vspace{3mm}

\noindent
\textbf{Singlet Channel:} For $\chi>1$, the super-conformal eigenfunction is given by
\begin{equation}
    \Upsilon^B_{\mp,h}(\chi,\zeta)=\left(1+{h\zeta\over \chi}\right) \widetilde{F}_{\mp,h}(\chi)
\end{equation}
where the conformal eigenfunction $\widetilde{F}_{\mp,h}(\chi)$ $(\chi>1)$ were found in~\cite{Maldacena:2016hyu,Peng:2017spg,Bulycheva:2017uqj}:
\begin{align}
    \widetilde{F}_{-,h}(\chi)=&{\Gamma\left({1\over 2}-{h\over 2}\right)\Gamma\left({h\over 2}\right)\over \sqrt{\pi}}\;\;{}_2F_1\left({h\over 2},{1-h\over 2}, {1\over 2}, {(\chi-2)^2\over \chi^2}\right)\ ,\label{eq: function m large chi}\\
    \widetilde{F}_{+,h}(\chi)=&-{2\Gamma\left(1- {h\over 2}\right)\Gamma\left({h\over 2}+{1\over 2} \right)\over \sqrt{\pi}}{\chi-2\over \chi}\;\;{}_2F_1\left({2-h\over 2},{h+1\over 2}, {3\over 2}, {(\chi-2)^2\over \chi^2}\right)\ . \label{eq: function p large chi}   
\end{align}
Using $\widetilde{F}_{-,h}(\chi)$, one can express \eqref{eq: four point function S expansion} as
\begin{align}
    &\left.{\mathcal{F}_{\singlet}\over \alpha_0}\right|_{\text{\tiny non-zero}}
    =-\sum_{n=1}^\infty \underset{h=2n}{\text{Res}}\; {\widetilde{F}_{-,h}(\chi) \over 2\pi \tan{\pi h\over 2}}\left(1+{h\zeta\over \chi}\right) {k^B_\singlet(h)\over 1-k^B_\singlet(h)}\cr
    &+\sum_{n=2}^\infty \underset{h=2n}{\text{Res}}\;  {\widetilde{F}_{-,h}(\chi)\over 2\pi \tan{\pi h\over 2}}\left(1+{(1-h)\zeta\over \chi}\right) {k^B_\singlet(1-h)\over 1-k^B_\singlet(1-h)}\cr
    &-{1\over 2\pi i}\left.\int_{\mathcal{C}} dh{\widetilde{F}_{-,h}(\chi)\over 2\pi\tan {\pi h\over 2}}\left[ \left(1+{h\zeta\over \chi}\right){k^B_\singlet(h)\over 1-k^B_\singlet(h)}-\left(1+{(1-h)\zeta\over \chi}\right){k^B_\singlet(1-h)\over 1-k^B_\singlet(1-h)} \right]\right|_{h={1\over 2}+is}
\end{align}
where we used
\begin{equation}
    {2\over \tan \pi h}={1\over \tan {\pi h\over 2}}-{1\over \tan {\pi (1-h)\over 2}}
\end{equation}
Now, one can replace $k^B_\singlet$'s in the residue sum with $k_{R,\singlet}$ by using \eqref{eq: kr identity1} and \eqref{eq: kr identity1}.
\begin{align}
    &\left.{\mathcal{F}_{\singlet}\over \alpha_0}\right|_{\text{\tiny non-zero}}=-\sum_{n=1}^\infty \underset{h=2n}{\text{Res}} \; {\widetilde{F}_{-,h}(\chi)\over 2\pi \tan{\pi h\over 2}} \left(1+{h\zeta\over \chi}\right) {k_{R,\singlet}(1-h)\over 1-k_{R,\singlet}(1-h)}\cr
    &-\sum_{n=2}^\infty \underset{h=2n}{\text{Res}} \;{\widetilde{F}_{-,h}(\chi)\over 2\pi \tan{\pi h\over 2}}\left(1+{(1-h)\zeta\over \chi}\right) {{{1\over q}-h\over 1-h-{1\over q}}k_{R,\singlet}(1-h)\over 1+{{1\over q}-h\over 1-h-{1\over q}}k_{R,\singlet}(1-h)}\cr
    &-{1\over 2\pi i}\int_{\mathcal{C}} dh{\widetilde{F}_{-,h}(\chi)\over 2\pi\tan {\pi h\over 2}}\left[ \left(1+{h\zeta\over \chi}\right){k^B_\singlet(h)\over 1-k^B_\singlet(h)}-\left(1+{(1-h)\zeta\over \chi}\right){k^B_\singlet(1-h)\over 1-k^B_\singlet(1-h)} \right]
\end{align}
We will pull the contour around poles in the residue sum to the contour $h={1\over 2}+is$. Note that $1-k_{R,\singlet}(1-h)$ in the denominator of the first term does not have zeros for $h>{1\over 2}$. On the other hand, $1+{{1\over q}-h\over 1-h-{1\over q} }k_{R,\singlet}(1-h)$ has one zero at $h=2$ so that the second term has double pole at $h=2$. Therefore, by pulling the contour, we have
\begin{align}
    &\left.{\mathcal{F}_{\singlet}\over \alpha_0}\right|_{\text{\tiny non-zero}}
    =-{1\over 2\pi i}\int_{\mathcal{C}} dh{\widetilde{F}_{-,h}(\chi)\over 2\pi\tan {\pi h\over 2}} \left(1+{h\zeta\over \chi}\right) \left[ {k^B_\singlet(h)\over 1-k^B_\singlet(h)}-{k_{R,\singlet}(1-h)\over 1-k_{R,\singlet}(1-h)}\right]\cr
    &+{1\over 2\pi i}\int_{\mathcal{C}} dh{\widetilde{F}_{-,h}(\chi)\over 2\pi\tan {\pi h\over 2}}\left(1+{(1-h)\zeta\over \chi}\right)\left[ {k^B_\singlet(1-h)\over 1-k^B_\singlet(1-h)} +{{{1\over q}-h\over 1-h-{1\over q}}k_{R,\singlet}(1-h)\over 1+{{1\over q}-h\over 1-h-{1\over q}}k_{R,\singlet}(1-h)} \right]\cr
    &+ \underset{h=2}{\text{Res}}\; {\widetilde{F}_{-,h}(\chi)\over 2\pi \tan{\pi h\over 2} }\left(1+{(1-h)\zeta\over \chi}\right) {{{1\over q}-h\over 1-h-{1\over q}}k_{R,\singlet}(1-h)\over 1+{{1\over q}-h\over 1-h-{1\over q}}k_{R,\singlet}(1-h)}\label{eq: singlet nonzero otoc}
\end{align}
By analytic continuation of $\widetilde{F}_{-,h}$ to the regime $\chi<1$, one can obtain its behavior at small~$\chi$:
\begin{equation}
    \widetilde{F}_{-,h}(\chi)\simeq {\Gamma\left({1\over 2} -{h\over 2}\right)\Gamma\left(h-{1\over 2} \right)\over 2^{1-h}\Gamma\left({h\over 2}\right)}(-i\chi)^{1-h}+\left(h\;\longrightarrow \; 1-h\right)\label{eq: ftilde m asympt}
\end{equation}
\begin{itemize}
    \item $\boldsymbol{\theta_3=\theta_4=0}:$ From~\eqref{eq: cross ratio asymp 1}, it is easy to see that the contour integral in~\eqref{eq: singlet nonzero otoc} does not contain the the exponentially growing terms as in the non-SUSY SYK model. The residue at the double pole $h=2$ gives a linear combination of $\widetilde{F}_{-,2}(\chi)$ and $\left.\partial_h\widetilde{F}_{-,h}(\chi)\right|_{h=2}$. The former leads to $e^{{2\pi \over \beta }t}$ which saturates the chaos bound. The contribution from the latter can be written as
\begin{align}
    \mathcal{F}_{\singlet,\text{non-zero}}\sim { \log {1\over -i\chi} \over -i \chi}\left(1-{\zeta\over \chi}\right)\sim t e^{{2\pi \over \beta}t }\left(1+{1\over 2}\theta_1\theta_2\right)
\end{align}
This seems to be faster than the exponential growth from the zero mode. However, recalling that the zero mode contribution is proportional to $\beta J$, this can be understood as the ${1\over \beta J}$ correction to the (leading) maximal Lyapunov exponent of $\langle \chi^i\chi^i \chi^j\chi^j \rangle$ and $\langle \chi^i\chi^i b^j b^j\rangle$ as explained in~\cite{Maldacena:2016hyu}. That is, the ${1\over \beta J}$ correction to a Lyapunov exponent $\lambda_L$ leads to the following ${1\over \beta J}$ sub-leading contribution to the exponential growth:
\begin{equation}
\lambda_L=\lambda_L^{(0)}+{1\over \beta J}\lambda_L^{(1)}+\cdots \hspace{3mm}\Longrightarrow \hspace{3mm} e^{\lambda_L t}=e^{\lambda_L^{(0)} t}+{\lambda_L^{(1)}\over  \beta J} \; t\;  e^{\lambda_L^{(0)}t}+\cdots
\end{equation}
This is what we have obtained in the non-zero mode contribution.
    
    \item $\boldsymbol{\theta_2=\theta_4=0}:$ From \eqref{eq: cross ratio asymp 5}, one can see that the contour integral does not grow exponentially. On the other hand, the residue at double pole $h=2$ is also a linear combination of $\widetilde{F}_{-,2}(\chi)$ and $\left.\partial_h\widetilde{F}_{-,h}(\chi)\right|_{h=2}$. First, the contribution from the former becomes
\begin{align}
    {1\over -i\chi }\left(1-{\zeta\over \chi}\right)\simeq{1\over 4}e^{{2\pi \over \beta}t} + { 1+i \over \sqrt{2} } {1\over 4}e^{{\pi \over \beta}t} \theta_1\theta_3+\cdots
\end{align}   
Note that the first term corresponds to the exponential growth of $\langle \chi^i\chi^i \chi^j\chi^j \rangle$ with the Lyapunov exponent ${2\pi \over \beta}$ while the second term comes from $\langle b^i\chi^i b^j\chi^j \rangle$ of which Lyapunov exponent is ${\pi\over \beta}$. On the other hand, the other contribution of the double pole can be evaluated by
\begin{equation}
    \partial_h\left[ (-i\chi)^{1-h}\left(1+{(1-h)\zeta\over \chi}\right)\right]_{h=2}
\end{equation}
Note that the derivative should be acted on both the exponent and the factor ${(1-h)\zeta\over \chi}$, which leads to the cancellation of the terms violating the chaos bound. Then, one can obtain
\begin{equation}
\mathcal{F}_{\singlet,\text{non-zero}}\sim te^{{2\pi\over \beta}t} + t e^{{\pi \over \beta}t} \theta_1\theta_3
\end{equation}
where we omit the numerical coefficients. Again, the first and the second term corresponds to the ${1\over \beta J}$ correction to the Lyapunov exponent ${2\pi \over \beta}$ of $\langle \chi^i\chi^i \chi^j\chi^j \rangle$ and to the Lyapunov exponent ${\pi\over \beta}$ of $\langle b^i\chi^i b^j\chi^j \rangle$, respectively.
\end{itemize}

\vspace{3mm}

\noindent
\textbf{Anti-symmetric Channel:} In the same way, one can evaulate the out-of-time-ordered correlators in the anti-symmetric channel. Using $\widetilde{F}_{+,h}$ in \eqref{eq: function p large chi}, we replace $k^B_{\anti}$ with $k_{R,\anti}$ in the sum by \eqref{eq: kr identity3} and \eqref{eq: kr identity4}, and we pull the contour around the poles to the contour $h={1\over 2}+is$. When pulling the contour, we pick up one double pole at $h=1$. Hence, we have
\begin{align}
    &{\mathcal{F}_{\anti}\over (q-1)\alpha_0}
    ={1\over 2\pi i}\int_{\mathcal{C}} dh{\widetilde{F}_{+,h}(\chi)\over 2\pi\cot {\pi h\over 2}} \left(1+{h\zeta\over \chi}\right) \left[ {k^B_\anti(h)\over 1-k^B_\anti(h)}+{{h-1+{1\over q}\over (q-1)(h-{1\over q})}k_{R,\anti}(1-h)\over 1-{h-1+{1\over q}\over (q-1)(h-{1\over q})}k_{R,\anti}(1-h)}\right]\cr
    &-{1\over 2\pi i}\int_{\mathcal{C}} dh{\widetilde{F}_{+,h}(\chi)\over 2\pi\cot {\pi h\over 2}}\left(1+{(1-h)\zeta\over \chi}\right)\left[ {k^B_\anti(1-h)\over 1-k^B_\anti(1-h)} -{{1\over q-1}k_{R,\anti}(1-h)\over 1-{1\over q-1}k_{R,\anti}(1-h)} \right]\cr
    &+ \underset{h=1}{\text{Res}}\; {\widetilde{F}_{+,h}(\chi)\over 2\pi \cot{\pi h\over 2} } \left(1+{(1-h)\zeta\over \chi}\right) {{1\over q-1}k_{R,\anti}(1-h)\over 1-{1\over q-1}k_{R,\anti}(1-h)}
\end{align}
where we used
\begin{equation}
    {2\over \tan \pi h}=-{1\over \cot {\pi h\over 2}}+{1\over \cot {\pi (1-h)\over 2}}
\end{equation}
Noting that the function $\widetilde{F}_{+,h}$ for small $\chi$ behaves as
\begin{equation}
    \widetilde{F}_{+,h}(\chi)\simeq   - i  {\Gamma\left(1-{h\over 2}\right) \Gamma\left(h-{1\over 2}\right)\over 2^{1-h}\Gamma\left({1\over 2}+{h\over 2}\right)} (-i\chi)^{h}+\left(h\longrightarrow 1-h\right)\ ,\label{eq: ftilde p small}
\end{equation}
one can immediately deduce that the contour integral does not grow exponentially because this is the almost same as that of the singlet channel. For the residue, we will also consider the two choices of the cross ratios. Because of the double pole, we have the following two contributions:
    \begin{equation}
        \widetilde{F}_{+,1}(\chi)\hspace{5mm},\hspace{5mm} \partial_h\left[\widetilde{F}_{+,h}(\chi)\left(1+{(1-h)\zeta\over \chi}\right)\right]_{h=1}
    \end{equation}
$\widetilde{F}_{+,1}(\chi)$ does not have any exponentially growing term for either $\theta_3=\theta_4=0$ or $\theta_2=\theta_4=0$. Hence, it is enough to consider the second contribution.
\begin{itemize}
    \item $\boldsymbol{\theta_3=\theta_4=0}:$ For this case, the conformal ratio $\chi$ is the same as one in the non-SUSY SYK model, and we have
\begin{equation}
    \partial_h \left[(-i\chi)^{1-h}\left(1+{(1-h)\zeta\over \chi}\right)\right]_{h=1}=-\log (-i\chi) -{\zeta\over \chi}\simeq {2\pi \over \beta}t+{1\over 2}\theta_1\theta_2+\cdots
\end{equation}    
Hence, there is only a linear growth in large $t$:
\begin{equation}
\mathcal{F}_{\anti,\text{non-zero}}\sim  t 
\end{equation}    
This is the linear growth in the anti-symmetric channel of $\langle \chi^i\chi^i \chi^j\chi^j \rangle$, which is consistent with the $\rho$ zero mode contribution in~\eqref{eq: zero mode contribution h}. Note that there is no growing term proportional to $\theta_1\theta_2$. This implies that the anti-symmetric channel of $\langle b^ib^i \chi^j\chi^j \rangle$ does not grow at large $t$, it is also consistent with the $\rho$ zero mode contribution in~\eqref{eq: zero mode contribution h}.

    \item $\boldsymbol{\theta_2=\theta_4=0}:$ Using \eqref{eq: cross ratio1 theta24} and \eqref{eq: cross ratio2 theta24}, one has
\begin{equation}
    \partial_h \left[(-i\chi)^{1-h}\left(1+{(1-h)\zeta\over \chi}\right)\right]_{h=1}\simeq {2\pi \over \beta}t -(1+i)e^{-{\pi \over \beta}t}\theta_1\theta_3+\cdots
\end{equation}
Hence, the growing term at large $t$ in this channel is 
\begin{equation}
\mathcal{F}_{\anti,\text{non-zero}}\sim t
\end{equation}    
Again, this is consistent with the $\rho$ and $k$ zero mode contributions.
    
\end{itemize}

\vspace{3mm}

\noindent
\textbf{Symmetric-traceless Channel:} The evaluation of the four point function of the symmetric-traceless channel is parallel to that of the singlet channel except for the eigenvalue of the kernel:
\begin{equation}
    k^B_\sym(h)=-{1\over q-1} k^B_\singlet(h)\ .
\end{equation}
Hence, its pole structure is different from that of the singlet channel. In particular, there is no zero mode in the symmetric-traceless channel. This also agrees with the fact that the zero modes in the effective action does not contribute to the symmetric-traceless channel. Therefore, without dropping any term in the four point function, we obtain
\begin{align}
    &{\mathcal{F}_{\sym}\over (q-1)\alpha_0}
    ={1\over 2\pi i}\int_{\mathcal{C}} dh{\widetilde{F}_{-,h}(\chi)\over 2\pi\tan {\pi h\over 2}} \left(1+{h\zeta\over \chi}\right) \left[ {k^B_\sym(h)\over 1-k^B_\sym(h)}-{k_{R,\sym}(1-h)\over 1-k_{R,\sym}(1-h)}\right]\cr
    &-{1\over 2\pi i}\int_{\mathcal{C}} dh{\widetilde{F}_{-,h}(\chi)\over 2\pi\tan {\pi h\over 2}}\left(1+{(1-h)\zeta\over \chi}\right)\left[ {k^B_\sym(1-h)\over 1-k^B_\sym(1-h)} +{{{1\over q}-h\over 1-h-{1\over q}}k_{R,\sym}(1-h)\over 1+{{1\over q}-h\over 1-h-{1\over q}}k_{R,\sym}(1-h)} \right]\cr
    &- \underset{h=h_\ast}{\text{Res}}\; {\widetilde{F}_{-,h}(\chi)\over 2\pi \tan{\pi h\over 2} }\left(1+{(1-h)\zeta\over \chi}\right) {{{1\over q}-h\over 1-h-{1\over q}}k_{R,\sym}(1-h)\over 1+{{1\over q}-h\over 1-h-{1\over q}}k_{R,\sym}(1-h)}
\end{align}
where we pick up the simple pole at $h=h_\ast$ which is a solution of
\begin{equation}
    1+{{1\over q}-h_\ast\over 1-h_\ast-{1\over q}}k_{R,\sym}(1-h_\ast)=0\hspace{10mm} ({1\over 2}<h_\ast<1)
\end{equation}
%
%
%
%
%
%
%
As in the singlet channel, the contour integrals do not contain growing term in large $t$. Since $h=h_\ast$ is a simple pole, one can consider the following two terms for the residue contribution:
\begin{equation}
    (-i\chi)^{1-h_\ast}\left(1+(1-h_\ast){\zeta\over \chi}\right) \hspace{5mm},\hspace{5mm}  (-i\chi)^{h_\ast}\left(1+(1-h_\ast){\zeta\over \chi}\right)\label{eq: st simple pole contribution}
\end{equation}
%
%
\begin{itemize}
    \item $\boldsymbol{\theta_3=\theta_4=0}:$ Since ${1\over 2}<h<1$, \eqref{eq: st simple pole contribution} is exponentially decreasing in $t$ (See \eqref{def: cross ratio chi 34} and \eqref{eq: cross ratio asymp 1} ).
    
    \item $\boldsymbol{\theta_2=\theta_4=0}:$ From \eqref{eq: cross ratio1 theta24} and \eqref{eq: cross ratio2 theta24}, we have
\begin{align}
    (-i\chi)^{h_\ast}\left(1+(1-h_\ast){\zeta\over \chi}\right)\sim& e^{-{2\pi \over \beta}h_\ast t} +i e^{-{2\pi \over \beta}\left(h_\ast-{1\over 2}\right)t}\theta_1\theta_3\\
     (-i\chi)^{1-h_\ast}\left(1+(1-h_\ast){\zeta\over \chi}\right)\sim& e^{-{2\pi\over \beta}(1-h_\ast)t}+i e^{-{2\pi\over \beta}\left({3\over 2}-h_\ast\right)t}\theta_1\theta_3
\end{align}
where we omitted the numerical factors. Since ${1\over 2}<h_\ast<1$, they are exponentially decreasing in $t$.
\end{itemize}
In both cases, there is no growth in large $t$, which is consistent with the zero mode contributions.

\vspace{3mm}

\noindent
\textbf{Summary:} We summarize the contribution of the non-zero modes to the large time behavior of the out-of-time-ordered correlators.

{
\renewcommand{\arraystretch}{2.3}
\begin{table}[H]
\centering
\begin{tabular}{c |c | c }
\multicolumn{2}{c|}{Channel}                  &\hspace{8mm} Non-zero Mode Contribution \hspace{8mm}   \\ \hline
\multirow{2}{*}{\hspace{3mm}Singlet\hspace{3mm}} & \hspace{2mm}${\scriptstyle\langle \chi^i\chi^i\chi^j\chi^j\rangle,\langle \chi^i\chi^ib^jb^j\rangle,\langle b^ib^ib^jb^j\rangle } $\hspace{2mm}  &       \hspace{6mm}$\displaystyle {t\over \beta} e^{{2\pi \over \beta}t}$\hspace{6mm}              \\ \cline{2-3} 
                         & ${\scriptstyle\langle b^i\chi^i b^j\chi^j\rangle}$ &    \hspace{6mm}  $\displaystyle {t\over \beta}  e^{{\pi \over \beta}t}$   \hspace{6mm}      \\ \hline
\multirow{2}{*}{Anti}    & ${\scriptstyle\langle \chi^i\chi^i\chi^j\chi^j\rangle } $  &       $\displaystyle{t\over \beta}$               \\ \cline{2-3} 
                         & ${\scriptstyle  \langle \chi^i\chi^ib^jb^j\rangle,\langle b^ib^ib^jb^j\rangle, \langle b^i\chi^i b^j\chi^j\rangle}$ & No Growth \\ \hline
\multicolumn{2}{c|}{Symmetric-traceless}                      & No Growth \\ \hline
\end{tabular}
\caption{Summary of the non-zero mode contribution to the out-of-time-ordered correlators at large $t$. We omitted the $SO(q)$ indices in the four point functions.}
\label{tab: summary non-zero mode}
\end{table}
}

\section{Conclusion}
\label{sec:conclusion}

In this work, we studied $\mathcal{N}=1$ supersymmetric SYK model with $\soq$ global symmetry. We showed that this model has the emergent super-reparametrization at strong coupling limit, and the $\soq$ global symmetry is enhanced to the $\hsoq$ local symmetry. Also, we demonstrated that the symmetry algebra is the semi-direct product of super-Virasoro algebra and super-Kac-Moody algebra. The emergent symmetries are spontaneously broken by the large $N$ classical solution. Furthermore, at finite coupling, the kinetic term breaks the emergent symmetries explicitly. This leads to Pseudo-Nambu-Goldstone bosons of which effective action is super-Schwarzian action plus an action of a super-particle on the $\soq$ group manifold. The bosonic zero mode of the super-Schwarzian effective action coupled to the bosonic bi-locals in the singlet channel to give the maximum Lyapunov exponent ${2\pi \over \beta}$ in the corresponding four point function. On the other hand, we showed that the fermionic zero mode of the super-Schwarzian action is coupled to fermionic bi-locals in the singlet channel, which leads to the ${\pi \over \beta}$ Lyapunov exponent in the out-of-time-ordered correlator $\langle b^i \chi^i b^j \chi^j \rangle$. Also, we saw that the bosonic zero mode from the broken $\hsoq$ symmetry gives a linear growth of the out-of-time-ordere correlator $\langle \chi^i \chi^i \chi^j \chi^j\rangle$ in the anti-symmetric channel at large time while fermionc zero mode does not contribute to any growth. We also evaluate the non-zero mode contributions, and we confirmed that they provide the ${1\over \beta J}$ correction to the zero mode contributions.

It is easy to consider $\mathcal{N}=2$ supersymmetry generalization of our model with $\soq$ global symmetry based on the $\mathcal{N}=2$ SUSY SYK model~\cite{Fu:2016vas,Yoon:2017gut} where the bi-local superfield is composed of one chiral superfield and one anti-chiral superfield and the corresponding supermatrix formulation was discussed in~\cite{Yoon:2017gut}. For the complex superfield, it would be more natural to consider $SU(q)$ global symmetry, but it is not straightforward because the matrix product of bi-locals is valid only between the superspace with the sam chirality while the flavor indices in the product should be conjugate to each other.

As pointed out in~\cite{Stanford:2017thb} for non-SUSY case, the coadjoint orbit method would be able to reproduce the low energy effective action for the broken $\hsoq$ symmetry. Furthermore, it would be interesting to study this effective action as a supermatrix model. Like the Marinari-Parisi supersymmetric matrix model~\cite{Marinari:1990jc}, one might be able to analyze it by eigenvalue distribution\footnote{I thank Robert de Mello Koch for pointing out this.}~\cite{Jevicki:1991yk,Rodrigues:1992by} (at least large $q$ limit), and it is interesting to show that this effective action is also one-loop exact.

It would be interesting to verify the large $N$, strong coupling results that we obtained in this work via exact diagonalization for sufficiently large systems. In particular in this work, we assumed that the large $N$ solution preserves supersymmetry which can then be checked via numerical methods. In fact, for the ${\mathcal N}=1$ SUSY SYK models~\cite{Fu:2016vas} checked that this was indeed the case\footnote{Their numerics also showed that the supersymmetry was broken non-perturbatively in the $1/N$ expansion. This was also consistent with the fact that Witten index for this model vanishes}.

In the maximally chaotic system, one would expect that a generic out-of-time-ordered correlator would saturate chaos bound. The linear growth in the anti-symmetric channel implies that the corresponding operators are not generic operator. Indeed, the bi-local field $\chi^i\chi^i$ in the anti-symmetric channel is related to the $\soq$ generator, and therefore, it might not be surprised that such a special operator does not grow with maximal Lyapunov exponent. On the other hand, supersymmetric SYK models including our model exhibit a new exponential growth with Lyapunov exponent ${\pi \over \beta}$. Although this does not violate the chaos bound, it is not clear how to understand this Lyapunov exponent from the usual bulk point of view. In the field theory, such a out-of-time-ordered correlated with Lyapunov exponent ${\pi \over \beta}$ does not get a contribution from the bosonic zero mode of the broken super-reparametrization because of the fermi statistics. This implies that there is an out-of-time-ordered correlator in the bulk which is not coupled to the boundary graviton but only to boundary gravitino. This type of correlators would not be captured by the previous geometrical computation by shock wave~\cite{Shenker:2013pqa}. It would be interesting to study the supersymmetric generalization of the shock wave method to reproduce the Lyapunov exponent ${\pi\over \beta}$.

\acknowledgments

We thank Antal Jevicki, Spenta Wadia, Robert de Mello Koch, Rajesh Gopakumar, Gautam Mandal, Dario Rosa and R. Loganayagam for extensive discussions. JY thank the Yukawa Institute for Theoretical Physics at Kyoto University for generous support during the course of this work, within the KIAS-YITP joint workshop 2017 ``Strings, Gravity and Cosmology'' (YITP-W-17-12). JY also thank Korea Institute for Advanced Study~(KIAS) for the hospitality during the completion of this work. JY would like to thank the organizers of the IPMU-KIAS-Kyunghee University joint workshop 2017 for giving an opportunity to present our work. We gratefully acknowledge support from International Centre for Theoretical Sciences~(ICTS), Tata institute of fundamental research, Bengaluru. We would also like to acknowledge our debt to the people of India for their steady and generous support to research in the basic sciences.

\appendix

\section{Notations and Conventions}
\label{app: notations}

\begin{table}[H]
{
\renewcommand{\arraystretch}{2.5}
\centering
\begin{tabular}{>{\centering\arraybackslash}m{10mm}|>{\centering\arraybackslash}m{140mm}}
{\tiny Matrix Product} & Definition \\ \hline
$\bistar$ &  $\displaystyle(A\bistar B)(\tau_1,\tau_2)=\int d\tau_3\; A(\tau_1,\tau_3)B(\tau_3,\tau_2)$         \\ \hline
$\mstar$ &    $\displaystyle(\maA\mstar \maB)^{\alpha_1\alpha_2}(\tau_1,\tau_2)=\sum_{\alpha_3=1}^q\int d\tau_3\; \maA^{\alpha_1\alpha_3}(\tau_1,\tau_3)\maB^{\alpha_3\alpha_2}(\tau_3,\tau_2)$        \\ \hline
$\sstar$ &  $\displaystyle(A\sstar B)(\tau_1,\theta_1;\tau_2,\theta_2)=\int  A(\tau_1,\theta_1;\tau_3,\theta_3)\; d\tau_3d\theta_3\;B(\tau_3,\theta_3;\tau_2,\theta_2)$          \\ \hline
$\msstar$ &    $\displaystyle(\maA\msstar \maB)^{\alpha_1\alpha_2}(\tau_1,\theta_1;\tau_2,\theta_2)=\sum_{\alpha_3=1}^q\int  \maA^{\alpha_1\alpha_3}(\tau_1,\theta_1;\tau_3,\theta_3)\;d\tau_3 d\theta_3\;\maB^{\alpha_3\alpha_2}(\tau_3,\theta_3;\tau_2,\theta_2)$        \\ \hline
\end{tabular}
\caption{Matrix Product}
\label{tab: matrix product}
}
\end{table}

\begin{table}[H]
{
\renewcommand{\arraystretch}{2.5}
\centering
\begin{tabular}{>{\centering\arraybackslash}m{15mm}|>{\centering\arraybackslash}m{80mm}}
Trace & Definition \\ \hline
$\tr$ &  $\displaystyle\tr (\maA)=\sum_{i=1}^q \maA^{ii}$          \\ \hline
$\bitr$ &    $\displaystyle\bitr(A)=\int d\tau\; A(\tau,\tau)$        \\ \hline
$\Tr$ &  $\displaystyle\Tr(\maA)=\sum_{i=1}^q\int d\tau\; \maA^{ii}(\tau,\tau)$          \\ \hline
$\str$ &    $\displaystyle\str(A)=\int d\tau d\theta\; A(\tau,\theta;\tau,\theta)$        \\ \hline
$\STr$ &    $\displaystyle\STr(\maA)=\sum_{i=1}^q \int d\tau d\theta\; \maA^{ii}(\tau,\theta;\tau,\theta)$     \\ \hline
\end{tabular}
\caption{Trace}
\label{tab: trace}
}
\end{table}

For any bi-local superfield, let us use the shorthand notation
\begin{equation}
    A(1,2)\equiv A(\tau_1,\theta_1;\tau_2,\theta_2)
\end{equation}
Then, the functional derivative of $S$ with respect to a bi-local superfield $A(1,2)$ is defined via 
\begin{equation}
 \delta S \equiv  \int d\mu_2 d\mu_1 \delta A(1,2) { \delta S \over \delta A(1,2) } \label{def: bilocal functional derivative}
\end{equation}
where $d\mu_i\equiv d\tau_i d\theta_i$ $(i=1,2)$. Note that the delta function $(\theta_1-\theta_2)\delta(\tau_{12})$ for the superspace is the identity matrix in the supermatrix formulation, and therefore, it is easy to see that
\begin{equation}
    A(1,2)=\int \theta_{13}\delta(\tau_{13})\; d\tau_3d\theta_3 \;A(3,2)=\int A(1,3)\; d\tau_3d\theta_3  \;\theta_{32}\delta(\tau_{32}) \label{eq: delta function superspace prop}
\end{equation}
where $\theta_{ij}\equiv \theta_i-\theta_j$. Using this, we will consider the functional derivative of (bosonic) bi-local superfields $A_-(1,2)$ and $B_+(1,2)$ which are anti-symmetric and symmetric in the bi-local superspace, respectively. \ie 
\begin{alignat}{2}
    A_-(1,2)=&-A_-(2,1)\hspace{8mm} &&(\;\mbox{\eg}\;\; \fluca_\singlet\;,\; \flucb_\sym\;)\\
    B_+(1,2)=&B_+(2,1) &&(\;\mbox{\eg}\;\;  \flucb_\anti\;)
\end{alignat}
Using \eqref{eq: delta function superspace prop}, the variation of $A_-(1,2)$ and $B_+(1,2)$ can be written as
\begin{align}
    \delta A_-(1,2)=&\int \theta_{13}\delta(\tau_{13})\; d\mu_3 \;\delta A_+(3,4)\;d\mu_4 \theta_{42}\delta(\tau_{42})\cr
    =&\int d\mu_4 d\mu_3 \delta A_+(3,4){1\over 2}\left[ \theta_{31}\theta_{42}\delta(\tau_{31})\delta(\tau_{42})-\theta_{32}\theta_{41}\delta(\tau_{31})\delta(\tau_{42})\right]\\
    \delta B_+(1,2)=&\int \theta_{13}\delta(\tau_{13})\; d\mu_3 \;\delta B_+(3,4)\;d\mu_4 \theta_{42}\delta(\tau_{42})\cr
    =&\int d\mu_4 d\mu_3 \delta B_+(3,4){1\over 2}\left[ \theta_{31}\theta_{42}\delta(\tau_{31})\delta(\tau_{42})+\theta_{32}\theta_{41}\delta(\tau_{31})\delta(\tau_{42})\right]
\end{align}
Based on our definition of the bi-local superfield in~\eqref{def: bilocal functional derivative}, one can read off the following functional derivatives:
\begin{align}
    {\delta A_-(1,2)\over \delta A_-(3,4)}={1\over 2}\left[ \theta_{31}\theta_{42}\delta(\tau_{31})\delta(\tau_{42})-\theta_{32}\theta_{41}\delta(\tau_{31})\delta(\tau_{42})\right]\\
    {\delta B_+(1,2)\over \delta B_+(3,4)}={1\over 2}\left[ \theta_{31}\theta_{42}\delta(\tau_{31})\delta(\tau_{42})+\theta_{32}\theta_{41}\delta(\tau_{31})\delta(\tau_{42})\right]
\end{align}
Also, note that the variation of $\STr \log \maPsi$ can be expressed as
\begin{equation}
    \delta [\STr \log \maPsi ]=\STr ( \maPsi^{-1}\msstar\delta \maPsi)=\int d\mu_2 \maPsi^{-1}(2,1)  d\mu_1\delta\maPsi(1,2)
\end{equation}
which leads to
\begin{equation}
    {\delta \STr \log \maPsi\over \delta \maPsi(1,2)}=\maPsi(2,1)
\end{equation}

\section{Shadow Representation}
\label{app: shadow representation}

In this appendix, we will briefly review the shadow representation for both non-SUSY and SUSY SYK models based on~\cite{Murugan:2017eto,Bulycheva:2017uqj}. We begin with the non-SUSY case first. For two decoupled CFTs, let us consider a four point function $\left\langle \mathcal{O}_1(\tau_1)\mathcal{O}_2(\tau_2)\mathcal{O}_3(\tau_3)\mathcal{O}_4(\tau_4) \right\rangle$ where $\mathcal{O}_1$ and $\mathcal{O}_2$ belong to one CFT and $\mathcal{O}_3$ and $\mathcal{O}_4$ belong to the other. Then, this correlation function is factorized into a product of two point functions. Now, let us assume that two CFTs are coupled by
\begin{equation}
    \epsilon \int dy \mathcal{V}_h(y) \mathcal{V}'_{1-h}(y) 
\end{equation}
where $\mathcal{V}_h$ and $\mathcal{V}_{1-h}$ is a primary operator in each CFT, respectively. Then, in addition to the disconnected diagram, the (connected) four point function is given by
\begin{equation}
    \left\langle \mathcal{O}_1(\tau_1)\mathcal{O}_2(\tau_2)\mathcal{O}_3(\tau_3)\mathcal{O}_4(\tau_4) \right\rangle_c =\epsilon\int dy \; \left\langle \mathcal{O}_1(\tau_1)\mathcal{O}_2(\tau_2)\mathcal{V}_h(y) \right\rangle\left\langle \mathcal{V}'_{1-h}(y)\mathcal{O}_3(\tau_3)\mathcal{O}_4(\tau_4) \right\rangle
\end{equation}
One can fix the three point functions to be
\begin{align}
    \left\langle \mathcal{O}_1(\tau_1)\mathcal{O}_2(\tau_2)\mathcal{V}_h(y) \right\rangle=&{\mathcal{M}(\tau_1,\tau_2,y)\over |\tau_1-\tau_2|^{2\Delta -h}|\tau_1-y|^{h}|\tau_2-y|^{h}}\label{eq: shadow rep three pt 1}\\
    \left\langle \mathcal{V}'_{1-h}(y)\mathcal{O}_3(\tau_3)\mathcal{O}_4(\tau_4) \right\rangle=&{\mathcal{M}(y,\tau_3,\tau_4)\over |\tau_3-\tau_4|^{2\Delta -1+h}|\tau_3-y|^{1-h}|\tau_4-y|^{1-h}}\label{eq: shadow rep three pt 2}
\end{align}
where $\mathcal{M}(\tau_1,\tau_2,\tau_3)$ is responsible for the symmetry of three point function.

For the non-SUSY SYK model, one can expect two bases for four point function: \ie anti-symmetric basis (\eg singlet and symmetric-traceless representation); symmetric basis (\eg anti-symmetric). Hence, one may write
\begin{align}
    \mathcal{M}_{-}(\tau_1,\tau_2,y)=&\sgn(\tau_{12})[a+b\;\sgn(\tau_1-y)\sgn(\tau_2-y)]\\
    \mathcal{M}_{+}(\tau_1,\tau_2,y)=&c+d\sgn(\tau_1-y)\sgn(\tau_2-y)
\end{align}
where $a,b,c,d$ are constants. In order to restrict $\mathcal{M}$ further, we will consider the symmetry of the four point functions. In particular, we will consider the normalized four point function of fermions in each channel:
\begin{equation}
    \boldsymbol{T}_{R}^{\alpha_1\alpha_2}\boldsymbol{T}_R^{\alpha_3\alpha_4}{\langle \psi_i^{\alpha_1}(\tau_1)\psi_i^{\alpha_2}(\tau_2)\psi_j^{\alpha_3}(\tau_3)\psi_j^{\alpha_4}(\tau_4) \rangle \over \langle \psi_i^{\gamma_1}(\tau_1)\psi_i^{\gamma_1}(\tau_2)\rangle\langle\psi_j^{\gamma_2}(\tau_3)\psi_j^{\gamma_2}(\tau_4) \rangle}
\end{equation}
Using $SL(2,\mathbb{R})$ symmetry, we can fix the points $\tau_1,\tau_2,\tau_3$ and $\tau_4$ to express it in term of the cross ratio $\chi={\tau_{12}\tau_{34}\over \tau_{13}\tau_{24}}$.
\begin{equation}
\tau_1=0\;\;,\;\;\tau_2=\chi\;\;,\;\;\tau_3=1\;\;,\;\;\tau_4=\infty
\end{equation}
Then, the shadow representation gives the following two bases:
\begin{align}
    \Phi_{-,h}(\chi)=&{1\over 2} \int_{-\infty}^\infty dy { |\chi|^h\mathcal{M}_-(0,\chi,y)\mathcal{M}_-(y,1,\infty) \over |y|^h |y-1|^{1-h}|y-\chi|^h }\label{def: nonsusy basis1}\\
    \Phi_{+,h}(\chi)=&{1\over 2} \int_{-\infty}^\infty dy { |\chi|^h\mathcal{M}_+(0,\chi,y)\mathcal{M}_+(y,1,\infty) \over |y|^h |y-1|^{1-h}|y-\chi|^h }\label{def: nonsusy basis2}
\end{align}
Recall that the exchagne of $t_1$ and $t_2$ corresponds to the following transformation of $\chi={\tau_{12}\tau_{34}\over \tau_{13}\tau_{24}}$:
\begin{equation}
    (\tau_1,\tau_2)\;\;\rightarrow\;\; (\tau_2,\tau_1)\qquad\Longrightarrow\qquad \chi\;\;\rightarrow \;\; {\chi\over \chi-1}
\end{equation}
and, similar for $(\tau_3,\tau_4)$. Hence, the symmetry\footnote{Due to the denominator of \eqref{def: nonsusy basis1} and \eqref{def: nonsusy basis1}, the symmetry of the basis is changed.} of the basis is given by
\begin{equation}
    \Phi_{h,\mp}(\chi)=\pm\Phi_{h,\pm}\left({\chi\over \chi-1}\right)
\end{equation}
%
%
%
%
%
%
%
%
%
%
In addition, another important symmetry of the basis is the exchange of $(\tau_1,\tau_2,\tau_3,\tau_4)\leftrightarrow (\tau_3,\tau_4,\tau_1,\tau_2)$. This symmetry corresponds to the following relation of basis.
\begin{equation}
    \Phi_{h,\mp}(\chi)=\Phi_{1-h,\mp}(\chi)
\end{equation}
%
%
%
%
%
%
From these two symmetries, one determine the basis completely:
\begin{align}
    \Phi_{h,-}(\chi)=&{1\over 2} \int_{-\infty}^\infty dy { |\chi|^h \over |y|^h |y-1|^{1-h}|y-\chi|^h }\\
    \Phi_{h, +}(\chi)=&{\sgn(\chi)\over 2} \int_{-\infty}^\infty dy { |\chi|^h\sgn(y)\sgn(y-1)\sgn(y-\chi) \over |y|^h |y-1|^{1-h}|y-\chi|^h }
\end{align}
For the SUSY SYK model, the three point function analogous to~\eqref{eq: shadow rep three pt 1} and~\eqref{eq: shadow rep three pt 2} can be written as
%
%
%
\begin{align}
    \boldsymbol{T}^R_{\alpha_1\alpha_2}\langle \psi_i^{\alpha_1}(\tau_1,\theta_1)\psi_i^{\alpha_2}(\tau_2,\theta_2)  \mathcal{V}^B_h(\tau_3,\theta_3)\rangle=& {\mathcal{M}_R(\tau_1,\tau_2,\tau_3)\over |\langle 1,2\rangle|^{2\Delta-h}|\langle 1,3\rangle|^{h}|\langle 2,3\rangle|^h}\\
    \boldsymbol{T}^R_{\alpha_1\alpha_2}\langle \psi_i^{\alpha_1}(\tau_1,\theta_1)\psi_i^{\alpha_2}(\tau_2,\theta_2)  \mathcal{V}^F_h(\tau_3,\theta_3)\rangle=&{\mathcal{M}_R(\tau_1,\tau_2,\tau_3)\sgn(\tau_{12})\over |\langle 1,2\rangle|^{2\Delta-h}|\langle 1,3\rangle|^{h}|\langle 2,3\rangle|^h}P(1,2,3)   
\end{align}
where the function $P(1,2,3)$ is defined by
\begin{equation}
    P(1,2,3)={ \theta_1(\tau_2-\tau_3)+\theta_2(\tau_3-\tau_1)+\theta_3(\tau_1-\tau_2)-2\theta_1\theta_2\theta_3 \over |\langle 1,2\rangle \langle 2,3\rangle \langle 3,1\rangle|^{1\over 2}}
\end{equation}
and the function $\mathcal{M}_R(\tau_1,\tau_2,\tau_3)$ is in general given by
\begin{align}
    \mathcal{M}_{\singlet/\sym}(\tau_1,\tau_2,\tau_3)\equiv&\sgn(\tau_{12})[a+b\sgn(\tau_{13})\sgn(\tau_{23})] \\
    \mathcal{M}_\anti(\tau_1,\tau_2,\tau_3)\equiv&a+b\sgn(\tau_{13})\sgn(\tau_{23})
\end{align}
As before, this leads to the basis for the four point functions:
%
%
\begin{align}
    \Upsilon^B_{-,h}=&{1\over 2}\int dy d\theta_y\;{ |\langle 1,2 \rangle|^h |\langle 3,4 \rangle|^{1/2-h}\sgn(\tau_{34})\widetilde{\mathcal{M}} P(3,4,y) \over |\langle 1,y \rangle|^h|\langle 2,y \rangle|^h|\langle 3,y \rangle|^{1/2-h} |\langle 4,y \rangle|^{1/2-h}}   \\
    \Upsilon^B_{+,h}=&{1\over 2}\int dy d\theta_y\;{ |\langle 1,2 \rangle|^h |\langle 3,4 \rangle|^{1/2-h}\sgn(\tau_{12})\widetilde{\mathcal{M}} P(3,4,y) \over |\langle 1,y \rangle|^h|\langle 2,y \rangle|^h|\langle 3,y \rangle|^{1/2-h} |\langle 4,y \rangle|^{1/2-h}} 
\end{align}
%
%
%
where $\Upsilon^B_{-,h}$ is the basis for the singlet and symmetric-traceless channel and $\Upsilon^B_{+,h}$ is for the anti-symmetric channel. The function $\widetilde{\mathcal{M}}$ is given by
\begin{align}
    \widetilde{\mathcal{M}}=&a_1+a_2\sgn(\tau_1-y)\sgn(\tau_2-y)+a_3\sgn(\tau_3-y)\sgn(\tau_4-y)\cr
    &+a_4\sgn(\tau_1-y)\sgn(\tau_2-y)\sgn(\tau_3-y)\sgn(\tau_4-y)
\end{align}
We want to also restrict the function $\widetilde{\mathcal{M}}$ by using the symmetry of four point function of the SUSY SYK model. Using $OSp(1|2)$, we choose
\begin{equation}
    \tau_1=0\;\;,\;\;\tau_2=\chi\;\;,\;\;\tau_3=1\;\;,\;\;\tau_4=\infty\;\;,\;\;\theta_3=0\;\;,\;\;\theta_4=0\;\;,\;\;\zeta=\theta_1\theta_2\ ,
\end{equation}
and, the basis $\upsilon^B_{\mp,h}$ is simplified to be
\begin{align}
    \Upsilon^B_{-,h}=&{1\over 2}\int dy d\theta_y\; {|\chi+\zeta|^h \widetilde{\mathcal{M}} \theta_y\over |y|^h |\chi-y|^h |1-y|^{1-h}}   \\
    \Upsilon^B_{+,h}=&{1\over 2}\int dy d\theta_y\;{|\chi+\zeta|^h \sgn(\chi)\widetilde{\mathcal{M}} \theta_y\over |y|^h |\chi-y|^h |1-y|^{1-h}}
\end{align}
where the function $\widetilde{\mathcal{M}}$ becomes
\begin{equation}
    \widetilde{\mathcal{M}}=a_1+a_2\sgn(y)\sgn(\chi-y)+a_3\sgn(1-y)+a_4\sgn(y)\sgn(\chi-y)\sgn(1-y)
\end{equation}
The symmetry of $\Upsilon^B_{\mp,h}$ discussed in~\eqref{eq: 12 exchange symmetry} leads to $a_3=a_4=0$ for $\Upsilon^B_{-,h}$ and $a_1=a_2=0$ for $\Upsilon^B_{+,h}$. However, unlike the non-SUSY SYK model, the exchange of $(\tau_1,\theta_1,\tau_2,\theta_2)$ and $(\tau_3,\theta_3,\tau_4,\theta_4)$ does not impose a constraint on the basis but gives the relation between $\Upsilon^B$ and $\Upsilon^F$. Hence, we could not determine the basis in this way. The basis $\Upsilon^B_{-,h}$ chosen in~\cite{Murugan:2017eto} for $\mathcal{N}=1$ SUSY SYK model is equivalent to the case where $a_2=a_3=a_4=0$. Hence, we take assumption\footnote{As in~\cite{Bulycheva:2017uqj}, one may claim that $\mathcal{V}_h^{B/F}$ is a composite operator of two fermions with derivative operators to restrict the form of the three point functions.} that the other basis $\Upsilon^B_{+,h}$ would correspond to $a_1=a_2=a_3=0$. Therefore, we have
\begin{align}
    \Upsilon^B_{-,h}(1,2,3,4)=&{1\over 2}\int dy d\theta_y\;{ |\langle 1,2 \rangle|^h |\langle 3,4 \rangle|^{1/2-h}\sgn(\tau_{34}) P(3,4,y) \over |\langle 1,y \rangle|^h|\langle 2,y \rangle|^h|\langle 3,y \rangle|^{1/2-h} |\langle 4,y \rangle|^{1/2-h}}\ ,\\
    \Upsilon^B_{+,h}(1,2,3,4)=&-{1\over 2}\int dy d\theta_y\;{ |\langle 1,2 \rangle|^h |\langle 3,4 \rangle|^{1/2-h}\sgn(\tau_{12}) P(3,4,y) \over |\langle 1,y \rangle|^h|\langle 2,y \rangle|^h|\langle 3,y \rangle|^{1/2-h} |\langle 4,y \rangle|^{1/2-h}}\cr
    &\hspace{12mm}\times\sgn(\tau_1-y)\sgn(\tau_2-y)\sgn(\tau_3-y)\sgn(\tau_4-y)
\end{align}

\section{Effective Action}
\label{app: Effective Action}

In this appendix, we will derive the effective action by $\epsilon$-expansion of $q$. We will consider the super-reparametrization and the $\hsoq$ local transformation of the large $N$ classical solution $\maPsi_{cl}$ in~\eqref{eq: classical solution supermatrix}
%
%
%
by
\begin{equation}
    \tau'=f(\tau,\theta)\hspace{5mm},\hspace{5mm} \theta'=y(\tau,\theta)\hspace{5mm},\hspace{5mm} \mg(\tau,\theta)\\
\end{equation}
Note that unlike~\eqref{def: super reparametrization parametrization} this parametrization of the super-reparametrization is constrained by
\begin{equation}
    \sD f= y \sD y\label{eq: constraint}
\end{equation} 
Taking 
\begin{equation}
    q={1\over 1-\epsilon}\hspace{10mm} (0<\epsilon<1)\ ,
\end{equation}
the $\epsilon$-expansion of the transformed classical solution can be written as
\begin{align}
    &\maPsi_{cl,[(f,y),\mg]}(1,2)=\Lambda\mg(1)\mg^{-1}(2) {\left[\sD_1 y_1 \sD_2 y_2\right]^{1\over q}\over | f_1-f_2-y_1y_2|^{1\over q} }\cr
    =&{1\over [(q-1)\alpha_0]^{1\over q}}\mg(1)\mg^{-1}(2){\sD_1 y_1 \sD_2 y_2\over | J(f_1-f_2-y_1y_2)| } \left[1-\epsilon\log \Omega+{1\over 2}\epsilon^2\left(\log \Omega\right)^2+\cdots\right]\label{eq: epsilon expansion of classical sol}
\end{align}
where $\alpha_0$ is defined in~\eqref{def: alpha0} and $\Omega$ is defined by
\begin{equation}
    \Omega\equiv{\sD_1 y_1 \sD_2 y_2\over |J(f_1-f_2-y_1y_2)| }
\end{equation}
where $f_i\equiv f(\tau_i,\theta_i)$ and $y_i\equiv y(\tau_i,\theta_i)$ ($i=1,2$). We will analyze $\STr(\smD\msstar\maPsi_{cl})$ of \eqref{eq: epsilon expansion of classical sol} order by order in $\epsilon$. For this, it is useful to consider the Taylor expansion of $f_1-f_2-y_1y_2$ and $\sD_1 y_1$. Using the constraint in~\eqref{eq: constraint}, we found
\begin{align}
    \sD_1 y_1=&\sum_{n=0}{1\over n!}\sD_2 \left[(\tau_{12}-\theta_1\theta_2)^n \partial_{\tau_2}^n y_2\right]\label{eq: taylor expansion1}\\
    f(\tau_1,\theta_1)=&f(\tau_2,\theta_2)+\sum_{n=1}^\infty {1\over n!} \sD_2\left[(\tau_{12}-\theta_1\theta_2)^n\partial_{\tau_2}^{n-1} \sD_2 f(\tau_2,\theta_2)\right]\label{eq: taylor expansion2}
\end{align}
In particular, \eqref{eq: taylor expansion2} can be simplified to be
\begin{align}
    &f_1-f_2-y_1y_2\cr
    =&(\tau_{12}-\theta_1\theta_2) [\sD y]^2+{1\over 2}\sD_2 \left[(\tau_{12}-\theta_1\theta_2)^2(\sD^2 y\sD y)\right]+{1\over 2}(\tau_{12}-\theta_1\theta_2)^2 \sD y \sD^3 y\cr
    &+{1\over 6}\sD_2 \left[(\tau_{12}-\theta_1\theta_2)^3(\sD y\sD^4 y+2\sD^2 y\sD^3 y)\right] +{1\over 6} (\tau_{12}-\theta_1\theta_2)^3\sD y \sD^5y+\cdots\label{eq: taylor expansion3}
\end{align}

\vspace{3mm}

\noindent
\textbf{Leading Contribution:} Let us consider the super-reparametrization for now. Using \eqref{eq: taylor expansion1} and \eqref{eq: taylor expansion3}, we have an expansion in $L$
\begin{align}
    &{\sD_1 y_1 \sD y_2\over f_1-f_2-y_1y_2}\cr
    =&{1\over L} +  {1\over 2} \theta_{12}{  \sD_2^4 y_2\over  \sD_2 y_2}- \theta_{12}{ \sD_2^2 y_2 \sD_2^3 y_2\over   [\sD_2 y_2]^2}+{1\over 6}L{ \sD_2^5 y_2\over  \sD_2 y_2} + {1\over 6}L{\sD_2^2 y_2\sD_2^4y_2\over  [\sD_2 y_2]^2}-{1\over 3}L{  \sD_2^3 y_2 \sD_2^3 y_2\over  [\sD_2 y_2]^2}+\cdots\label{eq: epsilon expansion leading}
\end{align}
where $L$ is defined by
\begin{equation}
    L\equiv \tau_{12}-\theta_1\theta_2
\end{equation}
To evaluate $\STr(\smD\msstar\maPsi_{cl})$, it is convenient to use supermatrix notation. Recall that a Grassmannian even bi-local superfield $\maA$ can be represented by the Grassmannian odd supermatrix:
\begin{equation}
    \maA=\begin{pmatrix}
    A_1 & A_3\\
    A_0 & A_2\\
    \end{pmatrix}
\end{equation}
Since the super-derivative matrix $\smD$ in~\eqref{def:bi-local superderivative} is a Grassmannian odd supermatrix, $\smD \msstar\maA$ is Grassmannian even supermatrix. Hence, its super-trace becomes
\begin{align}
     &\STr \smD\msstar \maA = \Tr[\partial_{\tau_1}A_0 - A_3]
\end{align}
From \eqref{eq: epsilon expansion leading}, one can read off the components of its supermatrix representation. In particular, we are interested in the following two components:
\begin{align}
    \left[{\sD_1 y_1 \sD y_2\over f_1-f_2-y_1y_2}\right]_0=&{1\over \tau_{12}}+\tau_{12}F(\tau_2)+\cdots\cr
    \left[{\sD_1 y_1 \sD y_2\over f_1-f_2-y_1y_2}\right]_3=&-{1\over \tau_{12}^2} - {1\over 2}\left[ {\sD_2^4 y_2\over  \sD_2 y_2}- 2{ \sD_2^2 y_2 \sD_2^3 y_2\over   [\sD_2 y_2]^2}\right]_2+F(\tau_2)+\cdots
\end{align}
where the ellipsis represents vanishing terms in the limit $\tau_{12}\rightarrow 0$. Then, we have
\begin{align}
     \STr \left[\sD{\sD_1 y_1 \sD y_2\over f_1-f_2-y_1y_2}\right]=&{1\over 2}\int d\tau \; \left[ {\sD^4 y\over  \sD y}- 2{ \sD^2 y \sD^3 y\over   [\sD y]^2}\right]_2 ={1\over 2}\int d\tau d\theta \;\left[ {\sD^4 y\over  \sD y}- 2{ \sD^2 y \sD^3 y\over   [\sD y]^2}\right]
\end{align}
For the $\hsoq$ transformation, we consider the following expansion:
\begin{align}
    { \mg(\tau_1,\theta_1)\mg^{-1}(\tau_2,\theta_2)  \over \tau_1-\tau_2-\theta_1\theta_2}
    =&\left[{1\over L  }\idm +{\theta_{12} \sD_2 \mg_2\cdot \mg_2^{-1}\over L}+ \sD_2^2 \mg_2\cdot\mg_2^{-1}+\theta_{12} \sD_2^3 \mg_2 \cdot \mg_2^{-1}\right.\cr
    &\left.+{1\over 2} L \sD_2^4 \mg_2 \cdot \mg_2^{-1}+{1\over 2}\theta_{12}L  \sD_2^4 \mg_2 \cdot \mg_2^{-1}+\cdots\right]\ .
\end{align}
From this expansion, one can easily read off the components 
\begin{align}
    \left[{ \mg(\tau_1,\theta_1)\mg^{-1}(\tau_2,\theta_2)  \over \tau_1-\tau_2-\theta_1\theta_2}\right]_0=&{1\over \tau_{12}}\idm +[\sD_2^2\mg_2\cdot \mg_2^{-1}]_1 +\tau_{12}\boldsymbol{F}'(\tau_2) +\cdots\cr
        \left[{ \mg(\tau_1,\theta_1)\mg^{-1}(\tau_2,\theta_2)  \over \tau_1-\tau_2-\theta_1\theta_2}\right]_3=&-{1\over \tau_{12}^2}\idm - {1\over \tau_{12}}[\sD_2\mg_2 \cdot \mg^{-1}_2]_2-\left[\sD_2^3\mg_2\cdot \mg_2^{-1}\right]_2+\boldsymbol{F}'(\tau_2)+\cdots
\end{align}
where the ellipsis denotes vanishing terms in the limit $\tau_{12}\rightarrow 0$. Then, the super-trace is given by
\begin{equation}
     \STr\left[\smD{ \mg(\tau_1,\theta_1)\mg^{-1}(\tau_2,\theta_2)  \over \tau_1-\tau_2-\theta_1\theta_2}\right]=\int d\tau d\theta \; \Tr[\sD_2^3\mg_2\cdot \mg_2^{-1}]
\end{equation}
In general, the super-reparametrization and the $\hsoq$ local transformation leads to 
\begin{align}
    &{\sD_1 y_1 \sD y_2\over f_1-f_2-y_1y_2}\mg(\tau_1,\theta_1)\mg^{-1}(\tau_2,\theta_2)\cr
    =&\left[{1\over L} +  {1\over 2} \theta_{12}{  \sD_2^4 y_2\over  \sD_2 y_2}- \theta_{12}{ \sD_2^2 y_2 \sD_2^3 y_2\over   [\sD_2 y_2]^2}+\cdots\right]\idm+\theta_{12}\left[{1\over L} +\cdots\right]\sD_2\mg_2\cdot \mg_2^{-1}+\left[1+\cdots\right]\sD_2^2 \mg_2\cdot\mg_2^{-1}\cr
    &+\theta_{12}\left[1 +\cdots\right] \sD_2^3 \mg_2 \cdot \mg_2^{-1}+{1\over 2}\left[ L +\cdots\right]\sD_2^4 \mg_2 \cdot \mg_2^{-1}
\end{align}
In the same way, we have
\begin{align}
    &\STr\left[\smD\msstar\left({\sD_1 y_1 \sD y_2\over f_1-f_2-y_1y_2}\mg(\tau_1,\theta_1)\mg^{-1}(\tau_2,\theta_2)\right)\right]\cr
    =&\int d\tau d\theta \; \left[{1\over 2}\left({\sD^4 y\over  \sD y}- 2{ \sD^2 y \sD^3 y\over   [\sD y]^2}\right)\tr \idm + \Tr [\sD_2^3\mg_2\cdot \mg_2^{-1}] \right]
\end{align}

\vspace{3mm}

\noindent
\textbf{Sub-leading Contribution:} Now, we will consider the sub-leading contribution of~\eqref{eq: epsilon expansion of classical sol}. The expansion of $\log \Omega$ in $L$ is given by
%
%
\begin{align}
    &\log \left({\sD_1 y_1 \sD y_2\over f_1-f_2-y_1y_2}\right)=-\log L +\log \left[1+{1\over 2} \theta_{12}L {  \sD_2^4 y_2\over  \sD_2 y_2}- \theta_{12}L{ \sD_2^2 y_2 \sD_2^3 y_2\over   [\sD_2 y_2]^2}+\cdots\right]\cr
    =&-\log L +{1\over 2} \theta_{12}L\left( {  \sD_2^4 y_2\over  \sD_2 y_2}- 2{ \sD_2^2 y_2 \sD_2^3 y_2\over   [\sD_2 y_2]^2}\right)+\cdots
\end{align}
Together with \eqref{eq: epsilon expansion leading} and the expansion $\mg_1 \mg_2^{-1}$, the expansion of the sub-leading contribution becomes
%
%
%
\begin{align}
    &{\sD_1 y_1 \sD y_2\over f_1-f_2-y_1y_2}\log \left({\sD_1 y_1 \sD y_2\over f_1-f_2-y_1y_2}\right)\mg(\tau_1,\theta_1)\mg^{-1}(\tau_2,\theta_2)\cr
    =&\left[-{\log L \over L}+{1\over 2} \theta_{12}\left( {  \sD_2^4 y_2\over  \sD_2 y_2}- 2{ \sD_2^2 y_2 \sD_2^3 y_2\over   [\sD_2 y_2]^2}\right)-{1\over 2}\theta_{12}\log L \left( {  \sD_2^4 y_2\over  \sD_2 y_2}- 2{ \sD_2^2 y_2 \sD_2^3 y_2\over   [\sD_2 y_2]^2}\right)+\cdots\right]\idm\cr
    &-{\log L \over L}\theta_{12} \sD_2 \mg_2\cdot \mg_2^{-1}-\theta_{12}\log L \sD_2^3\mg_2\cdot \mg_2^{-1}+\cdots
\end{align}
%
%
%
%
Reading off the components which are necessary in evaluating the effective action
\begin{align}
    &\left[{\sD_1 y_1 \sD y_2\over f_1-f_2-y_1y_2}\log \left({\sD_1 y_1 \sD y_2\over f_1-f_2-y_1y_2}\right)\right]_0\cr
    =&{\log \tau_{12}\over \tau_{12}}+\cdots\ ,\\
    &\left[{\sD_1 y_1 \sD y_2\over f_1-f_2-y_1y_2}\log \left({\sD_1 y_1 \sD y_2\over f_1-f_2-y_1y_2}\right)\right]_3\cr
    =&-{-1+\log \tau_{12} \over \tau_{12}^2}-{1\over 2} \left[ {  \sD_2^4 y_2\over  \sD_2 y_2}- 2{ \sD_2^2 y_2 \sD_2^3 y_2\over   [\sD_2 y_2]^2}\right]_2+{ \log\tau_{12}\over 2} \left[ {  \sD_2^4 y_2\over  \sD_2 y_2}- 2{ \sD_2^2 y_2 \sD_2^3 y_2\over   [\sD_2 y_2]^2}\right]_2+\cdots\ ,
\end{align}
we have
\begin{align}
    &-\epsilon \sD_1 \left[\smD\msstar\left\{{\sD_1 y_1 \sD y_2\over f_1-f_2-y_1y_2}\log \left({\sD_1 y_1 \sD y_2\over f_1-f_2-y_1y_2}\right)\right\}\right]\cr
    =&-\epsilon \int d\tau\; {1\over 2} \left[ {  \sD^4 y\over  \sD y}- 2{ \sD^2 y \sD^3 y\over   [\sD y]^2}\right]_2=-\epsilon \int d\tau d\theta\; {1\over 2} \left[ {  \sD^4 y\over  \sD y}- 2{ \sD^2 y \sD^3 y\over   [\sD y]^2}\right]
\end{align}

\vspace{3mm}

\noindent
\textbf{Sub-Sub-leading Contribution:} In the same way, we found
%
%
%
%
%
\begin{align}
    &{\sD_1 y_1 \sD y_2\over f_1-f_2-y_1y_2}\left[\log \left({\sD_1 y_1 \sD y_2\over f_1-f_2-y_1y_2}\right)\right]^2\mg(\tau_1,\theta_1)\mg^{-1}(\tau_2,\theta_2)\cr
    =&\left[{(-\log L)^2 \over L} -\theta_{12}\log L \left( {  \sD_2^4 y_2\over  \sD_2 y_2}- 2{ \sD_2^2 y_2 \sD_2^3 y_2\over   [\sD_2 y_2]^2}\right) +{1\over 2}\theta_{12}(-\log L)^2 \left( {  \sD_2^4 y_2\over  \sD_2 y_2}- 2{ \sD_2^2 y_2 \sD_2^3 y_2\over   [\sD_2 y_2]^2}\right)+\cdots\right]\idm\cr
    &+{(-\log L)^2 \over L}\theta_{12} \sD_2 \mg_2\cdot \mg_2^{-1}+L(-\log L)^2 \sD_2^2\mg_2\cdot \mg_2^{-1}+\theta_{12}(-\log L)^2 \sD_2^3\mg_2\cdot \mg_2^{-1}+\cdots
\end{align}
and the contribution to the effective action vanishes in the limit $\tau_{12}\rightarrow 0$:
\begin{equation}
    {1\over 2}(-\epsilon)^2\STr\left[ \smD \msstar \left({\sD_1 y_1 \sD y_2\over f_1-f_2-y_1y_2}\mg_1\mg_2^{-1}\left[\log \left({\sD_1 y_1 \sD y_2\over f_1-f_2-y_1y_2}\right)\right]^2\right)\right]=0
\end{equation}

\vspace{3mm}

\noindent
\textbf{Vanishing Divergence in the $\epsilon$-expansion:} In the higher order in $\epsilon$, one can easily see that the most terms vanish in the limit $\tau_{12}$ because they are proportional to $\tau_{12}^n(\log \tau_{12})^m$ ($n,m>0$). However, there are terms proportional to $(\log\tau_{12})^n$ ($m>0$), and it is not immediate to see whether they vanish or not until we evaluate the super-trace. Hence, we will collect this type of terms in all orders in $\epsilon$. First, there are such diverging terms proportional to $\sD_2 \mg_2\cdot \mg_2^{-1}$:
\begin{equation}
    \left({\sD_1 y_1 \sD y_2\over f_1-f_2-y_1y_2}\right)^{1\over q}\mg(\tau_1,\theta_1)\mg^{-1}(\tau_2,\theta_2)
    ={1\over L}\theta_{12}\sD_2 \mg_2\cdot \mg_2^{-1}\sum_{n=1}^\infty{(-\epsilon)^n\over n!}\left(-\log \tau_{12} \right)^n+\cdots
\end{equation}
which vanish because
\begin{equation}
    \tr \sD_2 \mg_2\cdot \mg_2^{-1}
    =0
\end{equation}
In addition, the rest of the divergent terms are summed up to be
%
%
%
%
\begin{align}
    &\left({\sD_1 y_1 \sD y_2\over f_1-f_2-y_1y_2}\right)^{1\over q}\mg(\tau_1,\theta_1)\mg^{-1}(\tau_2,\theta_2)\cr
    \Longrightarrow\hspace{3mm}&\left[{1\over 2} \theta_{12}\left( {  \sD_2^4 y_2\over  \sD_2 y_2}- 2{ \sD_2^2 y_2 \sD_2^3 y_2\over   [\sD_2 y_2]^2}\right)\idm+ \theta_{12}\sD_2^3\mg_2\cdot \mg_2^{-1}\right]e^{\epsilon\log \tau_{12} }+\cdots\cr
    &-\epsilon e^{\epsilon \log \tau_{12}} {1\over 2}\theta_{12}\left( {  \sD_2^4 y_2\over  \sD_2 y_2}- 2{ \sD_2^2 y_2 \sD_2^3 y_2\over   [\sD_2 y_2]^2}\right)\idm
\end{align}
Hence, only one component of the supermatrix gives a contribution to the super-trace (with super-derivative):
\begin{align}
    &\left[\left({\sD_1 y_1 \sD y_2\over f_1-f_2-y_1y_2}\right)^{1\over q}\mg(\tau_1,\theta_1)\mg^{-1}(\tau_2,\theta_2)\right]_3\cr
    =&-\left({1\over 2}\left[ {  \sD_2^4 y_2\over  \sD_2 y_2}- 2{ \sD_2^2 y_2 \sD_2^3 y_2\over   [\sD_2 y_2]^2}\right]_2 \idm +\left[\sD_2^3\mg_2\cdot \mg_2^{-1}\right]_2 \right)e^{\epsilon\log \tau_{12} }\cr
    &+{1\over 2}\left[ {  \sD_2^4 y_2\over  \sD_2 y_2}- 2{ \sD_2^2 y_2 \sD_2^3 y_2\over   [\sD_2 y_2]^2}\right]_2 \idm \epsilon e^{\epsilon\log \tau_{12} }
\end{align}
and, it vanishes in the limit $\tau_{12}\rightarrow 0$ because $\epsilon>0$:
\begin{align}
    &-\lim_{\tau_1\rightarrow \tau_2}\left({1\over 2}\left[ {  \sD_2^4 y_2\over  \sD_2 y_2}- 2{ \sD_2^2 y_2 \sD_2^3 y_2\over   [\sD_2 y_2]^2}\right]_2 \idm +\left[\sD_2^3\mg_2\cdot \mg_2^{-1}\right]_2 \right)e^{\epsilon\log \tau_{12} }=0\\
    &\lim_{\tau_1\rightarrow \tau_2}{1\over 2}\left[ {  \sD_2^4 y_2\over  \sD_2 y_2}- 2{ \sD_2^2 y_2 \sD_2^3 y_2\over   [\sD_2 y_2]^2}\right]_2 \idm \epsilon e^{\epsilon\log \tau_{12} }=0    
\end{align}

\section{Zero Mode Eigenfunctions}
\label{app: zero modes}

In Section~\ref{sec: leading contribution zero modes}, we evaluate the variation of the classical solution at finite temperature with respect to the infinitesimal super-reparametrization and the infinitesimal $\hsoq$ local transformation. In this analysis, we have not specify the form of each zero mode eigenfunction because the variation of the classical solution is enough to evaluate the contribution to the leading Lyapunov. In fact, the variation of the classical solution includes the zero mode eigenfunctions, and we present them with their normalization. From \eqref{eq: variation of classical solution zero mode1}, \eqref{eq: variation of classical solution zero mode2}, \eqref{eq: variation of classical solution zero mode3} and \eqref{eq: variation of classical solution zero mode4}, one can read off the zero mode eigenfunction $V_n^{(s)}$ ($s=2,{3\over 2},1, {1\over 2})$: 
\begin{itemize}
    \item \textbf{Bosonic Part of Super-reparametrization:}
\begin{equation}
    V_n^{(2)}\equiv e^{-in{\varphi_1+\varphi_2\over 2}}\left(1+{{1\over 2}\theta_1\theta_2 \over \sin\left({\varphi_1-\varphi_2\over 2}\right)}\right)\left[-n \cos \left(n{\varphi_1-\varphi_2\over 2}\right) +{\sin \left(n{\varphi_1-\varphi_2\over 2}\right) \over \tan \left({\varphi_1-\varphi_2\over 2}\right)}\right]\ .
\end{equation}

    \item \textbf{Fermionic Part of Super-reparametrization:}
\begin{align}
    V_n^{(3/2)}\equiv& e^{-in{\varphi_1+\varphi_2\over 2}}\left[ {\theta_1+\theta_2\over 2}\left(\cot{\varphi_{12}\over 4}\sin {n\varphi_{12}\over 2}-2n\cos {n\varphi_{12}\over 2} \right)\right.\cr
    &\hspace{20mm}\left.-  {\theta_1-\theta_2\over 2} i\left(\tan{\varphi_{12}\over 4}\cos {n\varphi_{12}\over 2}-2n\sin {n\varphi_{12}\over 2} \right)  \right]\ .
\end{align}

\item \textbf{Bosonic Part of $\hsoq$ Local Symmetry:}
\begin{equation}
    V_n^{(1)}\equiv e^{-in{\varphi_1+\varphi_2\over 2}}\sin \left(n{\varphi_1-\varphi_2\over 2}\right)\ .
\end{equation}

\item \textbf{Fermionic Part of $\hsoq$ Local Symmetry:}
\begin{equation}
    V_n^{(1/2)}\equiv e^{-in{\varphi_1+\varphi_2\over 2}}\left[\cos\left(n{\varphi_1-\varphi_2\over 2}\right){\theta_1-\theta_2\over 2}-i\sin\left(n{\varphi_1-\varphi_2\over 2}\right){\theta_1+\theta_2\over 2}\right] \ .
\end{equation}
\end{itemize}
Defining the inner product by
\begin{equation}
    \langle V^{(s)}_n,V^{(s')}_m \rangle\equiv \int {d\varphi_1d\varphi_2 d\theta_1 d\theta_2\over  \sin\left({\varphi_1-\varphi_2\over 2}\right)-{1\over 2}\theta_1\theta_2 } V^{(s)}_n(1,2)V^{(s')}_m(1,2)\ ,
\end{equation}
the inner product of the zero mode eigenfunctions are given by
\begin{align}
    \langle V^{(2)}_n,V^{(2)}_m\rangle=&2\pi^2 |n|(n^2-1)\delta_{n+m,0}\\
    \langle V^{({3\over 2})}_n,V^{({3\over 2})}_m\rangle=&-8\pi^2i \sgn(n) \left(n^2-{1\over 4}\right)\delta_{n+m,0}\\
    \langle V^{(1)}_n,V^{(1)}_m\rangle=&2\pi^2|n|\delta_{n+m,0}\\
    \langle V^{({1\over 2})}_n,V^{({1\over 2})}_m\rangle=&2\pi^2i \sgn(n) \delta_{n+m,0}
\end{align}
where others vanish.
\bibliographystyle{JHEP}
\bibliography{susycoloredsyk}

\end{document}